\documentclass[aps,physrev,reprint,superscriptaddress]{revtex4-2}

\usepackage{CJKutf8}
\usepackage[ruled]{algorithm2e}
\usepackage{bm}
\usepackage{amsmath} 
\usepackage{capt-of}
\usepackage{mathrsfs}
\usepackage{comment}
\usepackage{xcolor}
\usepackage[
    colorlinks=true,
    linkcolor={blue},
    citecolor={blue},
    urlcolor={blue},
]{hyperref}
\makeatletter
\renewcommand\NAT@def@citea{\def\@citea{\NAT@separator}}
\renewcommand\NAT@def@citea@space{\def\@citea{\NAT@separator}}
\renewcommand\NAT@def@citea@close{\def\@citea{\NAT@@close\NAT@separator}}
\renewcommand\NAT@def@citea@box{\def\@citea{\NAT@mbox{\NAT@@close}\NAT@separator}}
\makeatother
\usepackage{orcidlink}
\usepackage{graphicx}
\newcommand{\im}{\mathrm{i}}
\newcommand{\Cx}{C^x}
\newcommand{\Cphi}{C^\phi}
\newcommand{\diff}{\mathrm{d}}
\newcommand{\chix}{\chi^x}
\newcommand{\chiphi}{\chi^\phi}
\newcommand{\taurec}{\tau^{\mathrm{rec}}}
\newcommand{\taucon}{\tau^{\mathrm{con}}}
\newcommand{\taudiv}{\tau^{\mathrm{div}}}
\newcommand{\tauchn}{\tau^{\mathrm{chn}}}
\newcommand{\llangle}{\left\langle \mkern-4mu \left\langle}
\newcommand{\rrangle}{\right\rangle \mkern-4mu \right\rangle}
\newcommand{\calO}{\mathcal{O}}
\newcommand{\calD}{\mathcal{D}}
\newcommand{\calC}{\mathcal{C}}
\newcommand{\rI}{(1)}
\newcommand{\rII}{(2)}
\newcommand{\avgAlpha}[1]{\left\langle #1 \right\rangle_\alpha}
\newcommand{\refpanel}[2]{Fig.~\hyperref[#1]{\ref*{#1}(#2)}}
\makeatletter
\newcommand{\refpanels}[2]{%
  Fig.~%
  \def\refpanelsep{}%
  \@for\refpanelitem:=#2\do{%
    \refpanelsep
    \hyperref[#1]{\ref*{#1}(\refpanelitem)}%
    \def\refpanelsep{, }%
  }%
}
\makeatother
\begin{document}

\title{Nonlinear dynamics of random neural networks with second-order synaptic motifs}

\author{Jun Yang (\begin{CJK*}{UTF8}{gkai}杨骏\end{CJK*})\,\orcidlink{0000-0002-2484-2494}}
\email{junkyang@gatech.edu}
\affiliation{Interdisciplinary Graduate Program in Quantitative Biosciences, Georgia Institute of Technology, Atlanta, Georgia 30332, USA}
\affiliation{School of Mathematics, Georgia Institute of Technology, Atlanta, Georgia 30332, USA}

\author{Hannah Choi\,\orcidlink{0000-0002-8192-1121}}
\email{hannahch@gatech.edu}
\affiliation{School of Mathematics, Georgia Institute of Technology, Atlanta, Georgia 30332, USA}

\date{September 12, 2026}

\begin{abstract}
Classical theories of random neural networks typically assume independent connectivity, overlooking the local motif structures prevalent in biological circuits. Here, we investigate how four second-order synaptic motifs---chain, reciprocal, convergent, and divergent---shape the dynamics of nonlinear firing-rate networks. While previous studies have established that chain correlations generate outlier eigenvalues, we demonstrate that these motifs also jointly reshape the Jacobian eigenvalue bulk. Using the path-integral formalism, we derive a dynamic mean-field theory which reveals that the chain motif acts as a retarded feedback of the ensemble-mean activity through the response kernel, producing a rich repertoire of dynamical regimes, including ferromagnetic states and limit cycles. At sufficiently large magnitude, negative chain correlations produce a glassy, multistable regime that was previously mainly associated with partially symmetric networks. Our theory also distinguishes convergent from divergent motifs: divergent correlations primarily rescale temporal noise, while convergent correlations suppress temporal chaos by converting nonzero mean activity into quenched heterogeneity. Finally, analyses of the Lyapunov spectrum and participation-ratio dimension show that motif structure changes the geometry of chaotic activity, reducing entropy production and attractor dimensionality even when the effective spectral edge is held fixed. Together, these findings establish second-order motifs as a fundamental structural mechanism governing the dynamical regimes of local cortical circuits.
\end{abstract}

\maketitle

\section{Introduction}

Real-world networks are rarely fully random, and the brain is no exception. A common way to characterize such non-random structure is to quantify the frequency of specific connectivity patterns, or network motifs \cite{RN402}. Since the early 2000s, systematic over- or under-representation of motifs has been reported in synaptic connectivity across species, including \textit{C. elegans} \cite{RN407,RN408,RN410,RN411}, \textit{Drosophila} \cite{RN98,RN413}, zebrafish \cite{RN409,RN412}, mouse \cite{RN383,RN127,RN232,RN384,RN373,RN404,RN429}, rat \cite{RN91,RN92,RN385,RN305}, and human \cite{RN326}. Related non-random motif statistics have also been observed in functional connectivity inferred from neural activity \cite{RN3,RN19,RN15}.

Random neural networks are foundational models for neural dynamics and provide a theoretical basis for machine-learning frameworks such as reservoir computing (RC) \cite{RN416,RN418,RN417}. Classical models often assume dense i.i.d. Gaussian connectivity or sparse Erd\H{o}s--R\'enyi graphs \cite{RN294,RN312,RN131}. Recent work has shown, however, that introducing motif structures into these networks can strongly reshape the dynamics. Among motif classes, reciprocal correlations [\refpanel{fig:Eigenspectrum}{a}] have received the most theoretical attention \cite{RN40,RN266,RN277,RN275,RN372,RN358,RN202,RN278,RN270,RN423,RN426}, whereas the other three two-edge motifs---chain, convergent, and divergent correlations [\refpanel{fig:Eigenspectrum}{a}]---remain much less understood.

Some progress has been made for the remaining three second-order motifs. In addition to numerical evidence that these motifs can influence synchronization \cite{RN365,RN339}, a series of studies have developed analytical theories linking motif statistics to correlations, response properties, and activity dimensionality in spiking networks reducible to effective linear-filter descriptions \cite{RN9,RN419,RN25,RN26}. Other studies that incorporated all four second-order motifs have only focused on linear firing-rate networks \cite{RN232,RN271}. These approaches do not address how motif correlations interact with nonlinear recurrence, where the same structural perturbations can induce symmetry breaking, oscillations, chaos, and multistability. An analytical theory describing how all four second-order motifs and nonlinearity jointly shape the dynamics in a neural network model has not yet been established. 

To bridge this gap, we extend the classical random recurrent neural network \cite{RN294} by incorporating chain, reciprocal, convergent, and divergent motifs as second-order correlations in the coupling statistics. Previous work has shown that chain and reciprocal correlations can generate outlier eigenvalues in the eigenspectrum of the connectivity matrix \cite{RN232,RN316,RN271}. Here, we show that these motifs also reshape the eigenvalue bulk of the Jacobian, and, through nonlinear dynamics, generate regimes absent from the classical i.i.d. model. Using the dynamic mean-field theory (DMFT), we further show that positive chain correlations renormalize the effective mean coupling in a stationary state, breaking up-down symmetry to produce ferromagnetic states. This indicates that local connectivity structure, not population activity alone, can participate in excitatory-inhibitory balance. Negative chain correlations can instead destabilize a complex-conjugate outlier pair and generate limit cycles (LCs). When they become strong enough, they modulate the eigenvalue bulk like positive reciprocal correlations, slowing the dynamics and producing a glassy regime accompanied by a proliferation of fixed points.

Beyond the single-site mean-field description, Lyapunov-spectrum-based measures show that motif structure also reshapes the collective geometry of chaos. Negative chain and positive reciprocal correlations suppress chaos, while convergent and divergent motifs reduce attractor dimensionality even when the single-site mean-field dynamics remain unchanged. We provide a theoretical explanation for this dimensionality reduction by computing the participation ratio of the covariance spectrum, which shows that three-neuron motif correlations increase off-diagonal activity covariance. Altogether, our results demonstrate that second-order motifs and single-neuron nonlinearity jointly shape neural dynamics at the mean-field level and also exert effects beyond the single-site level on the collective network activity.

\section{Network model}

\subsection{Dynamics and connectivity}
\begin{figure}[htb]
\includegraphics{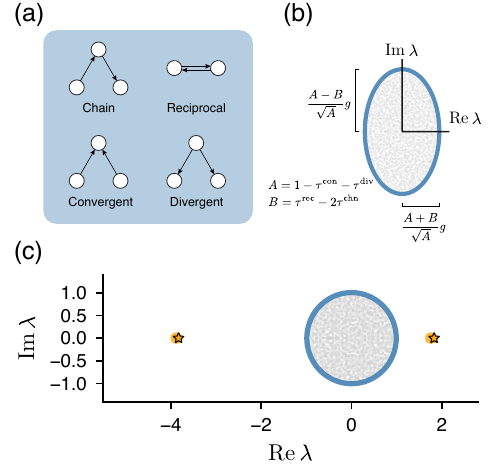}
\caption{\label{fig:Eigenspectrum}Second-order synaptic motifs and the eigenspectrum of the connectivity matrix.
(a) Schematic of the four second-order motifs.
(b) The eigenvalue bulk forms an ellipse whose size and aspect ratio are determined by the gain $g$ and motif strengths. Parameters: $N=2000$, $J_0=0$, $\taucon=\taudiv=\tauchn=0.1$, $\taurec=0$ and $g=1$.
(c) Positive chain correlations can additionally generate a pair of real outlier eigenvalues outside the bulk. Orange dots indicate the outliers obtained numerically from the sampled connectivity matrix, and stars mark the corresponding theoretical predictions. Parameters: $N=2000$, $J_0=-2$, $N\tauchn=7$, $\taucon=\taudiv=\tauchn$, $\taurec=0$, and $g=1$. In (b)\&(c), gray dots show eigenvalues from one numerical realization, the blue curve shows the theoretical bulk boundary.}
\end{figure}

We consider the Sompolinsky--Crisanti--Sommers (SCS) model \cite{RN294}
\begin{equation}\label{eq:SCSDynamics}
\dot{x}_i = -x_i + \sum_{j=1}^N J_{ij} \phi(x_j),
\end{equation}
where $x_i$ represents the input current to neuron $i$ and $\phi$ is a nonlinear activation function. Unless otherwise specified, we focus on odd-symmetric functions. In numerical simulations, we choose $\phi=\tanh$. In the original SCS model, the synaptic coupling strengths $J_{ij}\sim \mathcal{N}(0,g^2/N)$ are i.i.d. Gaussian random variables. We generalize this connectivity by incorporating a nonzero mean and four second-order motif correlations. We assume the marginal statistics of the connectivity satisfy 
\begin{equation}
\left\langle J_{ij} \right\rangle= \frac{J_0}{N}, \qquad \llangle J_{ij}^2\rrangle = \left\langle J_{ij}^2 \right\rangle
-\left\langle J_{ij} \right\rangle^2
=\frac{g^2}{N},
\end{equation}
where $\langle \cdot \rangle$ denotes the disorder average and $\llangle \cdot \rrangle$ denotes the cumulant. The strengths of the second-order motifs are parameterized by four Pearson correlation coefficients. 
For distinct neuron indices $i,j,k$, we define
\begin{align*}
\tauchn &:= \rho\!\left(J_{ij},J_{jk}\right),
&\taurec &:= \rho\!\left(J_{ij},J_{ji}\right),\\
\taucon &:= \rho\!\left(J_{ij},J_{ik}\right),
&\taudiv &:= \rho\!\left(J_{ik},J_{jk}\right).
\end{align*}
These correlation coefficients are subject to positivity constraints \cite{RN25, RN316}; in particular, realizing a nonzero chain correlation requires $|\tauchn|\le\sqrt{\taucon\taudiv}$, so convergent and divergent correlations cannot both vanish (see Appendix~\ref{app:NetworkGeneration} for network generation). When we study the effect of chain correlations, we therefore use the minimal symmetric choice $\taucon=\taudiv=|\tauchn|$, which saturates this bound. We assume that the connectivity entries are jointly Gaussian. 

This generalized Gaussian ensemble can be viewed as a dense approximation to
sparse second-order networks (SONETs) \cite{RN365} with matched first- and
second-order connectivity statistics \cite{RN271}. In SONETs, connectivity is
specified by connection probabilities: a baseline connection
probability $p$, together with joint occurrence probabilities for specific
motifs, such as $p^{\mathrm{chn}}$ for chain connections. In the Gaussian
ensemble, these discrete motif probabilities are replaced by Pearson
correlation coefficients that measure the over- or under-representation of the
corresponding local structures relative to an independent Erd\H{o}s--R\'enyi
graph. For example, the chain correlation is related to the sparse motif
probability by
\begin{equation}
\tauchn = \frac{p^{\mathrm{chn}}-p^2}{p(1-p)} .
\end{equation}
When the number of inputs contributing to each recurrent current $\sum_{ij}J_{ij}\phi(x_j)$ is large, so
that a central-limit approximation is appropriate, the resulting Gaussian
ensemble is expected to capture the macroscopic dynamical effects of these
second-order topological features. 

\subsection{Eigenspectrum of the connectivity matrix}
To gain some intuition about the network dynamics, we first examine the eigenspectrum of the connectivity matrix. Linearizing Eq.~\eqref{eq:SCSDynamics} about the trivial fixed point $\bm{x}=0$ yields the Jacobian $-I+\phi'(0)J$, which equals $-I+J$ for our odd activation function with $\phi'(0)=1$. The zero state therefore loses stability when an eigenvalue of $J$ crosses the line $\mathrm{Re}\,\lambda=1$, so the bulk edge and the outliers of $J$ directly control the onset of nontrivial dynamics. For large random matrices, the eigenspectrum usually consists of outliers and a bulk. The chain motif can induce two outliers, as shown in Refs.~\onlinecite{RN232, RN316, RN271}
\begin{equation}\label{eq:Outliers}
\lambda_{\pm}=\frac12 \left(J_0\pm\sqrt{J_0^2+4g^2(N\tauchn+\taurec)}\right),
\end{equation}
where we have assumed the three-neuron motif correlations are $\calO\left(N^{-1}\right)$: 
\begin{equation}\label{eq:MotifScaling}
\tau^\clubsuit \sim \frac1N \quad \text{for } \clubsuit \in
\{\mathrm{chn},\mathrm{con},\mathrm{div}\}.
\end{equation}
This scaling keeps the three-neuron motifs finite in aggregate: because each edge belongs to $\calO(N)$ chain, convergent, and divergent motifs but only one reciprocal pair, their correlations must be $\calO(N^{-1})$ for an $\calO(1)$ effect, which is why $N\tauchn$ rather than $\tauchn$ is the natural control parameter.
We use this weak scaling in the DMFT derivation below, while the $\calO(1)$ scaling provides a useful spectral description of strong motif effects and finite-size simulations.
If we set $\taurec=0$, positive $\tauchn$ gives rise to two real outliers
[\refpanel{fig:Eigenspectrum}{c}], whereas negative $\tauchn$ can give rise
to a pair of complex-conjugate outliers. Under the scaling in
Eq.~\eqref{eq:MotifScaling}, the three-neuron motif correlations are
subleading, so the eigenvalue bulk reduces to the classical elliptic law
\cite{RN318, RN316}, with semi-axes $g(1\pm \taurec)$ in the large-$N$
limit. If instead these correlations are $\calO(1)$, they modify the bulk at
leading order. The bulk remains elliptical, but its semi-axes become
$g(A\pm B)/\sqrt{A}$ [\refpanel{fig:Eigenspectrum}{b}], where, following
Ref.~\onlinecite{RN232}, we have introduced
\begin{subequations}
\begin{align}
A& = 1-\taucon-\taudiv,\\
B& = \taurec-2\tauchn.
\end{align}
\end{subequations}
The derivation is given in Appendix~\ref{app:EigenBulk}. The right edge of the bulk along the real axis controls the linear stability of the $\bm{x}=0$ state and defines an effective gain
\begin{equation}\label{eq:geff}
g_{\mathrm{eff}}=\frac{A+B}{\sqrt{A}}\,g .
\end{equation}

This spectral picture suggests two distinct dynamical effects of chain
correlations. Positive chain correlations can generate a positive outlier mode,
which may destabilize the zero-mean state and induce up-down symmetry breaking,
analogous to the effect of positive mean connectivity $J_0$ in an i.i.d.
network \cite{RN275, RN211, RN336}. Negative chain correlations, by contrast,
can generate a complex-conjugate outlier pair, suggesting an oscillatory
instability that may give rise to limit cycles. For strong $\calO(1)$
negative chain correlations, the same correlations also enter the bulk through
the parameter $B$, in the same way as positive reciprocal correlations. We
therefore expect them to slow the dynamics and, for sufficiently strong
correlations, potentially lead to glassy behavior \cite{RN277, RN202}. In
Secs.~\ref{sec:DMFT} and \ref{sec:NegativeChain}, we develop a DMFT that supports and refines these spectral predictions.

\section{Dynamic mean-field theory} \label{sec:DMFT}
Using the path-integral formalism and the saddle-point method (Appendix~\ref{app:DMFT}), we obtain the generic form of the single-site DMFT equation
\begin{align}\label{eq:GenericDMFTEq}
  \dot{x}(t)&=-x(t)+J_0 \langle{\phi(t)}\rangle+ g^2B \int_{-\infty}^t \chiphi(t,t')\phi(t') \,\diff t'\nonumber \\ &\quad+g^2  N\tauchn \int_{-\infty}^t \chiphi(t,t')\left\langle \phi(t')\right \rangle \diff t' + \eta(t),
\end{align}
where $\eta(t)$ is a zero-mean Gaussian process with autocorrelation 
\begin{equation} \label{eq:NoiseAuto}
\left\langle\eta(t) \eta(t')\right\rangle=Ag^2 \Cphi\left(t,t'\right)+ g^2N\taucon\langle\phi(t)\rangle\langle\phi(t')\rangle.
\end{equation}
Here, $\langle \cdot \rangle$ denotes the average over the effective
Gaussian process $\eta$. We have introduced the autocorrelation \begin{equation}
\Cphi(t,t'):=\left\langle \phi(t)\phi(t')\right\rangle
\end{equation}
and the response function
\begin{equation}
\chiphi(t,t')=\left\langle\frac{\delta \phi(x(t))}{\delta \eta(t')}\right\rangle.
\end{equation}
The mean-field theory also assumes the scaling \eqref{eq:MotifScaling} so that the chain and convergent motif terms remain finite in the large-$N$ limit. In what follows, we use this single-site
effective equation to analyze the effects of each motif.

\subsection{Long-time dynamics}
We first analyze the long-time dynamics reached after the initial transient, treating each motif class in turn. In the stationary regimes the ensemble mean $\langle\phi\rangle$ is time-independent and the chain feedback reduces to a static term, whereas negative chain correlations can destabilize the stationary state and generate limit cycles, which are not time-translation invariant.

\subsubsection{Positive chain correlation} \label{sec:PosChain}

\begin{figure*}[t]
\includegraphics{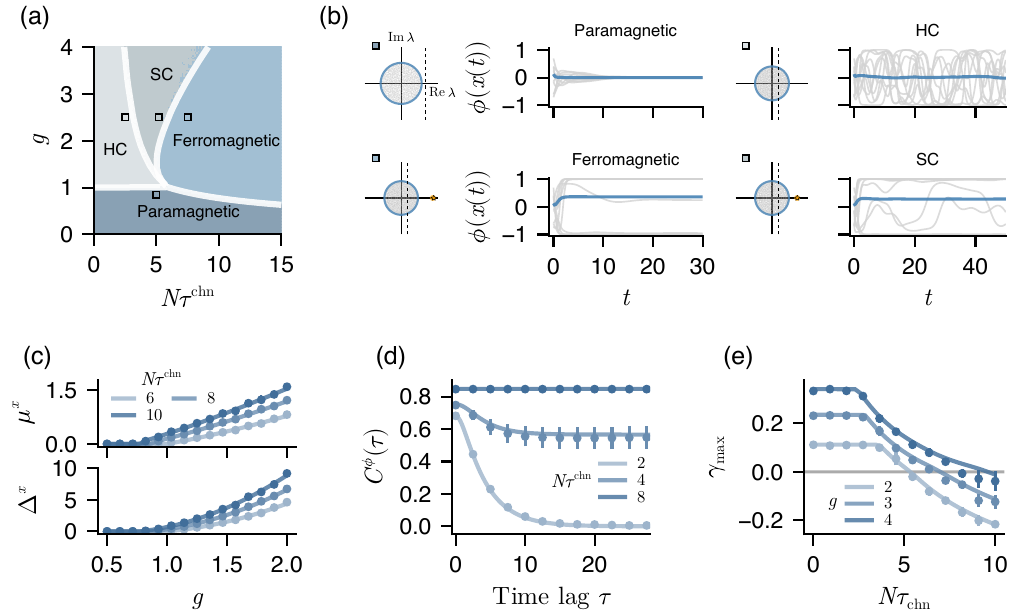}
\caption{\label{fig:PosChnPD}Positive chain correlations induce a ferromagnetic structure.
(a) Phase diagram in the $g$--$N\tauchn$ plane. HC and SC denote homogeneous chaos and structured chaos, respectively. Colored phase boundaries are obtained from simulations, and white curves show theoretical predictions. Square markers indicate the parameter points used in panel (b). (b) Representative eigenspectra and activity trajectories in the four phases. In the eigenvalue panels, black dashed lines mark the stability boundary, orange dots show measured outlier eigenvalues, and stars show the theoretical outlier prediction. In the trajectory panels, blue curves show the population-averaged activity $\langle \phi(x(t)) \rangle$, and gray curves show single-unit activities $\phi(x_i(t))$. (c) Fixed-point mean $\mu^x$ and variance $\Delta^x$ as functions of $g$ for several values of $N\tauchn$. (d) Stationary autocorrelation $C^\phi(\tau)$ for representative chain-correlation strengths. (e) Largest Lyapunov exponent as a function of $N\tauchn$ for several values of $g$. In (c)--(e), solid curves show theoretical predictions. At each parameter point, dots show the mean across 10 network realizations that converge to fixed points in (c), the median across 36 network realizations in (d), and the mean across 10 network realizations in (e). We use $N=2000$ in (a)--(d) and $N=4000$ in (e). In all panels, $J_0=-5$, $\taucon=\taudiv=\tauchn$, and $\taurec=0$. Error bars denote $\pm$SD.}
\end{figure*}

We first examine the effect of positive chain correlations. Positive $\tauchn$ generates a positive outlier in the eigenspectrum and acts in the DMFT equation as an effective positive feedback. To keep this feedback from trivially driving the network into saturation, we choose $J_0<0$. Although the mean-field theory applies to arbitrary combinations of motifs, here we set $\taurec=0$ for simplicity. Together with the scaling $\tauchn=\calO(N^{-1})$, this gives $B=\calO(N^{-1})$, so the retarded self-interaction term proportional to $B$ is subleading and vanishes in the large-$N$ limit. 

For a stationary state, define the stationary response and autocorrelation by \begin{align}
\chiphi(\tau)&=\left\langle \chiphi(t,t-\tau)\right\rangle_t,\\
\Cphi(\tau)&=\left\langle \phi(t)\phi(t+\tau)\right\rangle_t .
\end{align} 
Since $\langle \phi\rangle$ is time independent, the chain contribution becomes a static feedback term. With the self-interaction term neglected in the large-$N$ limit,
\begin{equation}
\int_0^{\infty}\chiphi(\tau)\,\diff \tau=\left\langle \phi'\right\rangle ,
\end{equation}
the DMFT equation reduces to
\begin{equation}\label{eq:DMFTPosChain}
\dot{x}(t)=-x(t)+\left(J_0+g^2N\tauchn \left\langle \phi' \right\rangle\right)
\left\langle \phi \right\rangle
+\eta(t),
\end{equation}
with noise autocorrelation
\begin{equation}\label{eq:CetaPosChain}
C^\eta(\tau)=Ag^2\Cphi(\tau)+g^2N\taucon\left\langle \phi\right\rangle^2 .
\end{equation}
Since $\taucon$ and $\taudiv$ are $\calO(N^{-1})$, the correction to $A=1-\taucon-\taudiv$ is subleading. By contrast, the static convergent contribution in Eq.~\eqref{eq:CetaPosChain} can remain finite when $\langle\phi\rangle\neq 0$. 

Equation~\eqref{eq:DMFTPosChain} shows that positive chain correlations act, to leading order in $1/N$, as a self-consistent enhancement of the mean connectivity,
\begin{equation}\label{eq:EffectiveJ0Chn}
J_0\;\longrightarrow\;J_0+g^2N\tauchn\left\langle \phi'\right\rangle .
\end{equation}
Thus, apart from the static convergent contribution to the noise, the positive-chain network has the same mean-field structure as an i.i.d. Gaussian network with a state-dependent effective mean connectivity, or equivalently an i.i.d. Gaussian network with an additional rank-one component.

The resulting phase diagram in the $g$--$N\tauchn$ plane contains four regimes (Fig.~\ref{fig:PosChnPD}). In the paramagnetic phase, $\bm{x}=0$ is the only stable fixed point. In the ferromagnetic phase, the network has two symmetric nonzero fixed points. In the homogeneous chaos (HC) phase, the dynamics is chaotic but the ensemble mean remains zero. Between HC and the ferromagnetic phase lies a structured-chaos (SC) phase, also called synchronous chaos or ferromagnetic chaos \cite{RN285,RN325,RN374}, in which the network remains chaotic but develops a nonzero ensemble mean.

The phase boundaries can be computed in the large-$N$ limit from two complementary criteria: the two boundaries of the paramagnetic phase follow from the linear stability of $\bm{x}=0$, while the HC--SC and SC--ferromagnetic boundaries require the nonlinear stationary statistics from DMFT [white curves in \refpanel{fig:PosChnPD}{a}]. Here, $\taurec=0$ and the three-neuron correlations scale as $\calO(N^{-1})$, so $A=1+\calO(N^{-1})$ and $B=\calO(N^{-1})$. Consequently, $g_{\mathrm{eff}}=g$ to leading order [Eq.~\eqref{eq:geff}], and the phase-boundary conditions below are expressed in terms of $g$:
\begin{enumerate}
\item Paramagnetic--HC. The eigenvalue bulk reaches the stability boundary before the positive outlier destabilizes the origin,
\begin{equation}
g=1,\qquad \lambda_+<1. 
\end{equation}
\item Paramagnetic--Ferromagnetic. The positive outlier reaches the stability boundary while the bulk remains stable, 
\begin{equation}
\lambda_+=1,\qquad g<1. 
\end{equation}
\item HC--SC. A perturbation analysis around the HC solution gives
\begin{equation}\label{eq:HCSCBoundaryB0}
\langle\phi'\rangle\lambda_+=1,
\end{equation}
where $\langle\phi'\rangle$ is evaluated in the HC state, 
\begin{equation}
    \langle\phi'\rangle=\int \calD z\, \phi'\left(\sqrt{\Cx(0)}\,z\right).
\end{equation}
Here, $C^x(\tau)=\langle x(t)x(t+\tau)\rangle$ is the stationary autocorrelation of the input current, and $C^x(0)=\langle x^2\rangle$ is its variance in the zero-mean HC state. We use $\calD z=\frac{\exp(-z^2/2)}{\sqrt{2\pi}}\,\diff z$ to represent the standard Gaussian measure. For the derivation, including the generic condition for non-negligible $B$, see Appendix~\ref{app:HCStability}.
\item SC--Ferromagnetic. The eigenvalue bulk of the Jacobian evaluated at the ferromagnetic fixed point $x_0$ reaches the stability boundary \cite{RN285, RN364}
\begin{equation}
-1+g\sqrt{\left\langle\phi'(x_0)^2\right\rangle_{x_0}}=0.
\end{equation}
The corresponding expression for non-negligible $B$ is given in Appendix~\ref{app:EigenBulk}.
\end{enumerate}
The HC--SC and SC--ferromagnetic boundaries depend on the stationary statistics of $x$. For the fixed-point regime, when the self-interaction term is negligible, the distribution of $x$ is Gaussian and is fully described by its mean $\mu^x$ and variance $\Delta^x$. The fixed-point equations obtained from Eqs.~\eqref{eq:DMFTPosChain} and \eqref{eq:CetaPosChain} are
\begin{subequations} \label{eq:B0FFPDist}
\begin{align}
    \mu^x&=\left(J_0+g^2N\tauchn \langle\phi'\rangle\right)\langle \phi\rangle,\\
    \Delta^x&=g^2 \left(\left\langle \phi^2 \right\rangle +N\taucon\langle\phi \rangle^2\right).
\end{align}
\end{subequations}
These equations predict the fixed-point statistics observed in simulations [\refpanel{fig:PosChnPD}{c}]. In the chaotic phases, the full stationary autocorrelation can be obtained by solving Eqs.~\eqref{eq:DMFTPosChain} and \eqref{eq:CetaPosChain} in the frequency domain, as described in Appendix~\ref{app:AlgorithmB0}. The resulting autocorrelation functions agree well with direct simulations [\refpanel{fig:PosChnPD}{d}].

Finally, the largest Lyapunov exponent (LLE) can be computed from the same mean-field statistics. For an i.i.d. Gaussian network with nonzero mean $J_0$, the LLE is determined by the stationary autocorrelation of $\phi'(x)$ \cite{RN211,RN217}. In Appendix~\ref{app:LLEB0}, we show that the three-neuron motifs considered here do not change the functional form of this result when $B$ is negligible. The motifs enter through the self-consistent stationary statistics, and the theoretical prediction agrees with direct simulations [\refpanel{fig:PosChnPD}{e}].

\subsubsection{Negative chain correlation}\label{sec:NegChnLC}

\begin{figure*}
\includegraphics{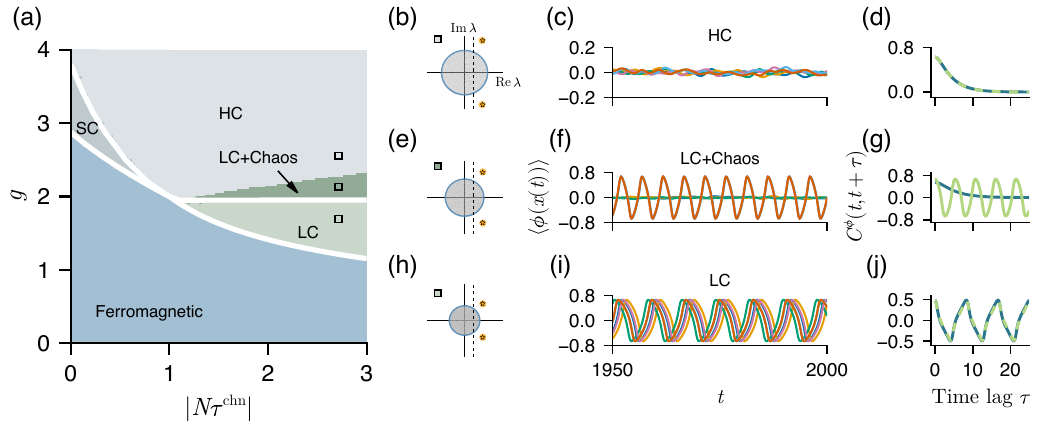}
\caption{\label{fig:NegChnPD}
Negative chain correlations can generate limit cycles (LCs).
(a) Phase diagram in the $g$--$|N\tau_{\mathrm{chn}}|$ plane. Colored regions show phases identified numerically, while white curves show theoretical stability boundaries.
(b,e,h) Example eigenspectra in the HC, LC+Chaos, and LC phases, respectively. Blue curves and stars indicate theoretical predictions. Black dashed lines mark the stability boundary.
(c,f,i) Population-averaged activity $\langle \phi(x(t))\rangle$ from six different initial conditions for the networks shown in (b,e,h).
(d,g,j) Numerical solutions of $C^\phi(\tau)$ obtained from the single-site DMFT equation at the corresponding parameter values. Blue curves denote solutions initialized near a small-mean state, whereas green curves denote solutions initialized near a large-mean state.
Each row corresponds to the square marker in (a).
Parameters: $J_0=4$, $N=4000$, $\tau_{\mathrm{con}}=\tau_{\mathrm{div}}=|\tau_{\mathrm{chn}}|$, and $\tau_{\mathrm{rec}}=0$.}
\end{figure*}

In Sec.~\ref{sec:PosChain}, we chose $J_0<0$ to balance the positive feedback generated by positive chain correlations. With this choice, sufficiently negative $\tauchn$ produces complex-conjugate outliers whose real part is $J_0/2<0$, so these modes remain stable and do not modify the stationary mean-field dynamics to leading order in $1/N$. Here we instead consider $J_0>0$, where the same complex-conjugate outliers can become unstable and drive an oscillatory instability. The nonlinear dynamics can then settle into stable LCs.

As in Sec.~\ref{sec:PosChain}, we set $\tau_{\mathrm{rec}}=0$ so that the retarded self-interaction term is negligible. When $g<1$, the homogeneous chaotic state is absent, and the network can only converge to fixed points or limit cycles. For $g>1$, however, the zero-mean chaotic state can coexist with oscillatory solutions. In Fig.~\ref{fig:NegChnPD}, we set $J_0=4$ and vary $g$ and $|N\tau_{\mathrm{chn}}|$. The resulting phase diagram contains five regimes. The ferromagnetic fixed-point, SC, and HC phases are the same as in Sec.~\ref{sec:PosChain}, whereas the LC and LC+Chaos phases are induced by the complex-conjugate outliers generated by negative chain correlations.

The white curve separating the LC and LC+Chaos regions is obtained from the local stability of the HC solution against oscillatory mean perturbations (Appendix~\ref{app:HCStability}). The instability occurs when
\begin{equation}\label{eq:HopfNegChn}
J_0 \langle\phi'\rangle = 2 .
\end{equation}
This is the complex-outlier version of the static HC--SC boundary, Eq.~\eqref{eq:HCSCBoundaryB0}: both impose $\langle\phi'\rangle\,\mathrm{Re}\, \lambda_+=1$, which reduces to Eq.~\eqref{eq:HopfNegChn} once negative chain correlations make the outliers complex. Thus, in the $B\sim N^{-1}$ regime, the dependence on $|N\tauchn|$ enters through the requirement $J_0^2+4g^2N\tauchn<0$ for a complex outlier pair, whereas the HC stability boundary itself is set by the real part of that pair. Below this boundary, the HC solution is unstable, and the stable attractor is a limit cycle. Above it, the HC solution is locally stable. In this region, the limit cycle can remain stable as well, giving rise to coexistence of HC and LC attractors. This coexistence appears as the LC+Chaos phase in \refpanel{fig:NegChnPD}{a}.

Because the stability of the LC solution is difficult to analyze analytically, we locate the coexistence region numerically. We solve the non-stationary DMFT equation from two initial conditions, one initialized at a small-mean state and the other at a large-mean state. In the LC+Chaos phase, the small-mean initial condition converges to the HC solution, whereas the large-mean initial condition converges to an LC solution [\refpanel{fig:NegChnPD}{g}]. In the pure HC and LC phases, both initial conditions converge to the same type of attractor [\refpanel{fig:NegChnPD}{d, j}]. The boundary between the LC and LC+Chaos phases obtained from this procedure agrees well with the theoretical stability condition in Eq.~\eqref{eq:HopfNegChn}. Direct network simulations confirm the same bistability: different initial conditions can converge either to a zero-mean chaotic trajectory or to a macroscopic oscillation [\refpanel{fig:NegChnPD}{f}].

\subsubsection{Distinct effects of convergent and divergent motifs}

The DMFT equation reveals a qualitative distinction between convergent and divergent motifs. Divergent correlations enter the effective dynamics only through the prefactor $A=1-\taucon-\taudiv$ multiplying the temporal noise autocorrelation. Convergent correlations, by contrast, generate an additional contribution dependent on the mean activity [Eq.~\eqref{eq:NoiseAuto}]. Thus, when $\langle\phi\rangle=0$, as in the HC phase with an odd activation function, the leading mean-dependent effect of convergent correlations vanishes. In this case, convergent and divergent motifs affect the single-site dynamics only through the finite-size correction to $A$.

The distinction becomes important once the activity has a nonzero mean. At stationarity, the static component of the effective input has variance $C^\eta(\infty)=Ag^2\Cphi(\infty)+g^2N\taucon\langle\phi\rangle^2$ [Eq.~\eqref{eq:CetaPosChain}], receiving contributions from both the long-time plateau of the activity autocorrelation and the time-independent convergent motif term. The effective noise can be decomposed into a quenched part and a zero-mean temporal part:
\begin{equation}
\eta(t)=g\sqrt{A\Cphi(\infty)+N\taucon\langle\phi\rangle^2}\,z+\zeta(t),
\end{equation}
where $z\sim\mathcal{N}(0,1)$ is a quenched Gaussian variable and $\zeta(t)$ has autocorrelation
\begin{equation}
C^\zeta(\tau)=Ag^2\left(\Cphi(\tau)-\Cphi(\infty)\right).
\end{equation}
In a large network, different neurons sample independent values of $z$. Convergent motifs therefore convert a nonzero population mean activity into quenched heterogeneity of the local field, whereas divergent motifs do not generate such a static field.

\begin{figure}[htb]
\includegraphics{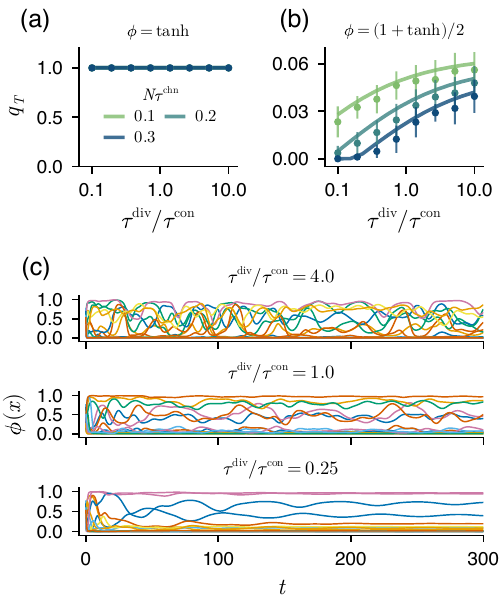}
\caption{\label{fig:ConDivq} Convergent motifs suppress temporal chaos.
(a) Fraction of temporal variance, $q_T$, as a function of the motif ratio $\taudiv/\taucon$ for the odd activation function $\phi=\tanh$. Solid curves show the DMFT prediction, while dots show the mean across realizations obtained from direct simulation. Each curve corresponds to a fixed value of $N\tauchn$. (b) Same as (a), but for the nonnegative activation function $\phi=(1+\tanh)/2$. (c) Representative trajectories of $\phi(x_i(t))$ for $\phi=(1+\tanh)/2$ and $N\tauchn=0.25$, illustrating the reduction of temporal chaos as convergent correlations become stronger. In (a)\&(b), $N=8000$, error bars represent $\pm$SD over 24 realizations. In (c), $N=2000$. Other parameters: $g=5$, $J_0=-5$, and convergent and divergent correlations satisfy $\taucon\taudiv=(\tauchn)^2$.}
\end{figure}

This effect is especially relevant for interpreting rate models as approximations to spiking neuronal populations. In the current-based form of Eq.~\eqref{eq:SCSDynamics}, $x_i$ represents the input current and $\phi(x_i)$ represents the firing rate \cite{TheoreticalNeuroscience,RN392,RN421}. We approximate a more realistic inhibitory neuronal population by using a nonnegative activation function, such as $\phi=(1+\tanh)/2$, together with a negative mean coupling $J_0<0$ \cite{RN217,RN211,RN285}. In this setting, any nontrivial active state has $\langle\phi\rangle>0$, so the static input generated by convergent correlations is present. A similar distinction between convergent and divergent motifs has also recently been found empirically in numerical simulations of spiking networks \cite{RN339}.

To quantify how much of the activity is temporally fluctuating rather than quenched across neurons, we introduce the fraction of temporal variance
\begin{equation}
q_T=\frac{\left\langle \llangle x_i^2 \rrangle_t \right\rangle_i}{
\left\langle \llangle x_i^2 \rrangle_t \right\rangle_i
+\llangle \left\langle x_i \right\rangle_t^2 \rrangle_i}.
\end{equation}
Here, $\llangle x_i^2 \rrangle_t=\left\langle x_i^2\right\rangle_t-\left\langle x_i\right\rangle_t^2$ is the temporal variance of neuron $i$, and $\llangle \left\langle x_i \right\rangle_t^2 \rrangle_i$ is the population variance of the time-averaged input current. Thus, $q_T=1$ indicates that the variability is purely temporal, whereas $q_T<1$ indicates that part of the variability manifests as static heterogeneity across neurons.

In the $B\approx 0$ regime considered here, the single-site input current is Gaussian, so the DMFT predicts $q_T$ directly from its stationary statistics,
\begin{equation}\label{eq:qTDMFT}
q_T=\frac{\Cx(0)-\Cx(\infty)}{\Cx(0)-\Cx(\infty)+g^2\left(A\,\Cphi(\infty)+N\taucon\langle\phi\rangle^2\right)},
\end{equation}
where $\Cx(0)-\Cx(\infty)$ is the population-averaged temporal variance, and the additional contribution $g^2\left(A\,\Cphi(\infty)+N\taucon\langle\phi\rangle^2\right)$ in the denominator is the population variance of the time-averaged input. For odd-symmetric $\phi$ in the HC phase, $\left\langle x_i\right\rangle_t=0$ up to finite-size fluctuations, so the quenched spatial term vanishes and $q_T\simeq 1$, independent of $\taucon$. For the nonnegative activation $\phi=(1+\tanh)/2$, increasing $\taucon$ strengthens the static convergent contribution and therefore suppresses the temporal fraction $q_T$. This prediction is confirmed by direct numerical simulations (Fig.~\ref{fig:ConDivq}).

\subsection{Transient dynamics}\label{sec:Transient}
 
\begin{figure}[htb]
\includegraphics{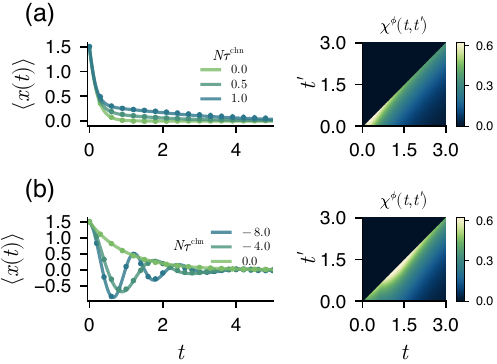}
\caption{\label{fig:Transient} Chain motifs alter transient dynamics.
(a) Positive chain correlations slow the relaxation of the mean input current in the HC phase. Left: relaxation dynamics of the population-averaged input $\langle x(t)\rangle$ for different values of $N\tau_{\mathrm{chn}}$. Right: Response kernel $\chi^\phi(t,t')$ for $N\tauchn=1$.
(b) Negative chain correlations induce damped oscillatory relaxation. Left: relaxation dynamics of $\langle x(t)\rangle$ for different values of $N\tauchn$. Right: Response kernel $\chi^\phi(t,t')$ for $N\tauchn=-8$.
Parameters: $N=2000$, $g_{\mathrm{eff}}=3$, $\taucon=\taudiv=|\tauchn|$, and $\taurec=2\tauchn$. We set $J_0=-5$ for positive $\tauchn$ in (a) and $J_0=0$ for negative $\tauchn$ in (b). Solid curves show DMFT predictions, and dots show the mean across 5 network realizations from direct numerical simulations for each parameter value in (a,b). Error bars indicate $\pm$SD and are smaller than the markers.
}
\end{figure}
Beyond shaping the long-time behavior, chain motifs also leave an imprint on the transient relaxation, which can reveal their presence even when the long-time statistics are identical to those of an i.i.d. network to leading order in $1/N$. For $B=0$, the retarded self-interaction vanishes. Nevertheless, the chain-induced feedback through the population mean remains finite because $N\tauchn=\calO(1)$. In the stationary HC and paramagnetic phases with an odd activation function, this contribution vanishes because $\langle\phi\rangle=0$. For initial conditions with nonzero mean activity, however, the time-dependent DMFT retains the feedback term
\begin{equation}
g^2N\tauchn
\int^t \chiphi(t,t')\langle \phi(t')\rangle\,\diff t',
\end{equation}
which modifies the transient relaxation while leaving the stationary HC solution unchanged.

To isolate the mean-feedback effect, we set $\taurec=2\tauchn$ so that $B=0$ and the retarded self-interaction term vanishes. We solve the time-dependent DMFT equation \eqref{eq:GenericDMFTEq} and compare the result with direct simulations of finite networks. Positive chain correlations produce a retarded positive feedback of the ensemble mean, leading to slower relaxation toward the zero-mean HC state [\refpanel{fig:Transient}{a}]. Negative chain correlations produce retarded negative feedback and generate damped oscillatory relaxation [\refpanel{fig:Transient}{b}]. In both cases, the DMFT predictions agree well with direct numerical simulations, showing that the mean-field theory captures the transient dynamics in addition to the long-time statistics.

\section{Negative chain correlation and glassy dynamics} \label{sec:NegativeChain}

\begin{figure*}[t]
\includegraphics{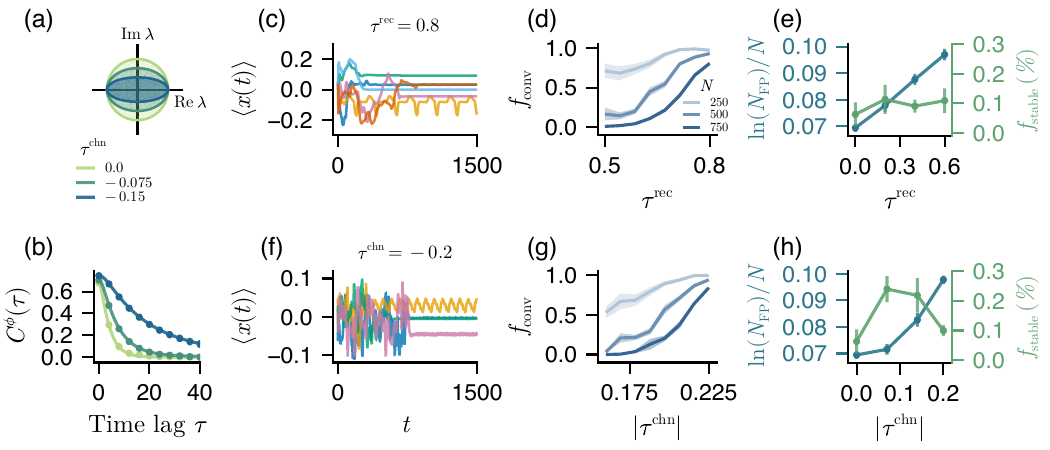}
\caption{\label{fig:GlassyPhase}Positive reciprocal and negative chain correlations slow down the dynamics and eventually produce a glassy phase. (a) Negative chain correlations reduce the imaginary semi-axis of the eigenvalue bulk when the spectral edge $g_{\mathrm{eff}}$ is held constant. Solid curves show the theoretical boundary, while dots show a numerical example with $N=2000$. (b) The autocorrelation timescale increases with negative chain correlation strength. Dots show the median across 36 network realizations from direct simulation for each parameter value, while solid curves show the DMFT prediction. Error bars representing $\pm$SD are smaller than the markers. Panels (c)--(e) correspond to positive reciprocal correlations, and (f)--(h) to negative chain correlations. (c,f) Example trajectories that settle into a non-chaotic state after a transient. (d,g) Fraction of trajectories that become non-chaotic within $T=1000$ time units. Shaded regions represent $\pm$SEM across $10$ network realizations. (e,h) The number of fixed points and fraction of stable fixed points as functions of motif strength. Dots show the mean, while error bars show $\pm$SEM over $7$ realizations. In all panels, $J_0=0$, $g_{\mathrm{eff}}=3$. In (c,f), $N=500$; in (e,h), $N=100$.}
\end{figure*}

The negative-chain effects discussed in Sec.~\ref{sec:NegChnLC} arise from $\calO(N^{-1})$ correlations that generate outlier modes. We now consider stronger negative chain correlations in finite-size networks, whose main effect is to reshape the spectral bulk through the self-interaction parameter $B$.
In the stationary HC regime, since our choice of an odd activation function implies $\langle \phi(t)\rangle=0$, the DMFT equation reduces to 
\begin{equation} \label{eq:HCDMFT}
\dot{x}(t)=-x(t)+g^2B\int_{-\infty}^{t} \chiphi(t-\tau)\phi(\tau) \, \diff \tau+\eta(t),
\end{equation}
where the Gaussian noise has autocorrelation $C^\eta(\tau)=Ag^2\Cphi(\tau)$. Apart from the rescaling of the noise correlation by the factor $A$, this equation has the same form as the DMFT equation for a network with isolated reciprocal correlations, with the replacement $\taurec\mapsto B$ \cite{RN277}. Thus negative chain correlations, for which $B=\taurec-2\tauchn$ is increased, act like positive reciprocal correlations at the level of the stationary HC single-site theory. Since the terms proportional to $N\tauchn$ and $N\taucon$ vanish in the HC regime, the DMFT equation \eqref{eq:HCDMFT} can formally remain well defined even when the motif strengths are scaled as $\calO(1)$. This formal limit should not be interpreted literally as an $N\to\infty$ network limit, however, because such motif strengths can produce outlier eigenvalues that diverge with system size.

This correspondence motivates the conjecture that negative $\tauchn$ can first slow the dynamics [\refpanel{fig:GlassyPhase}{b}] and, in finite-size networks, eventually induce a glassy regime analogous to that generated by positive $\taurec$. Here, glassy dynamics refers to multistability among many stable fixed points with small but positive spectral gaps, where the spectral gap is the distance between the Jacobian spectrum at a fixed point and the imaginary axis. In a finite-size network, trajectories from different initial conditions can settle into different attractors, so that long-time averages retain a dependence on the initial condition \cite{RN202}.

Numerical simulations support this glassy-dynamics conjecture. To isolate the effect of motifs from a trivial change in the bulk instability, we vary the motif strengths while holding the effective gain $g_{\mathrm{eff}}$ [Eq.~\eqref{eq:geff}] fixed, which keeps the predicted right edge of the eigenspectrum constant [\refpanel{fig:GlassyPhase}{a}]. We use positive reciprocal correlations as a reference case and compare them with negative chain correlations throughout Fig.~\ref{fig:GlassyPhase}. In both cases, when the correlation is sufficiently strong, trajectories initialized from different random initial conditions can converge to different stable non-chaotic attractors, including fixed points and, in some realizations, limit cycles [\refpanels{fig:GlassyPhase}{c,f}]. We quantify this tendency by the finite-time convergence fraction $f_{\mathrm{conv}}(T)$, defined as the fraction of initial conditions whose finite-time-averaged LLE becomes non-positive within a simulation time $T$. Larger values of $f_{\mathrm{conv}}(T)$ indicate that randomly initialized trajectories more readily leave the chaotic transient and settle into stable non-chaotic attractors. Consistent with a glassy picture, we also find that the time required to reach such attractors increases with network size $N$, suggesting that the relevant stable attractors become harder to find on finite simulation times as $N$ grows [\refpanels{fig:GlassyPhase}{d,g}].

To connect this dynamical behavior to the underlying attractor landscape, we next examine the number and stability of the network's fixed points. It has been shown that, in the large-$N$ limit, the canonical i.i.d. SCS model contains exponentially many fixed points in the chaotic regime, but that these fixed points are unstable saddles \cite{RN378,RN350}. For the SCS model with motif correlations, a corresponding large-$N$ theory for fixed-point counting has not yet
been derived. We therefore search for fixed points numerically in relatively small networks (Appendix~\ref{app:FPSearch}) and leave a more systematic theoretical analysis for future work (see Sec.~\ref{sec:Discussion} for more discussion).

This fixed-point search suggests that both strong positive reciprocal and negative chain correlations produce a mixed landscape: the total number of fixed points found increases rapidly, but most remain unstable saddles, while only a small subset becomes linearly stable [\refpanels{fig:GlassyPhase}{e,h}]. For example, in networks with $N=100$ and $\tauchn=-0.2$, we can find on the order of $10^4$ fixed points, but typically only $0.1\%$ of them are stable. Thus, the glassy regime is better understood as the stabilization of a small, dynamically accessible subset of an otherwise saddle-dominated fixed-point landscape.

Importantly, the proliferation of fixed points cannot be attributed to a trivial increase in the effective instability of the i.i.d. component. In the parameter regime considered here, keeping $g_{\mathrm{eff}}$ fixed requires decreasing the bare gain $g$ as the motif strength is increased. Since the number of fixed points of an i.i.d. Gaussian network in the chaotic regime increases with $g$ for $g>1$ \cite{RN350}, this decrease in $g$ would, by itself, reduce the number of fixed points. Therefore, the observed increase in the number of fixed points reflects the effect of motif correlations rather than a simple change in the marginal variance of the coupling strengths.

\section{Geometry of chaos}
\subsection{Lyapunov spectrum} \label{sec:LyapunovSpec}

\begin{figure*}[ht]
\includegraphics{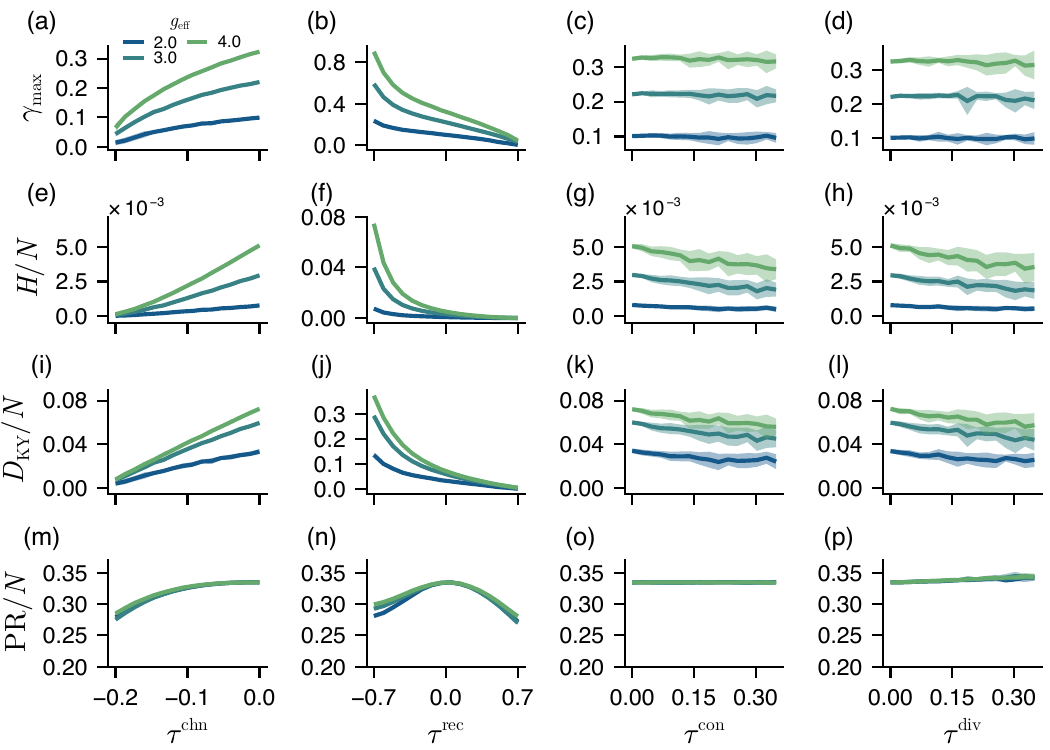}
\caption{\label{fig:LyapunovSpec}Lyapunov-spectrum-based measures across motif strengths. Rows show, from top to bottom, the LLE
$\gamma_{\max}$, the KS entropy per neuron $H/N$, the
KY dimension per neuron $D_{\rm KY}/N$, and the participation
ratio per neuron $\mathrm{PR}/N$ of the leading CLV.
Columns correspond to varying $\tauchn$, $\taurec$, $\taucon$, and
$\taudiv$, respectively. In the first column, we set $\taurec=0$ and
$\taucon=\taudiv=\left|\tauchn\right|$; in the other columns, all uninvolved
motif correlations are set to zero. Only nonnegative convergent and divergent correlations are considered, consistent with the network-generation scheme in Appendix~\ref{app:NetworkGeneration}. Curves show different values of
$g_{\mathrm{eff}}$. Parameters: $N=1000$ and $J_0=0$. Shaded regions denote $\pm$SD over network realizations (24 for $\tauchn$ and $\taurec$, 48 for $\taucon$ and $\taudiv$). Non-chaotic trajectories with non-positive LLEs are discarded.}
\end{figure*}

The single-site DMFT equation captures the macroscopic dynamics of the network in the large-$N$ limit, but it does not fully characterize how chaos is distributed across neurons. In particular, two networks can have similar single-site statistics while differing in the collective organization of their dynamics. For the chaotic phase with negligible self-interaction parameter $B$, the LLE can be obtained from single-site statistics through a Schr\"odinger operator \eqref{eq:ShrodingerOp} \cite{RN294, RN211}. When $B$ is non-negligible, the LLE is no longer given by such a simple expression, although it should still be determined by single-site DMFT quantities, including response and correlation functions \cite{RN372}. To probe dynamical features that are not visible in the single-site description, we therefore compute the Lyapunov spectrum numerically. From this spectrum, we extract the LLE, the Kolmogorov-Sinai (KS) entropy, and the Kaplan-Yorke (KY) dimension \cite{RN372, RN394, RN281}. The KS entropy quantifies the rate at which uncertainty is produced by the sensitive dependence of chaotic dynamics on initial conditions. The KY dimension can be interpreted as the interpolated dimension at which expansion along unstable directions is balanced by contraction along stable directions. We also compute the time-averaged participation ratio (PR) of the first covariant Lyapunov vector (CLV) $\bm{v}(t)$, defined as 
\begin{equation}
\operatorname{PR}=\left\langle\frac{\left(\sum_i{v_i^2(t)}\right)^2}{\sum_i v_i^4(t)}\right\rangle_t,
\end{equation}
which gives the effective number of neurons participating in the leading unstable direction \cite{RN281, RN272, RN203, RN250, RN261}.

In this section, we set the mean connection to zero, $J_0=0$, and focus on the high-dimensional HC phase, where $\langle \phi \rangle=0$. We study each motif class separately, corresponding to the four columns of Fig.~\ref{fig:LyapunovSpec}. For the chain-correlation sweep with $\taurec=0$, this restricts our analysis to negative chain correlations, $\tauchn<0$, for which the network does not develop the structured mean activity induced by positive chain motifs. Since we do not develop a large-$N$ theory for the full Lyapunov spectrum, our analysis is centered on finite-size networks. Accordingly, we extend the motif strengths beyond the weak-correlation scaling in Eq.~\eqref{eq:MotifScaling} used in the mean-field analysis and treat them as finite motif parameters.

One caveat is that these quantities should not necessarily be interpreted as intrinsic properties of a single chaotic attractor. As shown in Sec.~\ref{sec:NegativeChain}, networks with negative chain correlations or positive reciprocal correlations can exhibit coexistence of multiple attractors, and for many initial conditions the dynamics eventually relaxes to a non-chaotic state when simulated for sufficiently long times. However, for the system size considered here, $N\sim 1000$, the preceding chaotic transients can be very long. The Lyapunov quantities reported in this section should therefore be understood as finite-time, ensemble-averaged measures: they are averaged over network realizations and over trajectories that may correspond either to genuine chaotic attractors or to long-lived chaotic transients. This distinction is important because the infinite-time and large-$N$ limits need not commute in regimes with attractor coexistence. At fixed finite $N$, sufficiently long trajectories may eventually relax to non-chaotic attractors, whereas the lifetime of chaotic transients can grow with system size [\refpanels{fig:GlassyPhase}{d,g}]. Thus, the quantities reported here characterize the typical chaotic dynamics observed on the simulated time scale, rather than an asymptotic invariant measure of an individual network.

To isolate motif effects from a trivial change in the bulk instability, we fix $g_{\mathrm{eff}}$ while varying the motif strengths. We vary each type of motif correlation in isolation whenever possible. For chain correlations, which cannot be introduced without also generating convergent and divergent correlations, we set $\taucon=\taudiv=|\tauchn|$ so that the effects of convergent and divergent motifs remain minimal. For negative chain correlations and positive reciprocal correlations, we observe a sharp decrease in the LLE, KS entropy, and KY dimension, indicating that these motifs suppress chaos and drive the dynamics toward a lower-dimensional chaotic state [\refpanels{fig:LyapunovSpec}{a,b,e,f,i,j}]. These motifs also reduce the PR of the first CLV [\refpanels{fig:LyapunovSpec}{m,n}], suggesting that the leading chaotic fluctuations become concentrated in a smaller subset of neurons. By contrast, negative reciprocal correlations have the opposite effect on the Lyapunov spectrum: they increase the LLE, KS entropy, and KY dimension, indicating stronger and higher-dimensional chaos [\refpanels{fig:LyapunovSpec}{b,f,j}]. Interestingly, however, the PR still decreases, implying that the enhanced chaotic instability is also more spatially concentrated [\refpanel{fig:LyapunovSpec}{n}].

For convergent and divergent correlations, an important caveat is that random-matrix theory predicts only an eigenvalue bulk \cite{RN271}, but individual realizations can still develop dominant outlier-like modes due to sample-to-sample fluctuations in the eigenspectrum (Fig.~\ref{fig:ConDivOutlier}). Under the $\calO(1)$ motif scaling used here, these fluctuations can drive realization-specific symmetry breaking, producing SC-like or ferromagnetic fixed-point dynamics (Fig.~\ref{fig:DipTest}). We therefore restrict the analysis to realizations that remain in the unimodal HC regime (see Appendix~\ref{app:ConDivFilter}). Although the network generation algorithm allows $\taucon$ or $\taudiv$ up to $0.5$, we only report results up to $0.35$, because for stronger correlations a large fraction of generated networks contain dominant outlier eigenvalues and no longer display a typical HC regime.

As predicted by the DMFT, after filtering out realizations dominated by sample-specific outliers, the LLE remains essentially independent of convergent and divergent motif strengths [\refpanels{fig:LyapunovSpec}{c,d}]. This invariance follows from the Schr\"odinger operator in Eq.~\eqref{eq:ShrodingerOp}. In the HC phase with negligible self-interaction, the LLE is determined by the effective spectral edge $g_{\mathrm{eff}}$, which is held fixed in these comparisons. Consistent with this picture, the PR of the first CLV is also nearly unchanged [\refpanels{fig:LyapunovSpec}{o,p}]. Nevertheless, the KS entropy and KY dimension reveal differences that are not captured by the single-site LLE theory. Both convergent and divergent correlations reduce these quantities, indicating that even when the leading instability is unchanged, these motifs make the dynamics less chaotic and lower dimensional [\refpanels{fig:LyapunovSpec}{g,h,k,l}].

\subsection{Participation-ratio dimension}
\begin{figure*}[ht]
\includegraphics{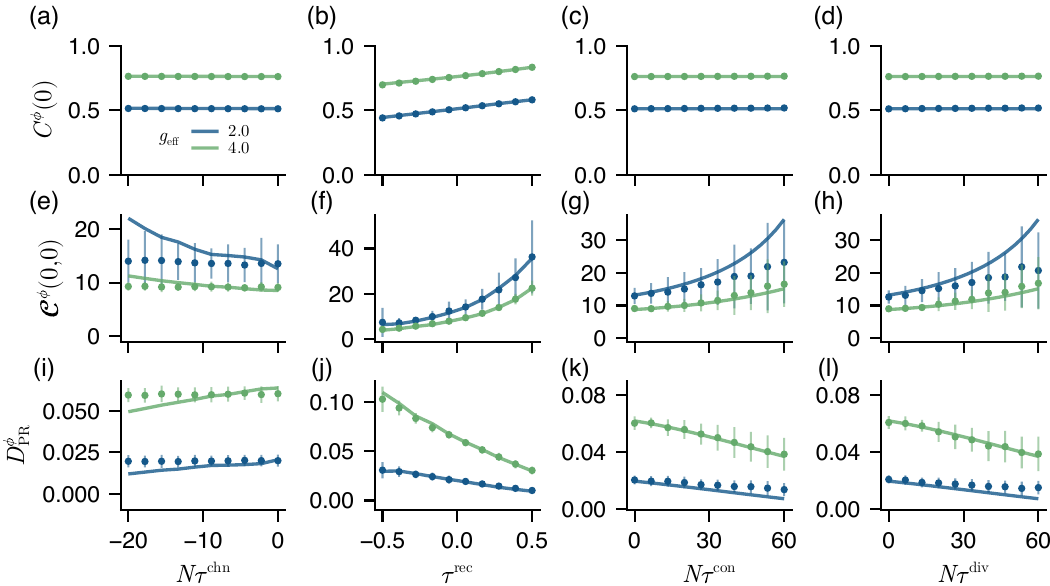}
\caption{\label{fig:AllMotifsPRD}Participation-ratio dimension across motif strengths. Rows show, from top to bottom, $\Cphi(0)$, $\calC^\phi(0,0)$, and the
participation-ratio dimension $D_{\mathrm{PR}}^\phi$. Columns correspond to
varying $\tauchn$, $\taurec$, $\taucon$, and $\taudiv$, respectively. In
panels (a), (e), and (i), we set $\taucon=\taudiv=\left|\tauchn\right|$. In all other panels, only the motif correlation shown on the horizontal axis is nonzero. Curves show the theoretical prediction, while dots show the mean across realizations from direct simulations. Other parameters are $N=2000$ and $J_0=0$. Error bars denote $\pm$SD over 128 independent realizations. }
\end{figure*}
The dimensionality of neural activity can also be quantified by the PR of the covariance spectrum \cite{RN26, RN278, RN232, RN247, RN353, RN281, RN345, RN40, RN388, RN325, RN390}. The PR dimensionality is closely related to principal-component analysis (PCA): the covariance eigenvalues measure the variance carried by each principal component, and the PR gives an effective number of principal components that participate in the population activity. Compared with the Lyapunov spectrum, the PR is more directly amenable to analytical treatment, and therefore provides a complementary explanation for the dimensionality reduction observed in Sec.~\ref{sec:LyapunovSpec}.

Let $\lambda_i^\phi$ denote the eigenvalues of the covariance matrix of the neural activity $\phi(x_i(t))$. The normalized PR dimension is
\begin{equation}
D_\mathrm{PR}^\phi=\frac{1}{N}\frac{\left(\sum_i \lambda_i^\phi\right)^2}{\sum_i \big(\lambda_i^\phi\big)^2}.
\end{equation}
In the HC phase, the stationary covariance matrix is
$C_{ij}^\phi(\tau)=\left\langle\phi(x_i(t))\phi(x_j(t+\tau))\right\rangle_t$.
Following Ref.~\onlinecite{RN278}, we introduce the self-averaging four-point correlator
\begin{equation}
\calC^\phi(\bm{\tau})= N\left\langle C_{ij}^\phi(\tau_1) C_{ij}^\phi(\tau_2) \right\rangle_J, \qquad  (i\neq j).
\end{equation}
The PR dimension can then be expressed in terms of the single-site autocorrelation and the off-diagonal covariance as
\begin{equation}
D_{\mathrm{PR}}^\phi = \frac{1}{1+\calC^\phi(0,0)/\Cphi(0)^2}.
\end{equation}

The four-point correlator was calculated using a two-cavity method for i.i.d. Gaussian networks and partially symmetric networks with $\taurec\neq0$ in Ref.~\onlinecite{RN278}. Here, we extend this calculation to include three-neuron motifs. We note that the assumptions underlying the two-cavity calculation require the three-neuron correlations to be scaled as in Eq.~\eqref{eq:MotifScaling}. With this scaling, the three-neuron correlations do not modify the leading $\calO(1)$ functional form of the four-point correlator $\calC^\phi$. Instead, their effects appear only as finite-size corrections in the $1/N$ expansion. The two-cavity method captures the first of these corrections, but does not determine the higher-order terms. The detailed derivation is given in Appendix~\ref{app:TwoCavity}.

The general expression for $\calC^\phi$ [Eq.~\eqref{eq:FourPointCorr}] is symmetric under $\taucon \leftrightarrow \taudiv$. That is, in the HC phase, convergent and divergent correlations are equivalent not only at the single-site mean-field level, but also in their leading effect on the four-point correlator. We first focus on the effect of isolated convergent (and, by symmetry, divergent) correlation. Introducing the frequency-domain shorthands
\[
 C_{12}=\Cphi(\omega_1)\Cphi(\omega_2),
 \qquad
 X_{12}=2\pi g^2\chiphi(\omega_1)\chiphi(\omega_2),
\]
the four-point correlator reduces to
\begin{equation}
\calC^\phi(\omega_1,\omega_2)=C_{12}
\left[
\left(\frac{1}{|1-X_{12}|^2}-1\right)
+\frac{N(\tau^{\mathrm{con}})^2|X_{12}|^2}{1-|X_{12}|^2}
\right].
\label{eq:C4Convergent}
\end{equation}
The first term is the contribution already present in an i.i.d. network. It depends on the complex phase of
$\chi^\phi(\omega_1)\chi^\phi(\omega_2)$. By contrast, the correction induced by convergent correlations depends only on
$|X_{12}|^2$, and is therefore insensitive to this phase.

When the retarded self-interaction proportional to $B$ is negligible, the response has the form
\begin{equation}
|X_{12}|^2=\frac{g^4 \langle \phi'\rangle^4}
{\left(1+\omega_1^2\right)\left(1+\omega_2^2\right)}
\leq g^4 \langle\phi'\rangle^4 .
\end{equation}
Thus, a sufficient condition for the convergent-correlation correction in Eq.~\eqref{eq:C4Convergent} to be positive is
$g\langle\phi'\rangle<1$. This condition is consistent with the HC regime considered here. Indeed, one can prove that $\sqrt{A}g\langle\phi'\rangle\leq 1$ \cite{RN353}. Since the motif scaling in Eq.~\eqref{eq:MotifScaling} gives $A\simeq 1$ for large networks, this bound suggests that $g\langle\phi'\rangle$ remains close to or below unity. Moreover, in the HC regime, $g\langle\phi'\rangle$ decreases with increasing $g$ \cite{RN353}. These results provide a useful intuition for why the denominator $1-|X_{12}|^2$ remains positive in the parameter regime considered here.

If $g_{\mathrm{eff}}$ is held fixed as in Secs.~\ref{sec:NegativeChain}\&\ref{sec:LyapunovSpec}, convergent and divergent correlations do not change the stationary single-site autocorrelation statistics [\refpanels{fig:AllMotifsPRD}{c,d}]. Therefore, their effect on
$D_{\mathrm{PR}}^\phi$ enters through the four-point correlator $\calC^\phi$. In the regime described above, isolated convergent and divergent correlations increase $\calC^\phi(0,0)$, and hence reduce the PR dimensionality of the population activity [\refpanels{fig:AllMotifsPRD}{g,k,h,l}].

When chain correlations are introduced, they must be accompanied by convergent and divergent correlations (Appendix~\ref{app:NetworkGeneration}), so the expression contains all three types of correlations simultaneously. Relative to an i.i.d. network, the $\calO(N^{-1})$ correction resulting from these three-neuron motifs reads
\begin{widetext}
\begin{align}\label{eq:ThreeNeuronCorrection}
\Delta \calC^\phi(\bm{\omega})&=\frac{C_{12}}{1-\left|X_{12}\right|^2}
\bigg[N\left(\left(\tauchn\right)^2+\taucon\taudiv\right)
\left(\frac{X_{12}^2}{1-X_{12}^2}+\frac{(X_{12}^*)^2}{1-(X_{12}^*)^2}
\right)+N\left(\left(\taucon\right)^2+\left(\taudiv\right)^2\right)\left|X_{12}\right|^2\nonumber \\ & \qquad +N\tauchn\left(\taucon+\taudiv\right)(X_{12}+X_{12}^*)(Y_{12}+Y_{12}^*)
+2N\left(\tauchn\right)^2\left(Y_{12}^2+(Y_{12}^*)^2\right)\bigg],
\end{align}
\end{widetext}
where we have defined $Y_{12}=2\pi g^2 \chiphi(\omega_1)^* \chiphi(\omega_2)$. This expression shows that the three-neuron motif corrections are not simply additive: convergent and divergent motifs couple through the mixed term $\taucon\taudiv$, while chain motifs couple to them through $\tauchn(\taucon+\taudiv)$. In
Appendix~\ref{app:CorrectionNonnegative}, we show that, when $\left|X_{12}\right|<1$ (verified in Fig.~\ref{fig:ThreeNeuronX12Max}), these corrections can only increase the zero-lag four-point correlator $\calC^\phi(0,0)$.

We compare this prediction with direct simulations for networks with chain
correlations in \refpanels{fig:AllMotifsPRD}{a,e,i}. For large
$N|\tauchn|$, the theory overestimates the four-point correlator and therefore
underestimates the participation-ratio dimensionality. This discrepancy is
expected because Eq.~\eqref{eq:ThreeNeuronCorrection} is a 
finite-size correction; if the motif strengths are increased beyond its regime
of validity, the nominal $\calO(N^{-1})$ contribution can become comparable to,
or larger than, the leading i.i.d. term. Nevertheless, simulations over a wider
range of motif strengths confirm the qualitative prediction: stronger
three-neuron motif correlations further reduce $D^\phi_\mathrm{PR}$
(Fig.~\ref{fig:PRDFullRange}).

The $\calO(1)$ reciprocal-correlation case corresponds to the partially symmetric ensemble analyzed in Ref.~\onlinecite{RN278}. Unlike the three-neuron motifs above, $\taurec$ modifies the leading four-point correlator and does not vanish as $N\to\infty$. In the fixed-$g_{\mathrm{eff}}$ comparison used here, increasing $\taurec$ increases both the single-site variance $\Cphi(0)$ and the zero-lag four-point correlator $\calC^\phi(0,0)$ [\refpanels{fig:AllMotifsPRD}{b,f}]. The increase in $\calC^\phi(0,0)$ dominates, producing a monotonic decrease in $D^\phi_{\mathrm{PR}}$ [\refpanel{fig:AllMotifsPRD}{j}], consistent with the reduction of the KY dimension in \refpanel{fig:LyapunovSpec}{j}.

\section{Discussion} \label{sec:Discussion}
In this work, we extend classical DMFT theory for random neural networks \cite{RN294,RN211,RN280} to incorporate four second-order synaptic motifs. Depending on the mean coupling, gain, and motif correlations, even a single-population network can exhibit a rich repertoire of dynamical regimes, including ferromagnetic fixed points, limit cycles, and chaos. This is reminiscent of random neural networks with a low-rank structure \cite{RN320,RN298,RN288}, and in fact, in the absence of reciprocal motifs and assuming $\calO(N^{-1})$ three-neuron correlations [Eq.~\eqref{eq:MotifScaling}], the network model is a special case of a more generic rank-two model studied in Ref.~\onlinecite{RN320} (Appendix~\ref{app:MappingToLowRank}). We show that, in a stationary state, over-representation of chain motifs renormalizes the effective mean coupling strength [Eq.~\eqref{eq:EffectiveJ0Chn}]. This observation suggests that E-I balance in structured neuronal networks need not arise solely from the cancellation between population-averaged activity and the mean synaptic strength, as assumed in classical balanced-network theories with independent connectivity \cite{RN217,RN131,RN132,RN285}. Instead, local connectivity structure, such as chain motifs, can contribute an additional self-consistent mean-field feedback that participates in the balance. Conversely, under-representation of chain motifs can destabilize a complex-conjugate pair of outlier eigenvalues and generate limit cycles, providing a new structural mechanism for neural oscillations.

The DMFT also clarifies why convergent and divergent motifs need not be dynamically equivalent. Divergent correlations primarily rescale the temporal noise through the factor $A$, whereas convergent correlations generate a static component of the effective input proportional to $g^2N\taucon\langle\phi\rangle^2$. Thus, in nonzero-mean states, convergent motifs can convert population-mean activity into quenched heterogeneity and suppress temporal fluctuations. By contrast, in the zero-mean HC phase, convergent and divergent motifs are indistinguishable at the single-site DMFT level.

For strong negative chain correlations in the stationary HC regime, the DMFT predicts a slowdown of chaotic dynamics analogous to that produced by positive reciprocal correlations, because both types of correlations enter the DMFT equation through the coefficient $B=\taurec-2\tauchn$ of the retarded self-interaction term. Direct finite-size network simulations show that making $\tauchn$ more negative can eventually drive the network into a glassy regime characterized by multistability and dependence on initial conditions. This identifies a previously unreported route to glassy dynamics in nonlinear random networks: glassiness can arise from sufficiently negative chain correlations even in fully asymmetric networks ($\taurec=0$), whereas similar dynamics have previously been found mainly in networks with partially symmetric interactions \cite{RN202,RN393,RN423,RN374}. When positive reciprocal correlations are also present, they further facilitate this transition, because both positive $\taurec$ and negative $\tauchn$ increase $B$ and therefore reinforce the same tendency toward slow dynamics.

In the SCS model and a similar balanced random neural network model with i.i.d. Gaussian couplings, the exponential growth rate of the number of fixed points (i.e., topological complexity) has been computed analytically in the large-$N$ limit \cite{RN350,RN329}. Extending this type of Kac--Rice calculation to neural networks with second-order motif correlations is likely difficult. A complete theory of topological complexity and fixed-point geometry for the motif-structured networks studied here is therefore beyond the scope of this work. Nevertheless, recent progress in related models provides useful qualitative context. In the generalized Lotka--Volterra model with partially symmetric Gaussian interactions, the quenched topological complexity can be reduced analytically to a finite-dimensional saddle-point problem; within that model, increasing the reciprocal correlation increases the total topological complexity \cite{RN393}. This trend is qualitatively consistent with our numerical observation [\refpanel{fig:GlassyPhase}{e}]. More generally, other solvable Gaussian random-field models with mixed gradient and non-gradient interactions show that total complexity of the topologically nontrivial regime is dominated by unstable equilibria \cite{RN427,RN374}. Stable equilibria can also be exponentially numerous in parts of parameter space, especially when the gradient component is sufficiently strong, but their complexity is smaller than the total complexity, so their fraction among all equilibria is exponentially small in $N$. 

Beyond the single-site mean-field description, both the Lyapunov spectrum and the covariance-spectrum PR reveal that motifs reshape the collective geometry of chaotic activity. The Lyapunov analysis shows that negative chain and positive reciprocal correlations suppress chaos, reducing the LLE, the KS entropy, and the KY dimension, whereas convergent and divergent correlations lower the KS entropy and KY dimension even when the spectral edge $g_{\mathrm{eff}}$ is held fixed. To complement this numerical picture with an analytical one, we derived the PR dimension for networks with $\calO(N^{-1})$ three-neuron motif correlations using a two-cavity method. Together with direct simulations, this calculation shows that these motifs generally reduce dimensionality. This identifies local motif structure as a candidate circuit-level mechanism that can constrain the dimensionality of neural dynamics. An interesting comparison is the PR dimension of a linear network driven by Gaussian white noise, for which the PR dimension of noise-induced fluctuations takes the form $D_\mathrm{PR}^x=\left(1-g_{\mathrm{eff}}^2\right)^2$ \cite{RN232}. In that linear noise-driven setting, motif effects enter the dimensionality only through the effective spectral edge $g_{\mathrm{eff}}$. By contrast, in the autonomous chaotic nonlinear networks studied here, three-neuron motifs can reduce dimensionality even when $g_{\mathrm{eff}}$ is held fixed. In our theory, the $\calO(N^{-1})$ scaling of three-neuron motifs implies that their correction to dimensionality vanishes as $N\to\infty$. For convergent and divergent correlations, however, one can in principle take the motif strengths to be $\calO(1)$ without causing the spectrum of $J$ to diverge. In this regime, individual networks need not be self-averaging (Appendix~\ref{app:ConDivFilter}), so our disorder-averaged theory does not fully describe realization-specific dynamics. Fixed-connectivity approaches developed for i.i.d. Gaussian networks \cite{RN353,RN375,RN376} may provide useful starting points, although adapting them to motif-structured connectivity remains an open problem.

The Gaussian all-to-all coupling used in this article should be viewed as an approximation to networks with sufficiently large in-degree. In extremely sparse networks, where each neuron receives only a small number of inputs, the Gaussian approximation breaks down. In that regime, one must instead use recently developed DMFT approaches for sparse networks \cite{RN381}.

The firing-rate model studied here does not directly address how different motifs influence synchronization among spiking neurons. Numerical studies suggest that motif effects on synchrony can be cell-type dependent \cite{RN339}, and that only chain and convergent motifs have a significant effect on synchronization \cite{RN365}. Nevertheless, firing-rate models can be viewed as coarse-grained descriptions of asynchronous spiking networks in regimes where spiking fluctuations are averaged out \cite{RN422,RN217}. From this perspective, our finding that chain and convergent motifs quench temporal chaotic fluctuations provides a possible intuition for why these motifs have distinct effects on synchrony. Recent work has taken an important step toward more biologically structured rate models by deriving low-rank mean-field equations for multi-population nonlinear networks with cell-type-dependent connectivity statistics and heterogeneous chain motifs \cite{RN431}. An important remaining direction is to connect these complementary rate-level theories to spiking E-I networks and oscillator networks, where the relationship between motif structure and synchronization can be studied more directly.

While our DMFT analysis focuses on second-order motifs, higher-order structures can in principle be incorporated by retaining higher-order cumulants in Eq.~\eqref{eq:CumulantGF}, provided that their large-$N$ scaling yields an extensive contribution to the disorder-averaged action. For generic higher-order motifs, this expansion introduces non-quadratic vertices in the MSRJD action, precluding a closed single-site theory with purely Gaussian colored noise and linear retarded feedback. Directed cyclic motifs provide a more tractable special case. Generalizing the two-edge reciprocal motif, directed-cycle cumulants create higher-order retarded self-interactions involving successive convolutions of the response kernel $\chiphi(t,t')$. Thus, incorporating cyclic cumulants is a straightforward extension of the present DMFT framework beyond second-order motifs. Recent work has explored the dynamical consequences of cyclic motifs using random matrix theory and numerical simulations \cite{RN306,RN326}, but did not derive the corresponding DMFT equations.

In this work, we have asked how motif statistics shape network dynamics. A complementary question is how such motif structures arise in the first place. Recent experiments suggest that overrepresented second-order motifs emerge gradually during development in mouse CA3 \cite{RN373}. Theoretical work has examined how spike-timing-dependent plasticity in recurrent spiking networks drives the evolution of weighted motif strengths on a fixed connectivity graph \cite{RN11}. Within a firing-rate framework, one could also incorporate Hebbian or anti-Hebbian plasticity \cite{RN425} and study how the motif statistics considered here evolve dynamically.

Since the brain is fundamentally an information-processing system, it is natural to ask whether motif structure has functional consequences. Many studies have examined how specific motifs influence information transmission, memory, and related computational properties \cite{RN7,RN10,RN79,RN380,RN74,RN83,RN103,RN1,RN397,RN396,RN398,RN406,RN326,RN429,RN430}. However, most analytical results have been restricted to isolated or small motif circuits, leaving much less understood how these motifs shape network-level computational capacity when embedded in large recurrent networks. Recent work has begun to address this question in the RC setting. Takasu and Aoyagi showed that non-zero reciprocal correlations can increase the memory capacity of a discrete-time reservoir computer, an effect they attributed to reduced average correlations between neuronal activities \cite{RN346}. In a continuous-time linear reservoir network, Pachitariu et al. found that symmetric connectivity supports stable representations that can be read out by a time-independent decoder, whereas non-symmetric connectivity can retain comparable working memory but stores information in rotating modes that require a readout matched to the elapsed time \cite{RN426}. By contrast, in a similar continuous-time reservoir model, Peng et al. found that increasing reciprocal connectivity, as well as increasing the abundance of short directed cycles, degrades performance on a spoken-digit recognition task \cite{RN326}. These contrasting results suggest that the computational impact of motif structure can depend sensitively on the operating regime of the reservoir, the way network gain is controlled, and the task requirements. Developing an analytical theory for how different motif classes affect RC performance remains an interesting direction for future work.

In summary, our work establishes second-order synaptic motifs as a structural mechanism that, together with single-neuron nonlinearity, shapes the dynamical regimes of recurrent neural networks, providing an analytical framework for understanding how the local connectivity statistics of cortical circuits shape their collective dynamics.

\begin{acknowledgments}
We thank David Clark for sharing the source code used in Ref.~\onlinecite{RN278}. This work was supported by an Alfred P. Sloan Research Fellowship in Neuroscience to H.C.
\end{acknowledgments}

\section*{Data Availability}
Code for reproducing the results is publicly available on GitHub \cite{GitHubRepo}.

%

\appendix

\section{Derivation of the DMFT equation} \label{app:DMFT}

To use the Martin--Siggia--Rose--Janssen--de
Dominicis (MSRJD) path-integral formalism \cite{RN361, RN360}, we add a Gaussian white noise process $\bm{\xi}(t)$ and an external source $-\bm{\tilde{j}}(t)$ to the model:
\begin{equation} 
\dot{x}_i = -x_i + \sum_{j=1}^{N} J_{ij} \phi(x_j)-\tilde{j}_i(t)+ \xi_i(t). \end{equation}
The noise process statistics are given by $\left\langle \xi_i(t) \right\rangle=0$ and $\left\langle \xi_i(t) \xi_j(t') \right\rangle=D \delta_{ij}\delta(t-t'). $ The generating functional for this model is \cite{RN280, RN309, RN211}

\begin{align}   Z \left[\bm{j}, \bm{\tilde{j}}\right] &= \int \mathcal{D} \bm{x} \mathcal{D} \bm{\tilde{x}} \exp \bigg(S_0 [\bm{x}, \bm{\tilde{x}}] - \bm{\tilde{x}}^\mathrm{T} J \bm{\phi}  \nonumber \\ &\qquad+\bm{j}^\mathrm{T} \bm{x} + \bm{\tilde{j}}^\mathrm{T} \bm{\tilde{x}} \bigg), 
\end{align}
where $S_0 [\bm{x}, \bm{\tilde{x}}] = \bm{\tilde{x}}^\mathrm{T}(\partial_t + I) \bm{x} + \frac{D}{2} \bm{\tilde{x}}^\mathrm{T} \bm{\tilde{x}}$. Here, we have introduced some shorthand notation: 
\begin{subequations} \begin{align}  & \bm{\tilde{x}}^\mathrm{T}\bm{\tilde{x}}=\int  \sum_i \tilde{x}_i(t) \tilde{x}_i(t)\, \diff t, \\
&\bm{\tilde{x}}^\mathrm{T} J \bm{\phi} = \int \sum_{ij} J_{ij}\tilde{x}_{i} (t)  \phi\left(x_j(t)\right) \diff t.   
\end{align}
\end{subequations}
We average the generating functional over the disorder:
\begin{align}\label{eq:AverageZ}
\left\langle Z \right\rangle_J &= \int \mathcal{D} \bm{x} \mathcal{D} \bm{\tilde{x}} \exp \left(S_0 [\bm{x}, \bm{\tilde{x}}] + \bm{j}^\mathrm{T} \bm{x} + \bm{\tilde{j}}^\mathrm{T} \bm{\tilde{x}} \right) \nonumber \\ &\qquad \times \left\langle \mathrm{e}^{-\bm{\tilde{x}}^\mathrm{T} J \bm{\phi}} \right\rangle_J.
\end{align}

The disorder average can be simplified using the fact that $\ln \left\langle \mathrm{e}^{-\bm{\tilde{x}}^{\mathrm{T}} J \bm{\phi}} \right\rangle_{J}$ is the cumulant generating functional of the joint distribution of $J_{ij}$. For the Gaussian ensemble considered here, joint cumulants above second order vanish exactly. The same derivation applies to non-Gaussian ensembles provided that higher-order cumulants contribute only subleading terms to the disorder-averaged action as $N\to\infty$. Defining $y_{ij}=\int \tilde{x}_i(t) \phi(x_j(t)) \, \diff t $,
we have 
\begin{align}\label{eq:CumulantGF}
\ln \left\langle \exp \bigg( -\sum_{ij}J_{ij}y_{ij}\bigg)\right\rangle_J &= -\sum_{ij} \llangle J_{ij} \rrangle y_{ij}\nonumber\\ &+ \frac12 \sum_{ijkl} \llangle J_{ik}J_{jl} \rrangle y_{ik}y_{jl}, 
\end{align}
where $\llangle \cdot \rrangle$ denotes the cumulant. Plugging in the first- and second-order cumulants 
\begin{subequations}
\begin{align}
\llangle J_{ij}\rrangle & =\frac{J_0}{N}, \\
 \llangle J_{ik}J_{jl} \rrangle &= \frac{g^2}{N}\big(\delta_{ij}\delta_{kl}+\taudiv (1-\delta_{ij})\delta_{kl} \nonumber  \\  &  \qquad +\taucon (1-\delta_{kl})\delta_{ij} \nonumber \\  & \qquad  +\tauchn\left(\delta_{il}\left(1-\delta_{jk}\right) +\delta_{jk}(1-\delta_{il})\right)\nonumber \\ & \qquad + \taurec\delta_{il}\delta_{jk}(1-\delta_{ik})\big),
\end{align}
\end{subequations}
and using the abbreviations $A=1-\taucon-\taudiv$ and $B=\taurec-2\tauchn$, we obtain
\begin{widetext}
\begin{align}\label{eq:ExpDisorderAverage}
\left\langle \mathrm{e}^{-\bm{\tilde{x}}^{\mathrm{T}} J \bm{\phi}} \right\rangle_{J} =  & \;\exp\left( -\frac{J_0}{N} \int \diff t \left( \sum_{i} \tilde{x}_{i}(t) \right) \left( \sum_{j} \phi_{j}(t) \right) \right)\nonumber \\ & \times \exp\left( \frac{g^2}{2N} A \iint \diff t \diff t' \left( \sum_{i} \tilde{x}_{i}(t) \tilde{x}_{i}(t') \right) \left( \sum_{k} \phi_{k}(t) \phi_{k}(t') \right) \right) \nonumber \\ & \times \exp\left( \frac{g^2}{2N} B \iint \diff t \diff t' \left( \sum_{i} \tilde{x}_{i}(t) \phi_{i}(t') \right) \left( \sum_{k} \tilde{x}_{k}(t') \phi_{k}(t) \right) \right) \nonumber \\ & \times \exp\left( \frac{g^2}{2N} \taudiv \iint \diff t \diff t' \left( \sum_{k} \phi_{k}(t) \phi_{k}(t') \right) \left( \sum_{ij} \tilde{x}_{i}(t) \tilde{x}_{j}(t') \right) \right) \nonumber \\ & \times \exp\left( \frac{g^2}{2N} \taucon \iint \diff t \diff t' \left( \sum_{k} \tilde{x}_{k}(t) \tilde{x}_{k}(t') \right) \left( \sum_{ij} \phi_{i}(t) \phi_{j}(t') \right) \right)\nonumber  \\ & \times \exp\left( \frac{g^2}{N} \tauchn \iint \diff t \diff t' \left( \sum_{k} \tilde{x}_{k}(t') \phi_{k}(t) \right) \left( \sum_{ij} \tilde{x}_{i}(t) \phi_{j}(t') \right) \right),
\end{align}
where we have introduced the abbreviation $\phi_i(t) = \phi(x_i(t))$. To decouple the neurons and obtain the single-site effective dynamics, we need to introduce four auxiliary fields
\begin{subequations}\label{eq:AuxiliaryFields}
\begin{align}
&R(t)=\frac{1}{N}\sum_{k}\tilde{x}_{k}(t),\\
&S(t)=\frac{1}{N}\sum_{k}\phi_{k}(t),\\
&Q(t,t^{\prime})=\frac{g^{2}}{N}\sum_{k}\phi_{k}(t)\phi_{k}(t^{\prime}),\\
&P(t,t^{\prime})=\frac{g^{2}}{N}\sum_{k}\tilde{x}_{k}(t)\phi_{k}(t^{\prime}).
\end{align}
\end{subequations}
We then insert $\delta$ functionals in Eq.~\eqref{eq:ExpDisorderAverage} to enforce Eq.~\eqref{eq:AuxiliaryFields} and introduce conjugate fields $\widetilde{R}(t)$, $\widetilde{S}(t)$, $\widetilde{Q}(t,t')$, and $\widetilde{P}(t,t')$. For example, 
\begin{align}\label{eq:InsertingDelta1} &\exp\left(\frac{g^2}{2N}A\iint\mathrm{d}t\mathrm{d}t^{\prime}\left(\sum_i\tilde{x}_i(t)\tilde{x}_i(t^{\prime})\right)\left(\sum_k\phi_k(t)\phi_k(t^{\prime})\right)\right)\nonumber\\
&\qquad = \int\mathcal{D}Q\exp\left(\frac{A}{2}\bm{\tilde{x}}^{\mathrm{T}}Q\bm{\tilde{x}}\right)\times\delta\left(-\frac{N}{g^{2}}Q+\sum_{k}\phi_{k}(t)\phi_{k}(t^{\prime})\right)\nonumber \\
&\qquad = \int\mathcal{D}Q\mathcal{D}\widetilde{Q}\exp\left(-\frac{N}{g^2}\widetilde{Q}^\mathrm{T}Q+\frac{A}{2}\bm{\tilde{x}}^\mathrm{T}Q\bm{\tilde{x}}+\bm{\phi}^\mathrm{T}\widetilde{Q}\bm{\phi}\right),
\end{align} 
where we have introduced shorthand
\begin{subequations}
\begin{align}
& \widetilde{Q}^\mathrm{T}Q = \iint \tilde{Q}(t,t')Q(t,t') \, \diff t\diff t' \\ & \bm{\tilde{x}}^\mathrm{T} Q\bm{\tilde{x}}=\sum_k \iint \tilde{x}_k(t) Q(t,t') \tilde{x}_k(t')\, \diff t \diff t'.
\end{align}
\end{subequations}
Strictly speaking, the insertion of the functional delta constraint produces a normalization factor, but since our goal is to use the saddle-point method to retrieve the mean-field equation, the normalization factor does not influence the result and can be absorbed into the functional measure. A similar calculation yields 
\begin{align} \label{eq:InsertingDelta2}
&\exp\left( -\frac{J_0}{N} \int \diff t \left( \sum_{i} \tilde{x}_{i}(t) \right) \left( \sum_{j} \phi_{j}(t) \right) \right) = \int \mathcal{D}R\mathcal{D}\widetilde{R}\mathcal{D}S\mathcal{D}\widetilde{S}\, \exp\left(-N\left(\widetilde{R}^\mathrm{T}R+\widetilde{S}^\mathrm{T}S+J_0 R^\mathrm{T}S\right)+\widetilde{R}^\mathrm{T}\bm{\tilde{x}}+\widetilde{S}^\mathrm{T} \bm{\phi}\right),\\ \label{eq:InsertingDelta3}
&\exp\left( \frac{g^2}{2N} B \iint \diff t \diff t' \left( \sum_{i} \tilde{x}_{i}(t) \phi_{i}(t') \right) \left( \sum_{k} \tilde{x}_{k}(t') \phi_{k}(t) \right) \right)=\int \mathcal{D}P\mathcal{D}\widetilde{P}
\, \exp\left( -\frac{N}{g^2}\widetilde{P}^\mathrm{T}P+\frac{B}{2}\bm{\phi}^\mathrm{T}P\bm{\tilde{x}}+\bm{\tilde{x}}^\mathrm{T}\widetilde{P}\bm{\phi} \right),
\end{align}
where we have used shorthand  
$\widetilde{R}^\mathrm{T}\bm{\tilde{x}}=\sum_k \int \widetilde{R}(t)\tilde{x}_k(t) \, \diff t.$ Substituting Eqs.~\eqref{eq:ExpDisorderAverage}--\eqref{eq:InsertingDelta3} back into the disorder-averaged generating functional Eq.~\eqref{eq:AverageZ}, we obtain
\begin{equation}\label{eq:ReducedAverageZ}
\left\langle Z \right\rangle_J = \int \calD R\calD \widetilde{R}\calD  S \calD \widetilde{S}\calD Q\calD \widetilde{Q} \calD P\calD \widetilde{P}\exp\left(N(\Phi+\ln \Psi)\right),
\end{equation}
where 
\begin{align}
& \Phi=-\widetilde{R}^{\mathrm{T}}R-\widetilde{S}^{\mathrm{T}}S-\frac{1}{g^{2}}\widetilde{Q}^{\mathrm{T}}Q-\frac{1}{g^{2}}\widetilde{P}^{\mathrm{T}}P-J_{0}R^{\mathrm{T}}S, \\ 
& \Psi =\int\calD x \calD\tilde{x}\exp\left(S_0[x,\tilde{x}]+j^\mathrm{T}x+\tilde{j}^\mathrm{T}\tilde{x}+\widetilde{R}^\mathrm{T}\tilde{x}+\widetilde{S}^\mathrm{T}\phi+\frac{A}{2}\tilde{x}^\mathrm{T}Q\tilde{x}+\phi^\mathrm{T}\widetilde{Q}\phi+\frac{B}{2}\phi^\mathrm{T}P\tilde{x}+\tilde{x}^\mathrm{T}\widetilde{P}\phi+\frac{Ng^2}{2}\tau^\mathrm{div}(R\phi)^\mathrm{T}(R\phi) \right. \nonumber \\
&\qquad \left.+\frac{N g^2}{2}\tau^{\mathrm{con}}(S\tilde{x})^{\mathrm{T}}(S\tilde{x})+Ng^2\tau^{\mathrm{chn}}(R\phi)^{\mathrm{T}}(S\tilde{x})\right).
\end{align}
For the saddle-point action to remain extensive, we must assume three-neuron correlations are $\calO\left(N^{-1}\right)$ [Eq.~\eqref{eq:MotifScaling}] while allowing $\taurec=\mathcal{O}(1)$. Setting the functional derivatives of $\Phi + \ln \Psi$ with respect to each auxiliary field to
zero, we obtain the saddle point of this system: 
\begin{eqnarray*}
&&R^\star(t)=\langle\tilde{x}(t)\rangle, \qquad S^\star(t)=\langle \phi(t) \rangle, \qquad Q^\star(t,t')=g^2\left\langle \phi(t)\phi(t') \right\rangle, \qquad P^\star (t,t')=g^2\left\langle \tilde{x}(t)\phi(t') \right\rangle,\\
&&\widetilde{R}^{\star}(t)=-J_0\langle\phi(t)\rangle+Ng^2\left(\tau^{\mathrm{div}}\left\langle\phi(t)\int\phi(t^{\prime})\langle\tilde{x}(t^{\prime})\rangle\diff t^{\prime}\right\rangle+\tauchn\left\langle\phi(t)\int \tilde{x}(t^{\prime})\langle\phi(t^{\prime})\rangle\diff t^{\prime}\right\rangle\right),\\
&&\widetilde{S}^\star(t)=-J_0\langle\tilde{x}(t)\rangle +Ng^2\left(\taucon\left\langle\tilde{x}(t)\int\tilde{x}(t^{\prime})\langle\phi(t^{\prime})\rangle\diff t^{\prime} \right\rangle+\tauchn\left\langle\tilde{x}(t)\int\phi(t^{\prime})\langle\tilde{x}(t^{\prime})\rangle\mathrm{d}t^{\prime}\right\rangle\right),\\
&&\widetilde{Q}^\star(t,t')=\frac{Ag^2}{2}\left\langle\tilde{x}(t)\tilde{x}(t') \right\rangle, \qquad \widetilde{P}^\star(t,t')=\frac{Bg^2}{2}\left\langle\phi(t)\tilde{x}(t')\right\rangle.
\end{eqnarray*}
Moments that involve only the response field are zero $\langle \tilde{x}\rangle = \langle\tilde{x}(t)\tilde{x}(t') \rangle=0$. We can also identify the autocorrelation $\langle\phi(t)\phi(t')\rangle=:C^\phi(t,t')$ and the response function $ \langle\tilde{x}(t)\phi(t')\rangle= \frac{ \delta \langle\phi(t') \rangle}{\delta \tilde{j}(t)}  =:-\chiphi(t',t)$. Thus, the saddle point simplifies to
\begin{subequations}\label{eq:ReducedSaddle}
\begin{align}
&R^\star(t) = 0, & &\widetilde{R}^{\star}(t) = -J_0\langle\phi(t)\rangle
- Ng^2\tauchn \int_{-\infty}^t \chiphi(t,t')\langle\phi(t')\rangle \,\diff t',\\
&S^\star(t) = \langle \phi(t) \rangle, & &\widetilde{S}^\star(t) = 0, \\
&Q^\star(t,t') = g^2 \Cphi(t,t'), & &\widetilde{Q}^\star(t,t') = 0, \\
&P^\star(t,t') = -g^2 \chiphi(t',t),  & &\widetilde{P}^\star(t,t') = -\frac{Bg^2}{2}\chiphi(t,t').
\end{align}
\end{subequations}
Substituting the saddle point Eqs.~\eqref{eq:ReducedSaddle} into Eq.~\eqref{eq:ReducedAverageZ} yields $\langle Z \rangle_J=Z_0^N$ where 
\begin{equation}
Z_0 = \iint\calD x\calD \tilde{x}\exp\left(S_0[x,\tilde{x}]+j^\mathrm{T}x+\tilde{j}^\mathrm{T}\tilde{x}+\tilde{x}^\mathrm{T}\widetilde{R}^\star+\frac{Ag^2}{2}\tilde{x}^\mathrm{T}C^\phi\tilde{x}-Bg^2\tilde{x}^\mathrm{T}(\chiphi\phi)+\frac{Ng^2}{2}\tau^\mathrm{con}(S^\star\tilde{x})^\mathrm{T}(S^\star\tilde{x})\right)
\end{equation}
is the effective single-site generating functional. The corresponding stochastic differential equation for this generating functional is 
\begin{equation}
\dot{x}(t)=-x(t)+J_0 \langle{\phi(t)}\rangle+ g^2B \int_{-\infty}^t \chiphi(t,t')\phi(t') \,\diff t' +g^2  N\tauchn \int_{-\infty}^t \chiphi(t,t')\left \langle \phi(t')\right \rangle \diff t' -\tilde{j}(t) + \eta(t),
\end{equation}
where $\eta(t)$ is a zero-mean Gaussian process with $ \langle \eta(t)\eta(t')\rangle = D\delta(t-t')+Ag^2\Cphi (t,t')+g^2 N\taucon\left\langle\phi(t)\right\rangle\!\left\langle\phi(t')\right\rangle $. Setting $D=0$ and $\tilde{j}(t)=0$, we recover Eq.~\eqref{eq:GenericDMFTEq} in the main text.

A similar path-integral calculation for the Lotka-Volterra model, including a derivation using the dynamic single-cavity method, can be found in Ref.~\onlinecite{RN338}.
\end{widetext}
\section{The eigenvalue bulk} \label{app:EigenBulk}

Our derivation follows the classic approach of Sommers et al.~\cite{RN318}, with the modifications required for the present setting. We consider a more general scaled matrix
\[
J^\phi_{ij}=J_{ij}\phi'(x_j),
\]
which relates to the Jacobian evaluated at a nontrivial fixed point constrained by $\bm{x}=J\phi(\bm{x})$. The spectral density $\rho(x,y)$ in the complex plane is obtained from the potential $\varphi(x,y)$ via the Poisson equation
\begin{equation}
\rho(x,y)=-\frac{1}{4\pi}\nabla^2 \varphi(x,y).
\end{equation}
The potential reads \cite{RN318}
\begin{equation}
\varphi(\omega)=-\frac{1}{N}\left\langle \ln \det \left[\left(I\omega^{*}-(J^\phi)^{\mathrm{T}}\right)\left(I\omega-J^\phi\right)\right]\right\rangle_{J},
\end{equation}
where $\omega=x+\im y$ is a complex spectral parameter. As has been noted in previous works \cite{RN318, RN316, RN366, RN367}, one can rewrite the averaged logarithm as the logarithm of an averaged replicated determinant in the large-$N$ limit, so that
\begin{equation}
\varphi(\omega)=\lim_{\epsilon\to 0^+}\frac{1}{N}\ln \left\langle \left(\det\left(M^\dagger M+\epsilon I\right)\right)^{-1}\right\rangle,
\end{equation}
with $M=\omega I-J^\phi$.
Here, a positive regulator $\epsilon I$ has been introduced to ensure that the matrix inside the determinant is positive definite. For simplicity, we introduce a shorthand $\alpha_j:=\phi'(x_j)$ in the rest of this section.

The determinant can be recast into a Gaussian integral
\begin{align}
 \mathcal{Z}:&=\left(\det\left(M^\dagger M+\epsilon I\right)\right)^{-1} \nonumber \\ 
 & =\int \prod_i \frac{\mathrm{d}^2 z_i}{\pi}\exp\left(-\boldsymbol{z^\dagger}M^\dagger M \boldsymbol{z}- \epsilon \boldsymbol{z^\dagger}\boldsymbol{z}\right),
\end{align}
and we apply a Hubbard--Stratonovich transformation to linearize the quadratic term
\begin{widetext}
\begin{align}
\mathcal{Z}\,&=\int\prod_{i}\frac{\mathrm{d}^{2} z_{i} \mathrm{d}^{2} y_{i}}{\pi^{2}} \exp\left(-\bm{y^{\dagger}y}-\epsilon\bm{ z^\dagger z}+\mathrm{i}\left(\bm{y^{\dagger}} M \bm{z}+\bm{z^{\dagger}}M^{\dagger}\bm{y}\right)\right) \nonumber \\ 
& = \int \calD x \calD \tilde{x} \calD y \calD y^* \calD z \calD z^*  \, \exp\left(
-|\bm{y}|^2 - \epsilon |\bm{z}|^2
+ \im \sum_i \tilde{x}_i \left(x_i - \sum_j J_{ij}\phi_j\right)  \right. \nonumber  \\ & \left. \qquad +\im \sum_i \left(\omega y_i^* z_i + \omega^* y_i z_i^*\right)
- \im \sum_{ij} J_{ij} \alpha_j \left(y_i^* z_j + y_i z_j^*\right)
\right),
\end{align}
where in the second line we have inserted a Dirac $\delta$ function to enforce the fixed-point condition. Defining 
\begin{equation}
h_{ij} = \tilde{x}_i \phi_j +\alpha_j (y_i^* z_j+y_i z_j^*),
\end{equation}
we can isolate the $J$-dependent term as
\begin{equation}
\label{eq:RMGeneratingFunction}
\mathcal{Z}\,=\int \calD x \calD \tilde{x} \calD y \calD y^* \calD z \calD z^* \, \exp\left(
-|\bm{y}|^2 - \epsilon |\bm{z}|^2
+ \im \sum_i \left(\tilde{x}_ix_i + \omega y_i^* z_i + \omega^* y_i z_i^*\right) \right)  \times \exp\left( -\im \sum_{ij} J_{ij}h_{ij}
\right).
\end{equation} 
The disorder average $\langle \mathcal{Z}\rangle$ has the same structure as Eq.~\eqref{eq:AverageZ}, thus a similar computation yields 
\begin{align}
\left\langle
\exp\left(-\im\sum_{ij}J_{ij} h_{ij}\right)
\right\rangle
&=\exp\left(-\frac{\im J_0}{N}\sum_{ij} h_{ij}
-\frac{g^2}{2N}
\left(A\sum_{ij}h_{ij}^2
+B\sum_{ij}h_{ij}h_{ji} \right. \right. \nonumber\\ 
& \qquad\left.  \left.  +\taudiv \sum_{ijk}h_{ik}h_{jk}
+\taucon \sum_{ijk}h_{ij}h_{ik}
+2\tauchn \sum_{ijk}h_{ij}h_{jk} \right) \right).
\end{align}

Before introducing macroscopic fields, we note that Eq.~\eqref{eq:RMGeneratingFunction} has a global $U(1)$ symmetry. That is, the microscopic action and measure are invariant under a global rotation 
\begin{equation}
y\to y\mathrm{e}^{\im \theta},\quad z\to z\mathrm{e}^{\im \theta}.
\end{equation} This symmetry simplifies the computation considerably. For example, the first quadratic term becomes
\begin{equation}
h_{ij}^2
=\tilde{x}_i^2\phi_j^2
+2\tilde{x}_i\phi_j\alpha_j\left(y_i^*z_j+y_i z_j^*\right) 
+ \alpha_j^2
\left(2|y_i|^2|z_j|^2 + (y_i^*)^2 z_j^2 + y_i^2 (z_j^*)^2\right).
\end{equation}
Decoupling its resolvent part would require introducing fields such as
\begin{equation}
R = \frac{g^2}{N}\sum_i \alpha_i^2 |z_i|^2\quad \text{and} \quad Q_{zz}= \frac{g^2}{N}\sum_i \alpha_i^2 z_i^2.
\end{equation}
The neutral field $R$ survives, whereas $Q_{zz}$ carries charge $+2$ under the global rotation and is therefore set to zero on the $U(1)$-symmetric saddle, as are the charge-$\pm 1$ fields generated by the mixed term $2\tilde{x}_i\phi_j\alpha_j(y_i^*z_j+y_iz_j^*)$. Dropping all such charged fields, the surviving neutral fields we need to introduce are, in summary,
\begin{align*}
& m_{\tilde{x}}=\frac{1}{N}\sum_i \tilde{x}_{i}, \quad m_\phi=\frac{1}{N}\sum_i \phi_i, \\ 
 &q_{\phi\phi}=\frac{g^2}{N}\sum_i \phi_i^2, \quad q_{\tilde{x}\phi}= \frac{g^2}{N}\sum_i \tilde{x}_i \phi_i, \quad q_{\tilde{x}\tilde{x}}=\frac{g^2}{N}\sum_i \tilde{x}_i^2, \\
&   R = \frac{g^2}{N} \sum_i \alpha_i^2 |z_i|^2,\quad P = \frac{g^2}{N}\sum_i \alpha_i y_i^* z_i ,\quad P^* =  \frac{g^2}{N}\sum_i \alpha_i y_i z_i^*.
\end{align*}
We have recycled notation $R$ and $P$, but they are different from those defined in Appendix~\ref{app:DMFT}. Substituting these reductions back gives the disorder average on the $U(1)$-symmetric saddle:
\begin{align}
\left\langle
\exp\left(-\im\sum_{ij}J_{ij}h_{ij}\right)
\right\rangle
&=\exp\bigg(-\im J_0 N m_{\tilde{x}}m_{\phi}
-AR\sum_i |y_i|^2
-\frac{N}{2g^2}Aq_{\tilde{x}\tilde{x}}q_{\phi\phi} \nonumber \\
& \qquad -\frac{N}{2g^2}B\,q_{\tilde{x}\phi}^2
-\frac{BN}{2g^2}P^2 -\frac{BN}{2g^2}(P^*)^2 \nonumber \\
& \qquad-\frac{N^2\taudiv}{2}m_{\tilde{x}}^2 q_{\phi\phi}
-\frac{N^2\taucon}{2}m_{\phi}^2 q_{\tilde{x}\tilde{x}} 
-N^2\tauchn m_{\tilde{x}}m_{\phi}q_{\tilde{x}\phi}
\bigg).
\end{align}

While the resolvent fields $\left\{R, P\right\}$ depend on the fixed-point fields through local susceptibility $\alpha_i$, the fixed-point fields $m$ and $q$ do not depend on the resolvent fields. Consequently, the saddle-point equations for the fixed-point sector remain identical to those of the unperturbed system. The integral over $x$ and $\tilde{x}$ simply reconstructs the normalization of the fixed-point distribution. This allows us to integrate out the fixed-point fields by replacing them with quenched random variables $x_i$ drawn from the theoretically obtained fixed-point distribution (see Appendix~\ref{app:FerroFPDist}). Thus, we write 
$\langle \mathcal{Z} \rangle= \langle \mathcal{Z}_1 \rangle_x$,
where
\begin{align}
 \mathcal{Z}_1= &\int \calD y \calD y^* \calD z \calD z^* \, \exp\bigg(
-\sum_i (1+AR)|y_i|^2 - \epsilon \sum_i |z_i|^2
+ \im \sum_i \left(\omega y_i^* z_i + \omega^* y_i z_i^*\right)\nonumber\\
&  \qquad \qquad -\frac{BN}{2 g^2} \left( P^2 + (P^*)^2\right)\bigg).
\end{align}
We then impose the definition of $R$ and $P$ by inserting $\delta$ functions
\begin{align}
& 1=\int\calD R \calD \widetilde{R}\, \exp\left(\widetilde{R}\left(\frac{N}{g^2}R- \sum_i \alpha_i^2|z_i|^2\right)\right), \\
& 1 = \int \calD P \,\calD P^* \,\calD \widetilde{P} \,\calD \widetilde{P}^*
\exp\left(
\widetilde{P}
\left(\frac{N}{g^2}P - \sum_i \alpha_i y_i^* z_i
\right)+ \operatorname{c.c.}
\right),
\end{align}
which leads to 
\begin{equation}
\mathcal{Z}_1 = \int \calD R \calD \widetilde{R} \calD P \calD P^* \calD \widetilde{P} \calD \widetilde{P}^* \exp\left( \frac{N}{g^2} \left(\widetilde{R} R + \widetilde{P} P +\widetilde{P}^*  P^*\right)  \right.
 \left.-\frac{BN}{2g^2}\left(P^2 + (P^*)^2\right) \right)\times \int \calD y \calD z \, \exp\left( S_{\text{micro}} \right),
\end{equation}
where the microscopic action $S_{\text{micro}}$ gathers all terms containing $y_i$ and $z_i$:
\begin{equation}
S_{\text{micro}} = \sum_i \left[ -(1+AR)|y_i|^2 - \left(\epsilon + \widetilde{R}\alpha_i^2 \right)|z_i|^2  +  \left(\im\omega - \alpha_i \widetilde{P}\right) y_i^* z_i + \left(\im\omega^* - \alpha_i \widetilde{P}^*\right) y_i z_i^* \right].
\end{equation}
\end{widetext}
Integrating out $y_i$ yields
\begin{equation}
\int \frac{\diff^2 y_i}{\pi}\exp(S_{\text{micro},i})
=\frac{1}{1+AR}
\exp\left(-\calC(\alpha_i)|z_i|^2\right),
\end{equation}
where
\begin{equation}
\calC(\alpha_i) =\epsilon +  \alpha_i^2\widetilde{R}
- \frac{\left(\im\omega - \alpha_i\widetilde{P} \right)\left(\im\omega^* -\alpha_i \widetilde{P}^* \right)}{1+AR}.
\end{equation}
Assuming $\mathrm{Re}\, \calC(\alpha_i) > 0$, the remaining integral is Gaussian
\begin{equation}
\int \frac{\diff^2 z_i}{\pi}\exp\left(-\calC(\alpha_i)|z_i|^2\right)
= \frac{1}{\calC(\alpha_i)}.
\end{equation}
The microscopic contribution is therefore
\begin{eqnarray}
\mathcal{Z}_{\mathrm{micro}}&&= 
\prod_i \frac{1}{(1+AR)\calC(\alpha_i)} \nonumber \\
&& \!\!\!\!=\exp\left(-N\ln(1+AR)- \sum_i \ln \calC(\alpha_i)\right).
\end{eqnarray}
Writing in a large-$N$ form, we have 
\begin{equation}
\mathcal{Z}_1 =
\int \calD R\,\calD \widetilde{R}\,\calD P\,\calD P^*\,\calD \widetilde{P}\,\calD \widetilde{P}^*
\exp\left(-N\Omega\right),
\end{equation}
where
\begin{eqnarray}
\label{eq:EigenBulkPhi}
\Omega&&=-\frac{\widetilde{R} R + \widetilde{P} P + \widetilde{P}^* P^*}{g^2} \nonumber \\ && \quad + \frac{B}{2g^2}\left(P^2 + \left(P^*\right)^2\right)+ \ln(1+AR)\nonumber\\  && \quad+ \frac{1}{N}\sum_i \ln \calC(\alpha_i).
\end{eqnarray}
Since $\langle\mathcal{Z}\rangle$ is evaluated by $\exp(-N\Omega)$ at the saddle point, the potential defined above satisfies $\varphi(\omega)=-\Omega(\omega)$ up to $\omega$-independent constants.

\subsection{Eigenvalue bulk for $B=0$} \label{app:B0Bulk}
If the effective parameter $B = 0$, the saddle-point condition
\begin{equation}
\label{eq:SaddleP}
\frac{\partial \Omega}{\partial P}=-\frac{\widetilde{P}}{g^2}=0, \qquad 
\frac{\partial \Omega}{\partial P^*}=-\frac{\widetilde{P}^*}{g^2}=0
\end{equation}
set $\widetilde{P}$ and $\widetilde{P}^*$ to zero as well. This reduces Eq.~\eqref{eq:EigenBulkPhi} to 
\begin{equation}
\Omega_R = - \frac{\widetilde{R}R}{g^2}+\ln(1+AR)+\avgAlpha{\ln\mathcal{C}_R(\alpha)},
\end{equation}
in the large-$N$ limit, where 
\begin{equation}
\calC_R(\alpha)=\epsilon+\alpha^2\widetilde{R}+\frac{|\omega|^2}{1+AR}.
\end{equation}
The saddle equations for $R$ and $\widetilde{R}$ are
\begin{eqnarray}
\frac{\partial \Omega_R}{\partial \widetilde{R}}
&&=-\frac{R}{g^2} + \avgAlpha{\frac{\alpha^2}{\calC_R(\alpha)}}= 0, \label{eq:B0RSaddle} \\
\frac{\partial \Omega_R}{\partial R} &&=-\frac{\widetilde{R}}{g^2}
+ \frac{A}{1+AR} \nonumber\\ &&\quad+ \avgAlpha{\frac{1}{\calC_R(\alpha)}\frac{\partial \calC_R(\alpha)}{\partial{R}}}
= 0. \label{eq:B0tildeRSaddleOriginal} 
\end{eqnarray}
Since 
\begin{equation}
\frac{\partial \calC_R}{\partial R}=\frac{A}{1+AR} \left(\epsilon+\alpha^2 \widetilde{R}-\calC_R\right),
\end{equation}
the second equation becomes
\begin{equation}
\label{eq:B0tildeRSaddle}
\widetilde{R}=\epsilon A g^2 \avgAlpha{\frac{1}{\calC_R(\alpha)}},
\end{equation}
where we have used Eq.~\eqref{eq:B0RSaddle}. 
Let $c = \widetilde{R}R$.
Now we introduce the rescaled denominator
\begin{equation}
K(\alpha):=AR\calC_R(\alpha)=\epsilon AR+\alpha^2 A c+\frac{AR}{1+AR}|\omega|^2,
\end{equation}
so that Eqs.~\eqref{eq:B0RSaddle}\&\eqref{eq:B0tildeRSaddle} become
\begin{eqnarray}
&&1 = Ag^2\avgAlpha{\frac{ \alpha^2}{K(\alpha)}}, \label{eq:B0S2}\\
& &c=\epsilon (ARg)^2 \avgAlpha{\frac{1}{K(\alpha)}}. \label{eq:B0S0}
\end{eqnarray}
Inside the eigenvalue bulk we seek the nontrivial branch on which $R\to\infty$ as
$\epsilon\to0^+$. We further assume (and later verify self-consistently) that
$\avgAlpha{K^{-1}(\alpha)}$ and $c$ remain finite in this limit. Since $R>0$ by
definition and convergence of the microscopic Gaussian integrals requires
$\operatorname{Re}\calC_R(\alpha)>0$, we have $\operatorname{Re}K(\alpha)>0$. Eq.~\eqref{eq:B0S0} then forces $\operatorname{Re}c>0$; Eq.~\eqref{eq:B0S2} requires $c$ to be real, hence $c>0$ on this bulk branch. Moreover,
Eq.~\eqref{eq:B0S0} implies
\begin{equation}
R\sim \epsilon^{-1/2}, \qquad \epsilon\to0^+,
\end{equation}
in agreement with the partially symmetric Gaussian ensemble studied in
Ref.~\onlinecite{RN318}. Using this scaling relationship in Eq.~\eqref{eq:B0S2} and taking
$\epsilon\to0^+$ yields
\begin{equation}
1 = g^2 \avgAlpha{\frac{\alpha^2}{\alpha^2 c+ A^{-1} |\omega|^2 }}.
\end{equation}
For fixed $\omega$, this self-consistency equation defines (when it exists) a
positive solution $c=c(\omega)>0$ corresponding to the bulk saddle. As $\omega$ is moved outward, this solution decreases and eventually disappears at the edge of the spectrum. Since the bulk branch is constrained to $c> 0$, the loss of the solution can only occur when it becomes marginal at $c\to 0^+$. Setting
$c=0$ in the limiting equation therefore gives the bulk boundary
\begin{equation}
|\omega|^2=x^2+y^2 = A g^2 \avgAlpha{\alpha^2}.
\end{equation}

\subsection{Eigenvalue bulk for $B \neq 0$}
If $B\neq 0$, the saddle condition Eq.~\eqref{eq:SaddleP} becomes
\begin{equation}
\frac{\partial \Omega}{\partial P}
=-\frac{\widetilde{P}}{g^2}+ \frac{B P}{g^2} = 0,\quad
\frac{\partial \Omega}{\partial P^*}
=-\frac{\widetilde{P}^*}{g^2}+ \frac{B P^*}{g^2} = 0.
\end{equation}
Substituting these back gives the reduced action
\begin{equation}
\label{eq:Bneq0FullAction}
\Omega=-\frac{\widetilde{R} R}{g^2}- \frac{\widetilde{P}^2 + \left(\widetilde{P}^*\right)^2}{2B g^2}+ \ln(1+AR)
+ \avgAlpha{\ln \calC(\alpha)}
\end{equation}
in the large-$N$ limit, where
\begin{equation}
\calC(\alpha)
=\epsilon + \alpha^2\widetilde{R} 
- \frac{\left(\im\omega -  \alpha\widetilde{P}\right)\left(\im\omega^* - \alpha\widetilde{P}^* \right)}{1+AR}.
\end{equation}
Differentiating with respect to $R$ and $\widetilde{R}$ as in Eqs.~\eqref{eq:B0RSaddle}\&\eqref{eq:B0tildeRSaddleOriginal} gives the saddle conditions in the same form as Eqs.~\eqref{eq:B0S2}\&\eqref{eq:B0S0}, with $K(\alpha)$ redefined as
\begin{eqnarray}
K(\alpha):&&=AR \calC(\alpha) \nonumber \\ &&=\epsilon A R+\alpha^2 A c \nonumber \\ &&\quad -\frac{AR}{1+AR}\left(\im\omega - \alpha\widetilde{P}\right)
\left(\im\omega^* - \alpha\widetilde{P}^*\right).
\end{eqnarray}
Substituting $K$ back into the action \eqref{eq:Bneq0FullAction} and taking the limit $R\to\infty$ on the bulk branch gives
\begin{equation}
\label{eq:Bneq0RInfAction}
\Omega=-\frac{\widetilde{R} R}{g^2}- \frac{\widetilde{P}^2 + \left(\widetilde{P}^*\right)^2}{2B g^2}
+ \avgAlpha{\ln K(\alpha)}
\end{equation}
with 
\begin{equation}
K(\alpha)=\epsilon A R+\alpha^2 A c-\left(\im\omega - \alpha\widetilde{P}\right)
\left(\im\omega^* - \alpha\widetilde{P}^*\right).
\end{equation}
Differentiating Eq.~\eqref{eq:Bneq0RInfAction} with respect to $\widetilde{P}$ and $\widetilde{P}^*$ yields
\begin{subequations}
\begin{eqnarray}
&&\widetilde{P}
=B g^2 \avgAlpha{\frac{\alpha\left(\im\omega^* -  \alpha\widetilde{P}^*\right)}{K(\alpha)}}, \\
&&\widetilde{P}^*
=B g^2 \avgAlpha{\frac{\alpha\left(\im\omega -  \alpha\widetilde{P}\right)}{K(\alpha)}}.
\end{eqnarray}
\end{subequations}
Defining 
\begin{equation}
S_n(x,y) = g^2\avgAlpha{\frac{\alpha^n}{K(\alpha)}}
\end{equation}
and using Eq.~\eqref{eq:B0S2} we obtain
\begin{eqnarray}
K(\alpha) &&= \epsilon AR+\alpha^2 Ac + x^2\left(1 - \frac{B S_1 \alpha}{1+B/A}\right)^2 \nonumber \\ && \qquad + y^2\left(1 + \frac{B S_1 \alpha}{1-B/A}\right)^2.
\end{eqnarray}
As we have discussed in the last subsection, $c$ remains finite and $R\sim \epsilon^{-1/2}$ as $\epsilon \to 0^+$. The boundary is obtained by setting $c=0$ in Eq.~\eqref{eq:B0S2}, which gives
\begin{equation}
\label{eq:GeneralBulkBoundary}
1 =Ag^2 \avgAlpha{\frac{\alpha^2}{ x^2\left(1 - \frac{B S_1 \alpha}{1+B/A}\right)^2 + y^2\left(1 + \frac{B S_1 \alpha}{1-B/A}\right)^2}}.
\end{equation}
Taking the limit $B\to 0$, we recover the result for $B=0$ in the last subsection; thus Eq.~\eqref{eq:GeneralBulkBoundary} is the general expression for any motif combination with $|B|<A$. 

Setting $\alpha=1$, we recover the bulk boundary of matrix $J$:
\begin{equation}
 \frac{x^2}{ g^2 (A + B)^2/A} + \frac{y^2}{g^2 (A- B)^2/A} = 1 ,
\end{equation}
which describes an ellipse with semi-axes $g(A\pm B)/\sqrt{A}$.
If we assume $\tauchn$, $\taucon$, and $\taudiv$ scale as $\mathcal{O} (N^{-1})$, this result can be viewed as a finite-size correction to a similar calculation in Ref.~\onlinecite{RN316} (although our bulk boundary calculation above does not require this scaling assumption).

\begin{figure*}[tbp]
\includegraphics{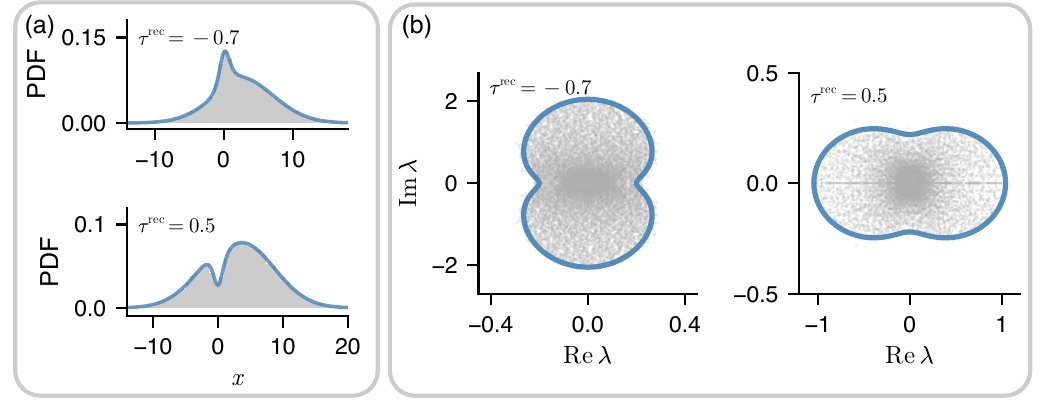}
\caption{\label{fig:FPStatisticsandJacEigen}
Fixed-point distribution and eigenspectrum bulk of $J^\phi$ at ferromagnetic fixed points. (a) Distribution of the fixed-point coordinates $x_i^\ast$ for large $|B|$. Gray histograms show the empirical distribution across ten network realizations, and solid blue curves show the self-consistent theoretical density from Eq.~\eqref{eq:FerroFPDistGeneral}. (b) Example eigenspectrum bulks at the
same parameter values. Gray dots are eigenvalues numerically obtained from ten
$J^\phi$ samples, and blue curves show the theoretical bulk boundary
Eq.~\eqref{eq:GeneralBulkBoundary} evaluated using the theoretical fixed-point
distribution in (a). For each $\taurec$ value, we aggregated results from ten
independent samples. Other parameters: $N=4000$, $J_0=-5$, $\tauchn=\taucon=\taudiv=10/N$, $g=3$.
}
\end{figure*}

This result highlights an interesting distinction: when $B=0$, the bulk boundary of $J^\phi$ can be obtained simply by scaling the circular boundary of the matrix $J$ by the effective gain $\sqrt{\langle\phi'(x)^2\rangle}$. While Harish and Hansel \cite{RN285} noted this based on numerical evidence, it is also consistent with the formal random-matrix analysis in Ref.~\onlinecite{RN364}, which justifies the same scaling for a quenched diagonal gain profile with a prescribed distribution. However, this simple scaling relationship breaks down when $B\neq0$. Because Eq.~\eqref{eq:GeneralBulkBoundary} no longer describes an ellipse in this regime, the boundary undergoes a fundamental change in shape rather than a simple rescaling (Fig.~\ref{fig:FPStatisticsandJacEigen}).

\subsection{The distribution of the ferromagnetic fixed points for $B\neq 0$}
\label{app:FerroFPDist}

In Sec.~\ref{sec:PosChain}, we considered networks with negligible $B$.
In that case, the ferromagnetic fixed points have Gaussian distributions
with mean and variance given by Eq.~\eqref{eq:B0FFPDist}. When $B\neq 0$,
the fixed-point distribution is no longer Gaussian because the retarded
self-interaction term is a static nonlinear feedback.

At a fixed point, the DMFT equation \eqref{eq:GenericDMFTEq} reduces to
\begin{equation}
\label{eq:FerroFPStaticEq}
x=\left(J_0+g^2N\tauchn \chiphi_{\mathrm{int}}\right)
\langle \phi\rangle+g^2B\chiphi_{\mathrm{int}}\phi(x)
+\eta ,
\end{equation}
where $\chiphi_{\mathrm{int}}=\int_0^\infty \chiphi(\tau)\,\diff \tau$
is the integrated response. The self-consistency equation for
$\chiphi_{\mathrm{int}}$ follows by differentiating the static equation
with respect to the local Gaussian, which gives
\begin{equation}
\label{eq:FerroFPChiInt}
\chiphi_{\mathrm{int}}
=\left\langle\frac{\phi'(x)}{1-g^2B\chiphi_{\mathrm{int}}\phi'(x)}\right\rangle_x .
\end{equation}

The quenched Gaussian field $\eta$ has zero mean and variance
\begin{equation}
\Delta^\eta=g^2 A\left\langle \phi^2\right\rangle
+g^2N\taucon \langle \phi\rangle^2.
\end{equation}
Thus the Gaussian variable is not $x$, but rather
\begin{equation}
\label{eq:FerroFPFDef}
\eta=F(x):=x-g^2B\chiphi_{\mathrm{int}}\phi(x)
-\left(J_0+g^2N\tauchn\chiphi_{\mathrm{int}}\right)
\langle \phi\rangle .
\end{equation}
If $F$ is one-to-one, the fixed-point distribution follows from a change of
variables. Since $\eta$ is Gaussian, we obtain
\begin{equation}
\label{eq:FerroFPDistGeneral}
p(x)=\frac{\left|F'(x)\right|}{\sqrt{2\pi\Delta^\eta}}\exp\left(-\frac{F(x)^2}{2\Delta^\eta}
\right).
\end{equation}
The parameters appearing in this distribution, including
$\Delta^\eta$, $\langle \phi\rangle$, $\left\langle \phi^2\right\rangle$, and
$\chiphi_{\mathrm{int}}$, are determined self-consistently from $p(x)$.

The change-of-variables formula assumes that $F$ is strictly monotone.
A sufficient condition for $F$ to be strictly increasing is
\begin{equation}
g^2B\chiphi_{\mathrm{int}}\sup_x \phi'(x)<1 .
\end{equation}
In particular, when $B\chiphi_{\mathrm{int}}<0$, this condition is
automatically satisfied for monotonically increasing $\phi$. When this
condition fails, the scalar fixed-point equation can become multi-valued, and
the one-to-one change-of-variables formula above no longer applies directly.
This breakdown is consistent with the onset of the glassy regime.

\section{Stability of the homogeneous chaos phase} \label{app:HCStability}

In this section, we analyze the linear stability of the HC phase against
perturbations of the ensemble mean. Let
$\mu^x(t)=\langle x(t)\rangle$ and
$\mu^\phi(t)=\langle \phi(x(t))\rangle$. In the HC phase,
$\mu_0^x=\mu_0^\phi=0$. Let $\chiphi(\tau)$ denote the stationary linear response of
$\mu^\phi(t)$ to a weak external input in the HC phase. This
response includes the effect of the retarded self-interaction proportional to $B$.
The perturbation of the population mean generates the effective field
\begin{equation}
\delta h (t)= J_0 \delta\mu^\phi(t)  + g^2 N\tauchn\int_{-\infty}^t \chiphi\left(t-s\right) \delta\mu^\phi\left(s\right) \, \diff s.
\end{equation}
The induced perturbation of the mean firing rate satisfies
\begin{equation}
\delta\mu^\phi(t)
=\int_{-\infty}^t \chiphi(t-t')\delta h(t')\,\diff t' .
\end{equation}
Setting $\delta \mu^\phi = m_0 \mathrm{e}^{\lambda t}$ with $m_0\neq0$, we obtain
\begin{equation} \label{eq:HCStabilityCondition}
1= \widehat{\chiphi}(\lambda)\left[J_0+ g^2 N \tauchn \widehat{\chiphi}(\lambda)\right],
\end{equation}
where $\widehat{\chiphi}(\lambda)=\int_0^\infty \mathrm{e}^{-\lambda \tau} \chiphi(\tau) \, \diff \tau$. The HC phase loses stability when a solution of
Eq.~\eqref{eq:HCStabilityCondition} crosses into the right half-plane. A static instability corresponds to $\lambda=0$, giving the HC--SC boundary
\begin{equation}
1 = \widehat{\chiphi}(0)\left(J_0 +  g^2N\tauchn\widehat{\chiphi}(0)\right).
\end{equation}
An oscillatory instability corresponds instead to
$\lambda=\im\omega$ with $\omega\neq0$. The corresponding HC stability boundary is 
therefore determined by
\begin{equation}
1=\widehat{\chiphi}(\im\omega)\left(J_0+g^2N\tauchn \widehat{\chiphi}(\im\omega)\right),
\qquad\omega\neq0 .
\end{equation}

For $B\sim 1/N$, the characteristic equation can be simplified because the
stationary HC single-site process is Gaussian. In this limit, the response kernel of
$x$ is simply
\begin{equation}
    \chix(\tau)=\mathrm{e}^{-\tau}\Theta(\tau),
\end{equation}
where $\Theta(\cdot)$ is the Heaviside step function. Therefore the response
kernel of the firing rate is
\begin{equation}
\chiphi(\tau)=\left\langle \phi' \right\rangle\mathrm{e}^{-\tau}\Theta(\tau),
\end{equation}
and its Laplace transform is
\begin{equation}
\widehat{\chiphi}(\lambda)=\frac{\left\langle \phi' \right\rangle}{1+\lambda}.
\end{equation}
Substituting this expression into the characteristic equation \eqref{eq:HCStabilityCondition}
gives the growth rates of the mean-field perturbations 
\begin{equation}
    \lambda_{\pm}
    =-1+\frac{\left\langle \phi' \right\rangle}{2}
    \left(J_0\pm\sqrt{J_0^2+4g^2N\tauchn}\right).
\end{equation}
The HC--SC boundary corresponds to a static instability, $\lambda=0$,
which gives Eq.~\eqref{eq:HCSCBoundaryB0} in the main text. By contrast, the oscillatory branch of the HC stability boundary corresponds to a complex-conjugate pair crossing the imaginary axis. In
the present $B\sim 1/N$ regime, this requires
\begin{equation}
    J_0^2+4g^2N\tauchn<0
\end{equation}
and
\begin{equation}
    \operatorname{Re}\lambda_{\pm}=-1+\frac{J_0\left\langle \phi' \right\rangle}{2}=0 .
\end{equation}
Thus the Hopf-like boundary is Eq.~\eqref{eq:HopfNegChn}
with onset oscillation frequency
\begin{equation}
    \omega=\frac{\left\langle \phi' \right\rangle}{2}
    \sqrt{-J_0^2-4g^2N\tauchn}.
\end{equation}

\section{Largest Lyapunov exponent for $B=0$} \label{app:LLEB0}
In the $B=0$ limit, the single-site DMFT equations take a form analogous to those of the network with non-vanishing mean coupling treated in Ref.~\onlinecite{RN211} (see their Appendices E\&F). We recapitulate the essential derivation steps below to show that, for $B=0$, the LLE retains the same relation to the lowest eigenvalue of an effective Schrödinger operator, whose coefficients depend on motif correlations.

Let us consider two replicas of the system indexed by $\alpha, \beta\in \{\rI, \rII\}$. We consider fluctuations around the mean $\delta x^\alpha(t) = x^\alpha(t) - \langle x^\alpha \rangle$. The DMFT equations [similar to Eq.~(14) in \cite{RN211}] for fluctuations are 
\begin{equation}
\frac{\diff}{\diff t}{\delta x}^\alpha(t) = -\delta x^{\alpha}(t)+\eta^\alpha(t),
\end{equation} 
where the shared connectivity creates a cross-correlation for $\alpha\neq \beta$:
\begin{equation}
C^\eta_{\alpha \beta}(t,s) = Ag^2 C^\phi_{\alpha\beta}(t,s)+g^2 N \taucon \left\langle \phi^\alpha \right\rangle \left\langle \phi^\beta \right\rangle .
\end{equation}
Defining $K_{\alpha\beta}^x(t,s) = \left\langle \delta x^\alpha(t) \delta x^\beta(s)\right\rangle$, we have 
\begin{align} \label{eq:KPartialDerivative}
(1+\partial_t)(1+\partial_s)K_{\alpha \beta}^x(t,s) & =  Ag^2 C^\phi_{\alpha\beta}(t,s) \nonumber \\
& \qquad +g^2N \taucon \left\langle \phi^\alpha \right\rangle \left\langle \phi^\beta\right\rangle.
\end{align}
The raw autocorrelation can be computed by $C^x_{\alpha\beta}(\tau)=K_{\alpha\beta}^x(\tau) +\left\langle x^\alpha \right\rangle \left\langle x^\beta \right\rangle$. The average squared distance between the replicas is
\begin{align}
    d(t,s)&:=\left\langle \left|x^{\rI}(t)-x^{\rII}(s)\right|^2\right\rangle \nonumber\\
    &\approx C^{x}_{\rI\rI}(t,s)+C^{x}_{\rII\rII}(t,s) -C^{x}_{\rI\rII}(t,s)  \nonumber \\& \qquad- C^{x}_{\rII\rI}(t,s)\nonumber\\
    &=K^{x}_{\rI\rI}(t,s)+K^{x}_{\rII\rII}(t,s)-K^{x}_{\rI\rII}(t,s) \nonumber \\& \qquad-K^{x}_{\rII\rI}(t,s).
\end{align}
The last step uses the fact that the two replicas should have identical mean $\left\langle x^{\rI}\right\rangle=\left\langle x^{\rII}\right\rangle$.
Applying Eq.~\eqref{eq:KPartialDerivative}, we arrive at 
\begin{align}
&(1+\partial_{t})(1+\partial_{s})d(t,s) =\nonumber\\ &\qquad Ag^2\left(\Cphi_{\rI\rI}+\Cphi_{\rII\rII}-  \Cphi_{\rI\rII}-\Cphi_{\rII\rI}\right),
\end{align}
where we have again used the fact that the two replicas have identical mean. If we ignore the $\mathcal{O}(1/N)$ motif correlations (setting $A=1$), we recover the result obtained in Eq.~(C6) of Ref.~\onlinecite{RN211}. Thus, the LLE takes the same form as in the large-\(N\) network without motifs [see Eqs.~(C10)--(C13) in Ref.~\onlinecite{RN211}],
\begin{equation}
\gamma_{\max}=-1+\sqrt{1-E_0},
\end{equation}
where \(E_0\) is the lowest eigenvalue of the effective Schr\"odinger operator
\begin{equation}\label{eq:ShrodingerOp}
\mathscr{L}=-\partial_\tau^2+1-Ag^2 C^{\phi'}(\tau),
\end{equation}
acting on square-integrable functions of \(\tau\). Here
\begin{equation}
C^{\phi'}(\tau)=\bigl\langle \phi'(x(t))\phi'(x(t+\tau)) \bigr\rangle_t
\end{equation}
is the stationary autocorrelation function of \(\phi'(x)\).
The motif dependence enters through the factor \(A\) and through the self-consistent mean \(\langle x\rangle\) and autocorrelation \(\Cx(\tau)\).

\section{Participation ratio dimension $D^\phi_{\text{PR}}$} \label{app:TwoCavity}
In this section, we assume $J_0=0$ and focus on the HC phase with $\tauchn\le0$. We note that the SC phase is qualitatively different, as it is characterized by high-dimensional temporal fluctuations around a structured rank-one mode \cite{RN320}. We leave the dimensionality analysis of the SC phase for future work and do not discuss it further in this paper. The derivation largely follows the two-cavity calculation in Ref.~\onlinecite{RN278}. We review the key steps here for completeness.

\subsection{Notation and cavity setup}
We add two cavity units labeled by
$\mu,\nu \in \{0,0'\}$ to the original network of $N$ units labeled by
$i,j \in \{1,\dots,N\}$. Let $\phi_i(t)$ denote the activity of neuron $i$ in the
original $N$-unit system before cavity units are introduced. The dynamics of a
cavity unit can then be written as
\begin{align}
    \dot{x}_\mu(t)
    = &-x_\mu(t) + \eta_\mu(t)\nonumber
    \\ &+ \frac{1}{\sqrt{N}}
    \sum_{\nu\in\{0,0'\}}
    \int_{-\infty}^{t} \diff t'\,
    F_{\mu\nu}(t,t') \phi_\nu(t'),
\end{align}
where
\begin{equation}
    \eta_\mu(t) = \sum_{i=1}^N J_{\mu i}\phi_i(t)
\end{equation}
is the cavity input, and the kernel
\begin{equation}
    F_{\mu\nu}(t,t')
    =
    \sqrt{N}\left[
        \sum_{i,j=1}^N J_{\mu i}\chiphi_{ij}(t,t')J_{j\nu}
        + J_{\mu\nu}\delta(t-t')
    \right]
\end{equation}
is $\calO(1)$ in \(1/\sqrt{N}\). As in the DMFT derivation (Appendix~\ref{app:DMFT}), we assume the scaling in Eq.~\eqref{eq:MotifScaling} so that the contributions of three-neuron motifs remain finite as $N \to \infty$. Once nonlocal motifs are present, \(J_{\mu i}\) and \(\phi_i\) are no longer
strictly independent, unlike in networks with only reciprocal correlations.
The corresponding correction to the cavity
factorization is higher order in the $N^{-1/2}$ expansion and will be neglected
here, following Ref.~\onlinecite{RN338}.

\subsection{The four-point function}
To derive $C^\phi_{00'}$, we first solve the cavity-unit equation
perturbatively to first order in $N^{-1/2}$. The solution can be written as
\begin{align}
    \phi_\mu(t)&=\phi_\mu^{\mathrm{free}}(t)
    \\ &\quad +\frac{1}{\sqrt{N}}\int \diff s\, \chiphi(t,s)
    \int \diff p \sum_\nu F_{\mu\nu}(s,p)\phi_\nu^{\mathrm{free}}(p),
    \label{eq:CavitySolution}
\end{align}
where $\phi_\mu^{\mathrm{free}}$ is obtained from the free cavity dynamics
\begin{equation}
    \dot{x}_\mu^{\mathrm{free}}(t)
    =-x_\mu^{\mathrm{free}}(t)+\eta_\mu(t).
\end{equation}
We can also compute the corresponding cross-response function $\chiphi_{\mu\nu}(t,t')$ for
$\mu\neq \nu$ by taking the functional derivative of
Eq.~\eqref{eq:CavitySolution}:
\begin{equation}
    \chiphi_{\mu\nu}(t,t')
    = \frac{1}{\sqrt{N}}\int \diff s\, \chiphi(t,s)
    \int \diff p\, F_{\mu\nu}(s,p)\chiphi(p,t').
    \label{eq:CrossResponseTime}
\end{equation}
In the frequency domain and assuming stationarity, Eq.~\eqref{eq:CrossResponseTime}
becomes
\begin{equation}
    \chiphi_{\mu\nu}(\omega)
    =2\pi N^{-1/2}{\chiphi(\omega)}^2 F_{\mu\nu}(\omega),
    \qquad \mu\neq \nu.
    \label{eq:CrossResponse}
\end{equation}
This result will be useful in subsequent calculations.

Using the same expansion as in Ref.~\onlinecite{RN278}, one obtains
\begin{align}
    C^\phi_{00'}(\omega)&=
    2\pi\bigg[
        \left|\chiphi(\omega)\right|^2 C^\eta_{00'}(\omega)\nonumber \\
         & \!\!\!\!\!\!\!\! + \frac{1}{\sqrt{N}}
        \left({\left(F_{00'}(\omega)\chiphi(\omega)\right)}^*
            + F_{00'}(\omega)\chiphi(\omega)\right)\Cphi(\omega)
    \bigg],
    \label{eq:C00phi}
\end{align}
with
\begin{equation}
    C^\eta_{00'}(\tau)= \sum_{i,j=1}^N J_{0i}J_{0'j}C^\phi_{ij}(\tau).
    \label{eq:Ceta-def}
\end{equation}
Here $C^\phi_{ij}(\tau)=\langle \phi_i(t)\phi_j(t+\tau)\rangle_t$, and the
frequency-domain quantities in Eq.~\eqref{eq:C00phi} are the corresponding
unitary Fourier transforms. Since Eq.~\eqref{eq:C00phi} assumes $C^\eta_{00'} \sim N^{-1/2}$, we must check that the presence of three-neuron motifs does not change this scaling.

Split Eq.~\eqref{eq:Ceta-def} into diagonal and off-diagonal pieces:
\begin{equation}
    C^\eta_{00'}(\tau)
    =\sum_{i=1}^N J_{0i}J_{0'i}C^\phi_{ii}(\tau)
    +\sum_{i \ne j} J_{0i}J_{0'j}C^\phi_{ij}(\tau).
    \label{eq:Ceta-split}
\end{equation}
The diagonal term has a small mean of order $\taudiv$, which is negligible under
the scaling assumption $N\taudiv\sim 1$. Its sample-to-sample fluctuation is
$\calO\left(N^{-1/2}\right)$, so the diagonal contribution is
$\calO\left(N^{-1/2}\right)$. For the off-diagonal term, cross-covariances \(\Cphi_{ij}\) are \(\calO\left(N^{-1/2}\right)\). The connectivity $J_{0i}$ and $J_{0'j}$ are uncorrelated
for $i \neq j$, so the whole sum is again \(\calO\left(N^{-1/2}\right)\).
Combining the two pieces justifies the scaling assumption needed for Eq.~\eqref{eq:C00phi}.

To obtain the four-point function, we take the $J$-average of
$C^\phi_{00'}(\omega_1)C^\phi_{00'}(\omega_2)$. Inserting 
Eq.~\eqref{eq:C00phi}, we obtain
\begin{widetext}
\begin{align}
   \calC^\phi(\bm{\omega})= N\left\langle
        C^\phi_{00'}(\omega_1)C^\phi_{00'}(\omega_2)
    \right\rangle_J &=
    4\pi^2\Bigg[
        \left|\chiphi(\omega_1)\chiphi(\omega_2)\right|^2
        \Gamma_{C^\eta_{00'}C^\eta_{00'}}(\bm{\omega})
        \nonumber\\
        &\qquad\quad
        + \Cphi(\omega_1)\Cphi(\omega_2)\left(
            \chiphi(\omega_1)\chiphi(\omega_2)
            \Gamma_{F_{00'}F_{00'}}(\bm{\omega})
            + {\chiphi(\omega_1)}^*\chiphi(\omega_2)
            \Gamma_{F^*_{00'}F_{0'0}}(\bm{\omega})
        \right)
        \nonumber\\
        &\qquad\quad
        + \left|\chiphi(\omega_1)\right|^2 {\chiphi(\omega_2)}^*\Cphi(\omega_2)
        \Gamma_{C^\eta_{00'}F^*_{00'}}(\bm{\omega})
        \nonumber\\
        &\qquad\quad
        + \left|\chiphi(\omega_2)\right|^2 {\chiphi(\omega_1)}^*\Cphi(\omega_1)
        \Gamma_{F^*_{00'}C^\eta_{00'}}(\bm{\omega})
        + \operatorname{c.c.}
    \Bigg].
    \label{eq:Cphi4-preGamma}
\end{align}
\end{widetext}
The disorder average~\eqref{eq:Cphi4-preGamma} involves four distinct $\calO(1)$ cavity correlator functions:
\begin{subequations}
\begin{align}
    \Gamma_{C^\eta_{00'}C^\eta_{00'}}(\bm{\omega})
    &=N\left\langle
        C^\eta_{00'}(\omega_1)C^\eta_{00'}(\omega_2)\right\rangle_J, \\
    \Gamma_{F_{00'}F_{00'}}(\bm{\omega}) 
    &=\left\langle F_{00'}(\omega_1)F_{00'}(\omega_2)\right\rangle_J,\\
    \Gamma_{F^*_{00'}F_{0'0}}(\bm{\omega})
    &=\left\langle F^*_{00'}(\omega_1)F_{0'0}(\omega_2)\right\rangle_J,\\
    \Gamma_{F^*_{00'}C^\eta_{00'}}(\bm{\omega})
    &=\sqrt{N}\left\langle
        F^*_{00'}(\omega_1)C^\eta_{00'}(\omega_2)\right\rangle_J,
\end{align}
\end{subequations}
where \(F_{\mu\nu}(\tau)=\langle F_{\mu\nu}(t,t-\tau)\rangle_t\).
The other two functions can be obtained by symmetry.
Clark et al. \cite{RN278} have computed these cavity correlator functions when the network has a nonzero $\taurec$. Below, we will show that under the scaling assumption \eqref{eq:MotifScaling}, the other three motifs contribute to $\calO(N^{-1})$ finite-size corrections of these cavity correlator functions.

\subsection{Evaluation of the cavity correlator functions}
We first evaluate $\Gamma_{C^\eta_{00'}C^\eta_{00'}}(\bm{\omega})$. Using
Eq.~\eqref{eq:Ceta-def}, we have
\begin{equation}
    \Gamma_{C^\eta_{00'}C^\eta_{00'}}(\bm{\omega})
    =N \sum_{ijkl}
    \left\langle
        J_{0i}J_{0'j}J_{0k}J_{0'l}
    \right\rangle_J
    C^\phi_{ij}(\omega_1)C^\phi_{kl}(\omega_2).
\end{equation}
Since the connectivity is Gaussian, Wick's theorem gives
\begin{align}
    \left\langle
        J_{0i}J_{0'j}J_{0k}J_{0'l}
    \right\rangle_J=&\,
    \llangle J_{0i}J_{0'j}\rrangle
    \llangle J_{0k}J_{0'l}\rrangle
    \nonumber\\
    &+\llangle J_{0i}J_{0k}\rrangle
    \llangle J_{0'j}J_{0'l}\rrangle
    \nonumber\\
    &+\llangle J_{0i}J_{0'l}\rrangle
    \llangle J_{0'j}J_{0k}\rrangle.
\end{align}
We denote the three resulting contributions by $T_1$, $T_2$, and $T_3$.

For the first contraction, only the divergent cumulant contributes, hence
\begin{align}
    T_1&= N \sum_{ik}
    \llangle J_{0i}J_{0'i}\rrangle
    \llangle J_{0k}J_{0'k}\rrangle
    C^\phi_{ii}(\omega_1)C^\phi_{kk}(\omega_2)
    \nonumber\\
    &=N{\left(\taudiv\right)}^2 g^4 \Cphi(\omega_1)\Cphi(\omega_2)
    \sim \frac{1}{N}.
\end{align}

For the second contraction, we use
\begin{align*}
    &\llangle J_{0i}J_{0k}\rrangle
    =\frac{g^2}{N}\left[\delta_{ik}+\taucon(1-\delta_{ik})\right],
    \\
    &\llangle J_{0'j}J_{0'l}\rrangle
    =\frac{g^2}{N}\left[\delta_{jl}+\taucon(1-\delta_{jl})\right].
\end{align*}
Substituting these into the definition of $T_2$ gives
\begin{align}
    T_2=T_2^{(\delta\delta)}+T_2^{(\mathrm{mix})}+T_2^{(\tau\tau)},
\end{align}
where
\begin{subequations}
\begin{align}
    T_2^{(\delta\delta)} &:=\frac{g^4}{N}\sum_{ij}
    C^\phi_{ij}(\omega_1)C^\phi_{ij}(\omega_2),
    \\T_2^{(\mathrm{mix})}
    &:= \frac{\taucon g^4}{N} \nonumber \\ 
    &\, \times\sum_{ijkl} \left[\delta_{ik}(1-\delta_{jl})+(1-\delta_{ik})\delta_{jl}\right]
    C^\phi_{ij}(\omega_1)C^\phi_{kl}(\omega_2),\\
    T_2^{(\tau\tau)}
    &:=\frac{{\left(\taucon\right)}^2 g^4}{N}\sum_{ijkl}
    (1-\delta_{ik})(1-\delta_{jl})
    C^\phi_{ij}(\omega_1)C^\phi_{kl}(\omega_2).
\end{align}
\end{subequations}
The diagonal-diagonal piece is
\begin{align}
    T_2^{(\delta\delta)}
    &=\frac{g^4}{N}\sum_{ij}
    C^\phi_{ij}(\omega_1)C^\phi_{ij}(\omega_2)
    \nonumber\\
    &=g^4\left[
        \Cphi(\omega_1)\Cphi(\omega_2)+ \calC^\phi(\bm{\omega})
    \right].
\end{align}

For the mixed piece, one of the two covariance factors is forced onto a diagonal
index pair while the other remains off-diagonal. Thus
\begin{align}
    T_2^{(\mathrm{mix})}
    &=\frac{\taucon g^4}{N}\Bigg[
        \sum_{ijl}(1-\delta_{jl})
        C^\phi_{ij}(\omega_1)C^\phi_{il}(\omega_2)\nonumber\\
        &\qquad\qquad+\sum_{ijk}(1-\delta_{ik})
        C^\phi_{ij}(\omega_1)C^\phi_{kj}(\omega_2)\Bigg]\nonumber\\
    &\sim \frac{1}{N^{3/2}}.
\end{align}
The largest contribution comes from terms containing one diagonal and one
off-diagonal covariance, e.g.
\begin{equation}
    \sum_{i\neq k}
    C^\phi_{ii}(\omega_1)C^\phi_{ki}(\omega_2)
    \sim \sqrt{N^2}\,N^{-1/2}=\sqrt{N},
\end{equation}
while terms with two off-diagonal covariances are of smaller order.
Therefore $T_2^{(\mathrm{mix})}$ is subleading.

For the $\tau\tau$ piece, both factors come from the off-diagonal part of the
connectivity covariance:
\begin{align}
    T_2^{(\tau\tau)}
    &=\frac{{\left(\taucon\right)}^2 g^4}{N}
    \sum_{ijkl}
    (1-\delta_{ik})(1-\delta_{jl})
    C^\phi_{ij}(\omega_1)C^\phi_{kl}(\omega_2)
    \nonumber\\
    &=\frac{{\left(\taucon\right)}^2 g^4}{N}
    \left[N^2 \Cphi(\omega_1)\Cphi(\omega_2)+ \calO(N)
    \right]\nonumber\\
    &=N{\left(\taucon\right)}^2 g^4 \Cphi(\omega_1)\Cphi(\omega_2)
    \sim \frac{1}{N}.
\end{align}
Collecting the three pieces, we obtain
\begin{align}
    T_2
    &=
    g^4\left[
        \left(1+N{\left(\taucon\right)}^2\right)\Cphi(\omega_1)\Cphi(\omega_2)
        + \calC^\phi(\bm{\omega}) \right] \nonumber\\
    & \qquad + \calO\left(N^{-3/2}\right).
\end{align}

For the third contraction, again only divergent cumulants survive, thus
\begin{align}
    T_3&=N \sum_{ij}
    \llangle J_{0i}J_{0'i}\rrangle
    \llangle J_{0j}J_{0'j}\rrangle
    C^\phi_{ij}(\omega_1)C^\phi_{ji}(\omega_2)\nonumber\\
    &={\left(\taudiv\right)}^2 g^4
    \left[
        \Cphi(\omega_1)\Cphi(\omega_2)
        + \calC^\phi(\bm{\omega})
    \right]\sim \frac{1}{N^2}.
\end{align}

Collecting the three contributions, we obtain
\begin{align}\label{eq:GammaCC}
   & \Gamma_{C^\eta_{00'}C^\eta_{00'}}(\bm{\omega})= 
    g^4\bigg[\calC^\phi(\bm{\omega})  \nonumber \\  & \quad +
         \left(1+N\left({\left(\taucon\right)}^2 +{\left(\taudiv\right)}^2\right)\right)  
        \Cphi(\omega_1)\Cphi(\omega_2)
    \bigg]
\end{align}
to first order in $1/N$.

We next evaluate $\Gamma_{F_{00'}F_{00'}}(\bm{\omega})$. Using the definition of
$F_{00'}$, we have
\begin{align}
    \Gamma_{F_{00'}F_{00'}}(\bm{\omega})
    &=N\Bigg[
    \sum_{ijkl}
    \left\langle
        J_{0i}J_{j0'}J_{0k}J_{l0'}
    \right\rangle_J
    \chiphi_{ij}(\omega_1)\chiphi_{kl}(\omega_2) \nonumber \\ &\qquad
    +\frac{1}{2\pi}\left\langle J_{00'}^2\right\rangle_J
    \Bigg].
\end{align}
Wick's theorem gives
\begin{equation}
    \Gamma_{F_{00'}F_{00'}}(\bm{\omega})
    =
    \frac{g^2}{2\pi}+R_1+R_2+R_3,
\end{equation}
where
\begin{align}
    R_1&:=N\sum_{ijkl}
    \llangle J_{0i}J_{j0'}\rrangle
    \llangle J_{0k}J_{l0'}\rrangle
    \chiphi_{ij}(\omega_1)\chiphi_{kl}(\omega_2),\\
    R_2&:=N\sum_{ijkl}
    \llangle J_{0i}J_{0k}\rrangle
    \llangle J_{j0'}J_{l0'}\rrangle
    \chiphi_{ij}(\omega_1)\chiphi_{kl}(\omega_2),\\
    R_3&:=N\sum_{ijkl}
    \llangle J_{0i}J_{l0'}\rrangle
    \llangle J_{j0'}J_{0k}\rrangle
    \chiphi_{ij}(\omega_1)\chiphi_{kl}(\omega_2).
\end{align}
For the first contraction, only the chain cumulant contributes:
\begin{align}
    R_1
    &=N\sum_{ik}
    \llangle J_{0i}J_{i0'}\rrangle
    \llangle J_{0k}J_{k0'}\rrangle
    \chiphi_{ii}(\omega_1)\chiphi_{kk}(\omega_2)
    \nonumber\\
    &= N{\left(\tauchn\right)}^2 g^4
    \chiphi(\omega_1)\chiphi(\omega_2)
    \sim \frac{1}{N}.
\end{align}

For the second contraction, we decompose it as we did for $T_2$:
\begin{align}
    R_2=R_2^{(\delta\delta)}+R_2^{(\mathrm{mix})}+R_2^{(\tau\tau)},
\end{align}
where
\begin{subequations}
\begin{align}
    R_2^{(\delta\delta)}
    &:=
    \frac{g^4}{N}\sum_{ij}
    \chiphi_{ij}(\omega_1)\chiphi_{ij}(\omega_2),
    \\
    R_2^{(\mathrm{mix})}
    &:=\frac{g^4}{N}\sum_{ijkl}
    \left[
        \taudiv\,\delta_{ik}(1-\delta_{jl})
        + \taucon\,(1-\delta_{ik})\delta_{jl}
    \right]\nonumber \\ &\qquad \times
    \chiphi_{ij}(\omega_1)\chiphi_{kl}(\omega_2),
    \\
    R_2^{(\tau\tau)}
    &:=
    \frac{\taucon\taudiv\,g^4}{N}\sum_{ijkl}
    (1-\delta_{ik})(1-\delta_{jl})
    \chiphi_{ij}(\omega_1)\chiphi_{kl}(\omega_2).
\end{align}
\end{subequations}
The diagonal-diagonal piece is
\begin{align}
    R_2^{(\delta\delta)}
    &=
    \frac{g^4}{N}\sum_{ij}
    \chiphi_{ij}(\omega_1)\chiphi_{ij}(\omega_2)
    \nonumber\\
    &=
    g^4\left[
        \chiphi(\omega_1)\chiphi(\omega_2)
        + N\left\langle
            \chiphi_{00'}(\omega_1)\chiphi_{00'}(\omega_2)
        \right\rangle_J
    \right].
\end{align}

For the mixed piece, one of the two response factors is forced onto a diagonal
index pair while the other remains off-diagonal. Thus
\begin{align}
    R_2^{(\mathrm{mix})}
    &=
    \frac{g^4}{N}\Bigg[
        \taudiv\sum_{ijl}
        (1-\delta_{jl})
        \chiphi_{ij}(\omega_1)\chiphi_{il}(\omega_2)
        \nonumber\\
        &\qquad\qquad
        +
        \taucon\sum_{ijk}
        (1-\delta_{ik})
        \chiphi_{ij}(\omega_1)\chiphi_{kj}(\omega_2)
    \Bigg]
    \nonumber\\
    &\sim \frac{1}{N^{3/2}}.
\end{align}
The largest contribution comes from terms containing one diagonal and one
off-diagonal response, e.g.
\begin{equation}
    \sum_{i\neq k}
    \chiphi_{ii}(\omega_1)\chiphi_{ki}(\omega_2)
    \sim \sqrt{N^2}\,N^{-1/2}=\sqrt{N},
\end{equation}
while terms with two off-diagonal responses are of smaller order. Since
$N\taucon\sim 1$ and $N\taudiv\sim 1$, the mixed piece is subleading.

For the $\tau\tau$ piece, both factors come from the off-diagonal part of the
connectivity covariance:
\begin{align}
    R_2^{(\tau\tau)}
    &=
    \frac{\taucon\taudiv\,g^4}{N}\sum_{ijkl}
    (1-\delta_{ik})(1-\delta_{jl})
    \chiphi_{ij}(\omega_1)\chiphi_{kl}(\omega_2)
    \nonumber\\
    &=
    \frac{\taucon\taudiv\,g^4}{N}
    \left[
        N^2 \chiphi(\omega_1)\chiphi(\omega_2)
        + \calO(N)
    \right]
    \nonumber\\
    &=
    N\taucon\taudiv\,g^4
    \chiphi(\omega_1)\chiphi(\omega_2)
    \sim \frac{1}{N}.
\end{align}
Collecting the three pieces, we obtain
\begin{align}
    R_2
    &=g^4\bigg[
        \left(1+N\taucon\taudiv\right)
        \chiphi(\omega_1)\chiphi(\omega_2)
        \nonumber \\ 
        & \qquad+ N\left\langle
            \chiphi_{00'}(\omega_1)\chiphi_{00'}(\omega_2)
        \right\rangle_J
    \bigg]
    + \calO\left(N^{-3/2}\right).
\end{align}
For the third contraction, again only the chain cumulant contributes:
\begin{equation}
    R_3
    =\frac{{\left(\tauchn\right)}^2 g^4}{N}\sum_{ij}
    \chiphi_{ij}(\omega_1)\chiphi_{ji}(\omega_2).
\end{equation}
The diagonal part of the sum is $N\chiphi(\omega_1)\chiphi(\omega_2)$, while the
off-diagonal part contains two $\calO(N^{-1/2})$ responses and is at most
$\calO(N)$. Hence $R_3\sim N^{-2}$ is negligible to first order in $1/N$.

Collecting the three contractions, we obtain
\begin{widetext}
\begin{equation}
    \Gamma_{F_{00'}F_{00'}}(\bm{\omega})=
    g^4\left[
        \left(1+N\left({\left(\tauchn\right)}^2+\taucon\taudiv\right)\right)
        \chiphi(\omega_1)\chiphi(\omega_2)
        +N\left\langle
            \chiphi_{00'}(\omega_1)\chiphi_{00'}(\omega_2)
        \right\rangle_J
    \right]
    +\frac{g^2}{2\pi}.
\end{equation}
Using Eq.~\eqref{eq:CrossResponse}, we have
\begin{equation}
    \Gamma_{F_{00'}F_{00'}}(\bm{\omega})
    =\frac{g^2}{2\pi} \times 
     \frac{
        1+2\pi g^2\left(1+N\left({\left(\tauchn\right)}^2+\taucon\taudiv\right)\right)
        \chiphi(\omega_1)\chiphi(\omega_2)
    }{
        1-4\pi^2g^4{\chiphi(\omega_1)}^2{\chiphi(\omega_2)}^2
    }.
\end{equation}
An analogous computation gives the third cavity correlator function:
\begin{equation}
    \Gamma_{F^*_{00'}F_{0'0}}(\bm{\omega})
    =
    \frac{g^2}{2\pi} \times 
    \frac{
        \taurec
        +2\pi g^2\left(
            {\left(\taurec\right)}^2+2N{\left(\tauchn\right)}^2
        \right)
        {\chiphi(\omega_1)}^*\chiphi(\omega_2)
    }{
        1-4\pi^2{\left(\taurec\right)}^2g^4
        {\left({\chiphi(\omega_1)}^*\right)}^2{\chiphi(\omega_2)}^2
    }.
\end{equation}

Finally, using Wick's theorem and the same scaling analysis above, we obtain to first
order in $1/N$
\begin{equation}
    \Gamma_{F^*_{00'}C^\eta_{00'}}(\bm{\omega})
    =  \left(
        \taurec+N\tauchn\left(\taucon+\taudiv\right)
    \right)g^4{\chiphi(\omega_1)}^*\Cphi(\omega_2) 
 +\taurec g^4 N \left\langle
        {\chiphi_{00'}(\omega_1)}^* C^\phi_{00'}(\omega_2)
    \right\rangle_J.
\end{equation}
Using Eqs.~\eqref{eq:CrossResponse} and~\eqref{eq:C00phi}, we find
\begin{align}
    N\left\langle
        {\chiphi_{00'}(\omega_1)}^* C^\phi_{00'}(\omega_2)
    \right\rangle_J
    &=
    4\pi^2 {\left({\chiphi(\omega_1)}^*\right)}^2 \Bigg[
        \left|\chiphi(\omega_2)\right|^2
        \Gamma_{F^*_{00'}C^\eta_{00'}}(\bm{\omega})
        \nonumber\\
        &\qquad\qquad
        +
        \Cphi(\omega_2)\left(
            {\chiphi(\omega_2)}^*
            \Gamma_{F_{00'}F_{00'}}^*(\bm{\omega})
            +
            \chiphi(\omega_2)
            \Gamma_{F^*_{00'}F_{0'0}}(\bm{\omega})
        \right)
    \Bigg].
\end{align}
Substituting this back and solving for
$\Gamma_{F^*_{00'}C^\eta_{00'}}(\bm{\omega})$ gives
\begin{align}
    \Gamma_{F^*_{00'}C^\eta_{00'}}(\bm{\omega})
    &=
    \frac{g^4 {\chiphi(\omega_1)}^* \Cphi(\omega_2)}{
        1-4\pi^2\taurec g^4
        {\left({\chiphi(\omega_1)}^*\right)}^2
        \left|\chiphi(\omega_2)\right|^2
    } \nonumber\\
    &\quad \times
    \Big[
        \taurec+N\tauchn\left(\taucon+\taudiv\right)
        +4\pi^2 \taurec {\chiphi(\omega_1)}^*
        \left(
            {\chiphi(\omega_2)}^*
            \Gamma_{F_{00'}F_{00'}}^*(\bm{\omega})
            +\chiphi(\omega_2)
            \Gamma_{F^*_{00'}F_{0'0}}(\bm{\omega})
        \right)
    \Big].
\end{align}
Since the first cavity correlator function~\eqref{eq:GammaCC} contains the four-point function, we
substitute it back into Eq.~\eqref{eq:Cphi4-preGamma} and solve for $\calC^\phi(\bm{\omega})$.
This gives the final expression
\begin{align}\label{eq:FourPointCorr}
    \calC^\phi(\bm{\omega})
    &= \frac{4\pi^2}{
        1-4\pi^2 g^4
        \left|\chiphi(\omega_1)\chiphi(\omega_2)\right|^2
    }
    \nonumber\\
    &\quad \times
    \Bigg[g^4\left(
            1+N\left(
                {\left(\taucon\right)}^2+{\left(\taudiv\right)}^2
            \right)
        \right)
        \Cphi(\omega_1)\Cphi(\omega_2)
        \left|\chiphi(\omega_1)\chiphi(\omega_2)\right|^2
        +2 \operatorname{Re}H(\bm{\omega})\Bigg],
\end{align}
where
\begin{align}
    H(\bm{\omega})
    =&
    \Cphi(\omega_1)\Cphi(\omega_2)\left(
        \chiphi(\omega_1)\chiphi(\omega_2)
        \Gamma_{F_{00'}F_{00'}}(\bm{\omega})
        +{\chiphi(\omega_1)}^*\chiphi(\omega_2)
        \Gamma_{F^*_{00'}F_{0'0}}(\bm{\omega})
    \right)
    \nonumber\\
    &+
    \left|\chiphi(\omega_1)\right|^2
    {\chiphi(\omega_2)}^*\Cphi(\omega_2)
    \Gamma_{C^\eta_{00'}F^*_{00'}}(\bm{\omega})+
    \left|\chiphi(\omega_2)\right|^2
    {\chiphi(\omega_1)}^*\Cphi(\omega_1)
    \Gamma_{F^*_{00'}C^\eta_{00'}}(\bm{\omega}).
\end{align}

\end{widetext}

\section{Positivity of the three-neuron motif correction to $\calC^\phi(0,0)$}
\label{app:CorrectionNonnegative}

The correction to the zero-lag four-point correlator is obtained by integrating
$\Delta\calC^\phi(\bm{\omega})$ over the full frequency plane. We show that,
although $\Delta\calC^\phi(\bm{\omega})$ need not be nonnegative pointwise,
its integral over frequency is nonnegative. Let
\begin{equation}
 z(\omega)=\sqrt{2\pi} g \chiphi(\omega),
\end{equation}
so that
\begin{equation} \label{eq:X12Y12toz}
X_{12}=z(\omega_1) z(\omega_2), \qquad
Y_{12}=z(\omega_1)^*z(\omega_2).
\end{equation}
If $\left|X_{12}\right|<1$
(Fig.~\ref{fig:ThreeNeuronX12Max}) for all frequencies, then
\begin{equation} \label{eq:X12Expansion}
\frac{1}{1-\left|X_{12}\right|^2}
=\sum_{m=0}^\infty \left|X_{12}\right|^{2m},\qquad
\frac{X_{12}^2}{1-X_{12}^2}
=\sum_{n=1}^\infty X_{12}^{2n}.
\end{equation}
Since the time-domain response $\chiphi(t)$ is real,
$z(\omega)$ satisfies the symmetry $z(\omega)^*=z(-\omega)$.

Substituting Eqs.~\eqref{eq:X12Y12toz} and \eqref{eq:X12Expansion}
into the correction $\Delta\calC^\phi(\bm{\omega})$
[Eq.~\eqref{eq:ThreeNeuronCorrection}] and integrating over all frequencies,
we obtain
\begin{align} \label{eq:IntegratedDeltaCphi}
&\iint \diff \omega_1 \diff \omega_2\,
\Delta \calC^\phi(\bm{\omega})
 =N\sum_{m=0}^\infty
\bigg(\beta_1 M_m^2 \nonumber \\
&\qquad +4\beta_2 M_m L_{m0}
+2\beta_3 L_{m0}^2+2\beta_4 \sum_{n=1}^\infty L_{mn}^2\bigg),
\end{align}
where
\begin{subequations}
\begin{align}
M_m&=\int \diff\omega\, \Cphi(\omega) |z(\omega)|^{2(m+1)},\\
L_{mn}&=\int \diff \omega\, \Cphi(\omega) |z(\omega)|^{2m}z^{2(n+1)}(\omega).
\end{align}
\end{subequations}
The conjugate symmetry $z(\omega)^*=z(-\omega)$, together with the evenness
of $\Cphi(\omega)$, guarantees that $L_{mn}$ is real. The coefficients are
\begin{subequations}
\begin{align}
\beta_1 &= \left(\taucon\right)^2 + \left(\taudiv\right)^2,\\
\beta_2 &= \tauchn\left(\taucon+\taudiv\right),\\
\beta_3 &= 3\left(\tauchn\right)^2+\taucon\taudiv,\\
\beta_4 &= \left(\tauchn\right)^2+\taucon\taudiv.
\end{align}
\end{subequations}

The first three terms in Eq.~\eqref{eq:IntegratedDeltaCphi} form a quadratic
form in $M_m$ and $L_{m0}$ with matrix
\begin{equation}
W =\begin{pmatrix}
 \beta_1 & 2\beta_2 \\
 2\beta_2 & 2\beta_3
\end{pmatrix}.
\end{equation}
Its determinant is
\begin{align}
\det W
&=2\left(\taucon\taudiv-\left(\tauchn\right)^2\right)\left(\left(\taucon\right)^2+\left(\taudiv\right)^2\right) \nonumber \\ &\qquad+4\left(\tauchn\right)^2(\taucon-\taudiv)^2
\nonumber \\
&\ge 0,
\end{align}
where we used the constraint on motif correlations,
$\taucon\taudiv\ge \left(\tauchn\right)^2$ (see Appendix~\ref{app:NetworkGeneration}).
Moreover, $\beta_1\ge 0$, so $W$ is positive semi-definite. The remaining
term in Eq.~\eqref{eq:IntegratedDeltaCphi} is also non-negative because
$\beta_4=\left(\tauchn\right)^2+\taucon\taudiv\ge 0$.
Therefore, the three-neuron motif correction to the zero-lag four-point correlator 
\begin{equation}
\Delta\calC^\phi(0,0)
=\frac{1}{2\pi}
\iint \diff \omega_1 \diff \omega_2\,
\Delta\calC^\phi(\bm{\omega}) \ge 0.
\end{equation}

\section{Network generation} \label{app:NetworkGeneration}
We use the construction described in Refs.~\onlinecite{RN25,RN316,RN338} to generate the networks with motif correlations:
\begin{equation}
J_{ij}=\frac{J_0}{N}+\frac{g}{\sqrt{N}}
\left(\upsilon_i+\mu_j+\nu_{ij}\right).
\label{eq:J_construction_general}
\end{equation}
Here $\upsilon_i$, $\mu_j$, and $\nu_{ij}$ are zero-mean Gaussian random
variables with covariance matrices
\begin{equation}
\begin{pmatrix}
\left\langle \upsilon_i^2 \right\rangle & \langle \upsilon_i \mu_i\rangle \\
\langle \upsilon_i \mu_i\rangle & \left\langle \mu_i^2 \right\rangle
\end{pmatrix}
=\begin{pmatrix}
\taucon & \tauchn \\
\tauchn & \taudiv
\end{pmatrix},
\end{equation}
and
\begin{equation}\label{eq:GenerationABMatrix}
\begin{pmatrix}
\left\langle \nu_{ij}^2 \right\rangle & \langle \nu_{ij}\nu_{ji}\rangle \\
\langle \nu_{ij}\nu_{ji}\rangle & \left\langle \nu_{ji}^2\right\rangle
\end{pmatrix}
= \begin{pmatrix}
A & B \\
B & A
\end{pmatrix}.
\end{equation}
The covariance matrices must be positive semi-definite, which requires
\begin{equation}
\taucon \ge 0, \quad \taudiv \ge 0, \quad
\left|\tauchn\right|\le \sqrt{\taucon\,\taudiv},\quad |B| \le A.
\end{equation}
More general Gaussian ensembles can admit negative convergent or divergent correlations of order $1/N$, which are not excluded by the derivation in Appendix~\ref{app:DMFT} but are not sampled here.

In the boundary case $\left|\tauchn\right|=\sqrt{\taucon\taudiv}$, the
covariance of $(\upsilon_i,\mu_i)$ is rank one, so the two variables are
perfectly correlated. For the parameter range in which
\begin{equation}
1-2\taucon-2\taudiv-|\taurec|\ge 0,
\end{equation} this degenerate case can be sampled
using the equivalent explicit parametrization of Ref.~\onlinecite{RN271},
\begin{widetext}
\begin{align}
J_{ij} &= \frac{J_0}{N}
+ \frac{g}{\sqrt{N}}
\bigg(
\operatorname{sgn}\left(\tauchn\right)\sqrt{\taucon}\,\varepsilon_i
+\sqrt{\taudiv}\,\varepsilon_j
-\operatorname{sgn}\left(\tauchn\right)\sqrt{\taucon}\,\omega_{ij}
+\sqrt{\taudiv}\,\omega_{ji}   \nonumber \\ 
 &  \qquad +\operatorname{sgn}\left(\taurec\right)\sqrt{\frac{\left|\taurec\right|}{2}}\,\eta_{ij}
+\sqrt{\frac{|\taurec|}{2}}\,\eta_{ji}
+\sqrt{1-2\taucon-2\taudiv-|\taurec|}\,\zeta_{ij}
\bigg),\label{eq:J_construction_simple}
\end{align}
\end{widetext}
where $\varepsilon_i$, $\omega_{ij}$, $\eta_{ij}$, and $\zeta_{ij}$ are independent
standard Gaussian random variables.

\section{Mapping chain-motif networks to low-rank networks} \label{app:MappingToLowRank}
Mastrogiuseppe and Ostojic \cite{RN320} studied rate networks whose connectivity is the sum of an i.i.d. Gaussian random matrix and a low-rank component,
\begin{equation}
J_{ij}=G_{ij}+\frac{1}{N}\sum_{a=1}^r m_i^{(a)}n_j^{(a)} ,
\end{equation}
where the low-rank vectors are assumed to be independent of the random bulk. A concurrent study by Zhang et al. developed a related low-rank mean-field construction for multi-population nonlinear rate networks with heterogeneous chain-motif correlations \cite{RN431}. In this section, we consider the corresponding single-population $B=0$ limit and show that the low-rank mean-field theory yields the same DMFT equation.

Using the decomposition \eqref{eq:J_construction_general}--\eqref{eq:GenerationABMatrix}, we can write the motif network with $B=0$ as 
\begin{equation}
J=G+\frac{1}{N}\left(\bm{m} \bm{1}^\mathrm{T} + \bm{1}\bm{n}^{\mathrm{T}}\right),
\end{equation}
where $G$ is an i.i.d. Gaussian matrix with zero mean and variance $Ag^2/N$. The two rank-one terms contain two correlated Gaussian random vectors with statistics given by
\begin{subequations}
\begin{align}
    &\langle m_i\rangle = J_0,  \qquad \langle n_i \rangle = 0,  \\
    &\llangle m_i^2 \rrangle= \taucon g^2 N, \qquad  \llangle n_i^2 \rrangle= \taudiv g^2 N, \\
    &\llangle m_i n_i \rrangle=\tauchn g^2 N.
\end{align}
\end{subequations}
Substituting this connectivity into the rate dynamics gives
\begin{equation}
\dot{x}_i(t)=-x_i(t)
+\sum_j G_{ij}\phi(x_j(t))+m_i\mu^\phi(t)+\kappa(t),
\end{equation}
where
\begin{subequations}
\begin{align}
\mu^\phi(t)&=\frac{1}{N}\sum_j\phi(x_j(t)), \\
\kappa(t)&=\frac{1}{N}\sum_j n_j\phi(x_j(t)).
\end{align}
\end{subequations}
We separate the mean component of the connectivity vector, $m_i=J_0+\delta m_i$,
so that $\delta m_i$ is a quenched zero-mean Gaussian variable correlated with $n_i$. In the large-$N$ limit, the random bulk input becomes a centered Gaussian
process $\eta(t)$, independent of $\delta m$. The effective single-site
process for a neuron with quenched variable
$\delta m=\sqrt{g^2N \taucon}z$ is
\begin{equation}\label{eq:LowRankMF}
\dot{x}_z(t)=-x_z(t) +J_0\mu^\phi(t) + \sqrt{g^2N \taucon}z\mu^\phi(t) +\kappa(t)+\eta(t),
\end{equation}
where $z$ is a standard Gaussian random variable with $\llangle nz \rrangle=g\tauchn\sqrt{N/\taucon}$. The zero-mean Gaussian process $\eta$ satisfies $\left\langle \eta(t) \eta(t')\right\rangle=Ag^2 C^\phi\left(t,t'\right)$. The mean and autocorrelations are self-consistent
\begin{subequations}
\begin{align}
\mu^\phi(t)&=\int \calD z\, \left\langle \phi\left(x_z(t)\right) \right\rangle_\eta,\\
\Cx(t,t') &= \int \calD z \, \left\langle x_z(t) x_z(t') \right\rangle_\eta,\\
\Cphi(t,t') &= \int \calD z\, \left\langle \phi(x_z(t))\phi\left(x_z(t')\right) \right\rangle_\eta.
\end{align}
\end{subequations}

To close the mean-field equations, we express $\kappa(t)$ in terms of the mean activity and response function. Define
\begin{equation}
F_t^\phi(z)=\left\langle \phi(x_z(t)) \right\rangle_\eta,
\end{equation}
in the large-$N$ limit, the empirical average in
$\kappa(t)$ is replaced by an average over 
$(n,z)$:
\begin{equation}
\kappa(t)=\left\langle n F_t^\phi(z) \right\rangle_{n,z}.
\end{equation}
Since the joint distribution of $(n,z)$ is Gaussian, Stein's lemma gives
\begin{align}
\left\langle n F_t^\phi(z) \right\rangle_{n,z}
&=\llangle nz \rrangle\left\langle\frac{\partial F_t^\phi(z)}{\partial z}\right\rangle_z\nonumber \\ &=g\tauchn\sqrt{\frac{N}{\taucon}}\left\langle\frac{\partial \phi_z(t)}{\partial z}
\right\rangle_{\eta,z}. 
\end{align}
Interpreting $ h(t)= \delta m \mu^\phi(t)$ as an additive probe, the response function can be defined as 
\begin{equation}
\chiphi(t,t')= \left\langle\frac{\delta \phi(x_z(t))}{\delta h(t')}\right\rangle_{\eta,z}.
\end{equation}
Then
\begin{equation}
\left\langle
\frac{\partial \phi_z(t)}{\partial z}
\right\rangle_{\eta, z}=g \sqrt{N\taucon}\int_{-\infty}^{t}\chiphi(t,t')\mu^\phi(t') \, \diff t',
\end{equation}
and we identify 
\begin{equation}
\kappa(t)=g^2 N\tauchn \int_{-\infty}^t \chiphi(t,t') \langle \phi(t') \rangle_{\eta,z} \,\diff t'.
\end{equation}
Absorbing the static $z$ term into the Gaussian process $\eta(t)$, we recover the DMFT equation in the main text [Eq.~\eqref{eq:DMFTPosChain}].

\section{Numerical algorithm for the DMFT equation}

\subsection{Algorithm for $B=0$} \label{app:AlgorithmB0}

When the retarded self-interaction term in the DMFT equation vanishes, the stationary distribution of $x$ is Gaussian. At each iteration, the nonlinear autocorrelation $C^{\phi}(\tau)$ can be evaluated from $\langle x \rangle$ and $\Cx(\tau)$ by a two-dimensional Gaussian integral, using the joint normality of $x(t)$ and $x(t+\tau)$.
The updated power spectrum is then obtained from
\begin{equation}
S^{x}(\omega)=\frac{A g^{2}S^{\phi}(\omega)}{1+\omega^2}.
\end{equation}
The susceptibility is given analytically by
\begin{equation}
\chi^{\phi}(\omega)
= \frac{\langle\phi'\rangle}{\sqrt{2\pi}\,(1+\im\omega)} .
\end{equation}

For the non-stationary trajectories shown in Fig.~\ref{fig:NegChnPD}\&\ref{fig:Transient}, we used the same Gaussian closure, but without assuming time-translation invariance. The time-dependent mean and covariance of $x$ were propagated directly. At each step, the corresponding mean activity, nonlinear covariance, and local gain were evaluated by Gaussian quadrature. These quantities determine the covariance of the effective Gaussian input, which was then filtered through the linear single-neuron dynamics to update the covariance of $x$. The mean activity was updated separately. In this way, the transient DMFT dynamics can be solved deterministically, without sampling effective trajectories.

\subsection{General algorithm for $B\neq 0$}

When the self-interaction term is nonzero, we must integrate the single-site DMFT equation in the time domain. There are two different approaches for obtaining the response functions \cite{RN293}: a) using Novikov's theorem (e.g., \cite{RN278}), b) using a time-domain integration (e.g., \cite{RN290}). 

According to Novikov's theorem \cite{RN278, RN290}, in the stationary state we have  
\begin{equation} 
\chiphi(\omega)=C^{\eta \phi}(\omega)/C^{\eta}(\omega) 
\end{equation}  
in the frequency domain, where, in the time domain, we define $C^{\eta \phi}(\tau) = \left\langle \eta(t) \phi(t+\tau) \right\rangle_t$ and $C^\eta(\tau) = \langle\eta(t)\eta(t+\tau)\rangle_t$. 

The alternative route is to take the functional derivative of both sides of the DMFT equation \eqref{eq:GenericDMFTEq} with respect to the external drive $\eta(t')$ to obtain, to leading order in $1/N$, 
\begin{align} 
\frac{\partial}{\partial t}\chix_\eta(t,t')=&-\chix_\eta(t,t')  + g^2 B \int_{t'}^t \chiphi(t,s) \chiphi_\eta(s,t') \, \diff s \nonumber\\ &\qquad + \delta(t-t'), 
\end{align}
where $\chix_\eta(t,t')=\delta x(t)/\delta \eta(t')$ and $\chiphi_\eta(t,t')=\phi'(x(t))\chix_\eta(t,t') $ is the response along a single realization of the effective stochastic process. The ensemble-averaged response $ \chiphi(t,t') = \left\langle \phi'(t) \chix_\eta(t,t')  \right\rangle_\eta$ can be computed by averaging across simulated single-site trajectories. For $B\neq 0$, one cannot in general factor this as $\chiphi(t,t') = \left\langle \phi'(t) \right\rangle \left\langle \chix_\eta(t,t') \right\rangle$ because $\chix_\eta\left(t,t'\right)$ depends on the same realization of the process as $\phi'(t)$. 

\begin{figure}[ht]
\includegraphics{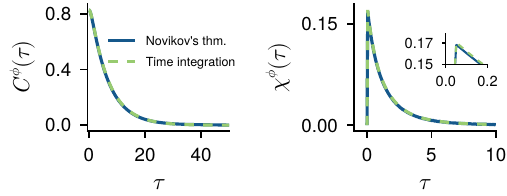}
\caption{\label{fig:NovikovVSTimeInteg}Comparison of the Novikov-theorem-based algorithm and the time integration algorithm. The inset shows that the time integration method can slightly overestimate $\chiphi$ for integration step size $\Delta t=0.05$. Parameters: $J_0=0.0$, $\taurec=0.2$, $g=4.0$, $\tauchn=\taucon=\taudiv=0.0$.}
\end{figure}

For stationary statistics, we use Novikov's theorem to obtain the response function (described in Algorithms~\ref{alg:dmft_main}\&\ref{alg:dmft_subs}), because it gives a more accurate estimate of the PR dimension with the same integration time step $\Delta t$. The difference between responses $\chiphi$ obtained from the two algorithms is visually small (Fig.~\ref{fig:NovikovVSTimeInteg}), but it is large enough to lead to a large deviation in the PR dimension. Nonetheless, using a smaller $\Delta t$ for the time integration method can also achieve higher accuracy. 

\begin{algorithm}[htp]
\caption{General stationary DMFT solver.\label{alg:dmft_main}}
\KwData{Coupling $g, J_0$; size $N$; motifs $(\tauchn,\taurec,\taucon,\taudiv)$; activation $\phi$; step $\Delta t$, horizon $T$; trajectories $N_{\text{traj}}$; damping factors $\lambda_C, \lambda_R$; tolerance $\varepsilon$}
\KwResult{Autocorrelations $C^x, C^\phi$, response $\chiphi$, means $\mu^x, \mu^\phi$}

$A \leftarrow 1 - \taucon - \taudiv,\quad B \leftarrow \taurec - 2\tauchn$\;
Initialize $C^x, C^\phi$ (decaying guess), $\chiphi, \mu^x, \mu^\phi$\;
$t_n = n\Delta t\;(n=0,\dots,N_t-1)$\;

\Repeat{$\delta < \varepsilon$}{
  $\{\eta_i(t_n)\} \leftarrow$ \textsc{SampleNoise}$(C^\phi, \mu^\phi)$\;
  $\{x_i(t_n)\} \leftarrow$ \textsc{IntegrateTrajectories}$(\{\eta_i\}, \chiphi, \mu^x)$\;
  $\mu^x_{\text{new}},\, \mu^\phi_{\text{new}},\, C^x_{\text{new}},\, C^\phi_{\text{new}} \leftarrow$ \textsc{MeasureObservables}$(\{x_i\})$\;
  $\chiphi_{\text{new}} \leftarrow$ \textsc{ComputeResponse}$(\{x_i\}, \{\eta_i\}, C^\phi, \mu^\phi)$\;
  $\delta \leftarrow \|C^\phi_{\text{new}} - C^\phi\| / \|C^\phi\|$\;
  $\mu^x \leftarrow \lambda_C\,\mu^x + (1-\lambda_C)\,\mu^x_{\text{new}}$,\quad
  $\mu^\phi \leftarrow \lambda_C\,\mu^\phi + (1-\lambda_C)\,\mu^\phi_{\text{new}}$\;
  $C^x \leftarrow \lambda_C\,C^x + (1-\lambda_C)\,C^x_{\text{new}}$,\quad
  $C^\phi \leftarrow \lambda_C\,C^\phi + (1-\lambda_C)\,C^\phi_{\text{new}}$\;
  $\chiphi \leftarrow \lambda_R\,\chiphi + (1-\lambda_R)\,\chiphi_{\text{new}}$\;
}
\end{algorithm}

\begin{algorithm}[htp]
\caption{Subroutines for Algorithm~\ref{alg:dmft_main}.\label{alg:dmft_subs}}

\SetKwProg{Proc}{Procedure}{}{}

\Proc{\textsc{SampleNoise}$(C^\phi, \mu^\phi)$}{
  $S^\eta_k \leftarrow A g^2 \operatorname{Re}\left[\mathrm{FFT}[C^\phi]_k\right]$\;
  $S^\eta_0 \leftarrow S^\eta_0 + N g^2 \taucon (\mu^\phi)^2 N_t$\;
  Sample $\{\eta_i(t)\}_{i=1}^{N_{\text{traj}}}$ from the circular power spectrum $\max(S^\eta_k, 0)$\;
  \Return $\{\eta_i(t)\}$\;
}

\Proc{\textsc{IntegrateTrajectories}$(\{\eta_i\}, \chiphi, \mu^x)$}{
    \tcc{Circular padding for warm-up}
  $N_{\text{pad}} \leftarrow \lfloor N_t / 2.5 \rfloor,\quad N_{\text{tot}} \leftarrow N_t + N_{\text{pad}}$\;   
  $\eta_i^{\text{pad}}(t) \leftarrow \begin{cases} \eta_i(t + N_t - N_{\text{pad}}) & t = 0,\dots,N_{\text{pad}}-1 \\ \eta_i(t - N_{\text{pad}}) & t = N_{\text{pad}},\dots,N_{\text{tot}}-1 \end{cases}$\;
  $x_i(0) \sim \mathcal{N}\!\left(\mu^x,\, C^x(0) - (\mu^x)^2\right)$\;
  \For{$t = 0,\dots,N_{\mathrm{tot}}-1$}{
    $K_t \leftarrow \min(t+1, N_t)$\;
    \For{$i = 1$ \KwTo $N_{\mathrm{traj}}$}{
      $f_i \leftarrow \eta_i^{\text{pad}}(t)
      + g^2 B \Delta t \sum_{s=0}^{K_t-1} \chiphi(s)\,\phi(x_i(t-s))
      + N g^2\tauchn \Delta t \sum_{s=0}^{K_t-1} \chiphi(s)\,\langle \phi(x_j(t-s))\rangle_j
      + J_0\langle \phi(x_j(t))\rangle_j$\;
      $x_i(t{+}1) \leftarrow \mathrm{e}^{-\Delta t} x_i(t) + \left(1 - \mathrm{e}^{-\Delta t}\right) f_i$\;
    }
  }
  \Return $\{x_i(t)\}_{t=N_{\text{pad}}}^{N_{\text{tot}}-1}$\;
}

\Proc{\textsc{MeasureObservables}$(\{x_i\})$}{
  $\mu^x \leftarrow \langle x_i(t)\rangle_{i,t},\quad \mu^\phi \leftarrow \langle \phi(x_i(t))\rangle_{i,t}$\;
  $C^x(\tau) \leftarrow \langle x_i(t)\,x_i((t{+}\tau)\!\mod N_t)\rangle_{i,t}$\;
  $C^\phi(\tau) \leftarrow \langle \phi(x_i(t))\,\phi(x_i((t{+}\tau)\!\mod N_t))\rangle_{i,t}$\;
  \Return $\mu^x, \mu^\phi, C^x, C^\phi$\;
}

\Proc{\textsc{ComputeResponse}$(\{x_i\}, \{\eta_i\}, C^\phi, \mu^\phi)$}{
  $C^\eta(\tau) \leftarrow Ag^2 C^\phi(\tau) + Ng^2\taucon(\mu^\phi)^2$\;
  $S^{\eta\phi}_k \leftarrow \big\langle \mathrm{FFT}[\phi(x_i)]_k \cdot \overline{\mathrm{FFT}[\eta_i]_k}\big\rangle_i$\;
  $\chiphi \leftarrow \mathrm{IFFT}\left[S^{\eta\phi} / \mathrm{FFT}[C^\eta]\right]$\;
  \Return $\chiphi$\;
}
\end{algorithm}

\section{Details of numerical simulation}

All direct network simulations in this paper were computed using the Tsitouras 5/4 Runge--Kutta method \cite{Tsit5}, as implemented in DifferentialEquations.jl \cite{DifferentialEquationsjl}. 

\subsection{Filtering outlier-dominated networks in the isolated convergent/divergent case} \label{app:ConDivFilter}

\begin{figure}[ht]
\includegraphics{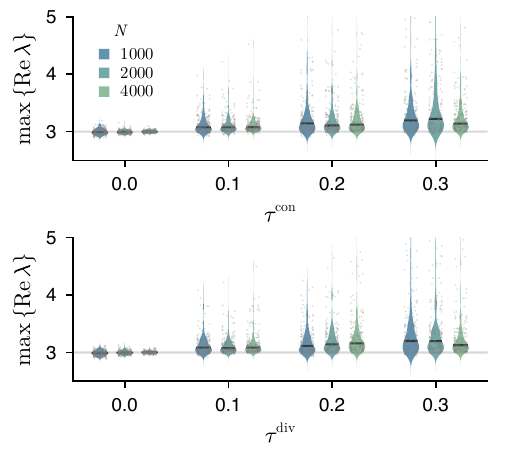}
\caption{\label{fig:ConDivOutlier}Distribution of the largest real part of the eigenspectrum for networks with convergent or divergent correlations. Top: only the convergent correlation $\taucon$ is nonzero. Bottom: only the divergent correlation $\taudiv$ is nonzero. Gray scatters show individual network realizations, and shaded violins indicate the corresponding distributions, each obtained from 200 network realizations for each combination of network size and motif strength. The gray horizontal line marks the theoretical bulk boundary, $g_{\mathrm{eff}}=3$. Other parameter: $J_0=0$.}
\end{figure}

For networks with isolated convergent and/or divergent correlations, random matrix theory predicts an eigenvalue bulk but no deterministic low-rank outlier \cite{RN271}. Nevertheless, individual network realizations can contain dominant outlier eigenvalues because of sample-to-sample fluctuations. These fluctuations do not vanish with system size under the $\calO(1)$ motif scaling, as illustrated by the broad tails in the distributions of the maximum real part of the eigenspectrum in Fig.~\ref{fig:ConDivOutlier}.

Such outlier-dominated realizations can produce strongly bimodal population activity, even though this behavior is not captured by the single-site mean-field theory. We therefore filter these realizations to focus on the unimodal HC phase. Specifically, we apply Hartigan's dip test \cite{diptest,hartigan1985dip} to the distribution of the population-averaged activity $\langle x(t)\rangle$ after discarding the initial transient. Networks with significant evidence for non-unimodality are excluded from the statistics. Example outlier-dominated networks and the corresponding increase in the fraction of non-unimodal realizations are shown in Fig.~\ref{fig:DipTest}.

Hartigan's dip test should not be applied when the effective parameter $B$ is non-negligible, because the self-interaction term can generate bimodal activity in the HC phase \cite{RN277,RN358}. In that case, bimodality is a genuine dynamical feature of the network rather than a signature of a realization-specific dominant outlier.

\subsection{Lyapunov exponents}

The Lyapunov exponents were computed by integrating the network dynamics
together with the variational equations and periodically reorthonormalizing
the tangent vectors using a QR decomposition. The exponents were estimated
from the time-averaged logarithms of the diagonal entries of the resulting
$R$ matrices \cite{RN399}. Because the leading CLV coincides with the first Gram–Schmidt vector, and we only need the leading direction, it suffices to track the first column of the $Q$ matrix \cite{RN261}. The reorthonormalization interval $\Delta t_{\mathrm{ONS}}$ and the averaging time $T_{\mathrm{LE}}$ were chosen adaptively according to the characteristic timescale of the network dynamics.
We used the effective self-interaction parameter $B$ as a guide, since larger
positive values of $B$ slow the relaxation and require longer averaging times.
For networks with only convergent or divergent correlations, where this
timescale is not modified through $B$, we used the reference values
$\Delta t_{\mathrm{ONS}}=0.1$ and $T_{\mathrm{LE}}=1000$. In all cases,
$\Delta t_{\mathrm{ONS}}$ and $T_{\mathrm{LE}}$ were bounded to avoid
excessively long tangent-vector evolution between QR steps or insufficient
sampling; specifically, we imposed $0.01\le\Delta t_{\mathrm{ONS}}\leq 0.25$ and
$100\le T_{\mathrm{LE}}\le 8000$.

Let $\gamma_i$ denote the Lyapunov exponents associated with the invariant
measure sampled by a trajectory after transients. For this measure, Ruelle's
inequality gives the upper bound
\begin{equation}
H \leq \sum_{\gamma_i>0} \gamma_i ,
\end{equation}
where $H$ is the KS entropy of the dynamics on the sampled
attractor \cite{RN164}. We therefore use this upper bound as an estimate of the KS entropy. Equality holds when Pesin's entropy formula applies, for example for smooth systems with an SRB measure on the corresponding attractor \cite{RN400}. The KY dimension \cite{RN401} is defined from the ordered Lyapunov spectrum $\gamma_1\geq \gamma_2\geq \cdots \geq \gamma_N$. Let
\begin{equation}
j=\max\left\{m:\sum_{i=1}^{m}\gamma_i\geq 0\right\}.
\end{equation}
If $j<N$, then
\begin{equation}
D_{\mathrm{KY}}
=
j+\frac{\sum_{i=1}^{j}\gamma_i}{|\gamma_{j+1}|}.
\end{equation}
If all partial sums are non-negative, $D_{\mathrm{KY}}=N$.

\subsection{Fixed-point search}\label{app:FPSearch}

In Sec.~\ref{sec:NegativeChain}, we searched for fixed points of the network
by brute force. Following the fixed-point search strategy of
Ref.~\onlinecite{RN350}, we initialized the solver from a large ensemble of random
initial conditions and solved $\dot{\bm{x}}=\bm{0}$ using the improved
Levenberg--Marquardt algorithm \cite{RN848} implemented in
NonlinearSolve.jl \cite{NonlinearSolvejl}. Specifically, each initial
condition was drawn from a hierarchical Gaussian ensemble:
$\mu\sim\mathcal{N}(0,1)$ and $x_i(0)\sim\mathcal{N}(\mu,200^2)$ independently across neurons. A converged solution was retained as a fixed-point candidate only if
$\|\dot{\bm{x}}\|/\sqrt{N}<10^{-10}$. The stability of each fixed point was
then determined from the Jacobian: a fixed point was classified as stable when the maximum real part of the Jacobian eigenvalues was smaller than $-10^{-6}$. Duplicate candidates were identified and removed.
As a convergence check for this sampling procedure, we monitored the cumulative
number of distinct fixed points as more initial conditions were searched,
stopping at the value $N_{\mathrm{IC}}=10^7$. The empirical
fixed-point complexity approached a plateau (Fig.~\ref{fig:FPbyIC}), providing
a practical guarantee that most fixed points accessible to this algorithm had
been found.

\onecolumngrid
\section{Supplementary figures}
\setcounter{figure}{0}
\renewcommand{\thefigure}{K\arabic{figure}}
\renewcommand{\theHfigure}{K.\arabic{figure}}

\begin{center}
\includegraphics{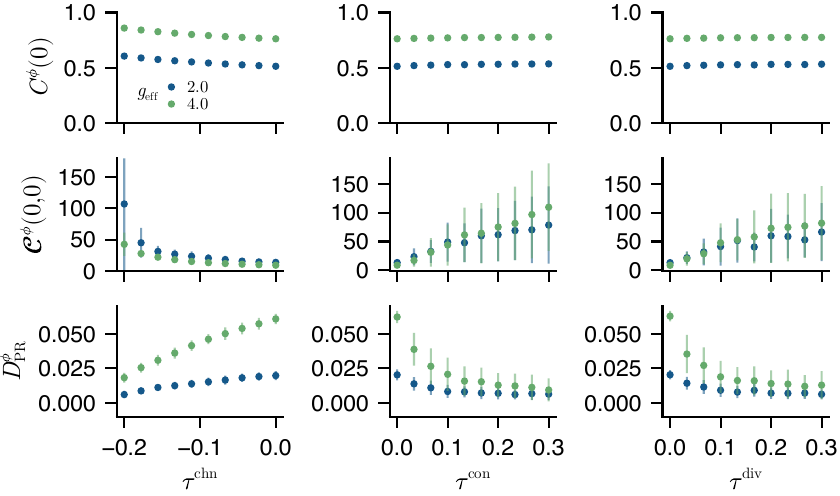}
\captionof{figure}{Numerical evaluation of the two-point correlation, four-point correlation, and participation ratio dimension across an extended range of motif correlations. This figure extends the numerical simulations presented in Fig.~\ref{fig:AllMotifsPRD}. Dots show the mean across realizations, and error bars show $\pm$SD of 128 realizations. Networks with non-unimodal activity are removed for the reason described in Appendix~\ref{app:ConDivFilter}.}
\label{fig:PRDFullRange}
\end{center}

\begin{center}
\includegraphics{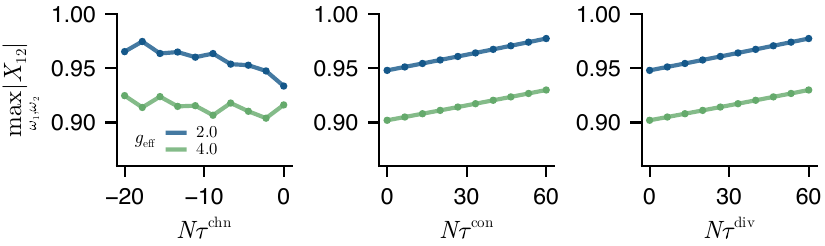}
\captionof{figure}{For the parameters in Fig.~\ref{fig:AllMotifsPRD}, we have $|X_{12}(\bm{\omega})|<1$ for networks with three-neuron motif correlations.}
\label{fig:ThreeNeuronX12Max}
\end{center}

\begin{center}
\includegraphics{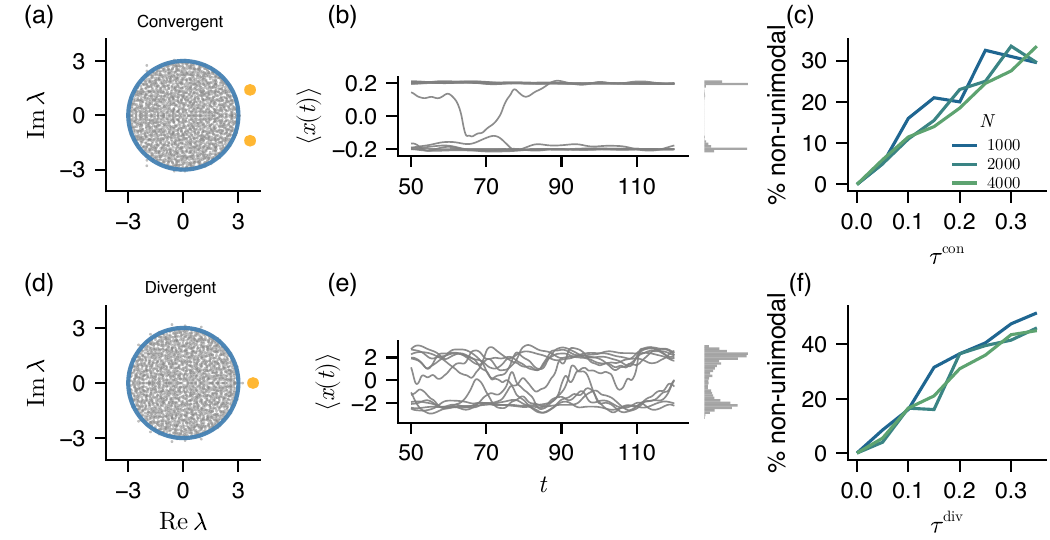}
\captionof{figure}{Large convergent and divergent motif correlations induce pronounced fluctuations in the eigenspectrum and may generate dominant low-rank modes. (a) Example eigenspectrum of a network with $\taucon=0.25$, with all other motif correlations set to zero. The eigenvalue with the largest real part is highlighted in orange, and the blue curve shows the theoretical bulk boundary. (b) Mean activity, $\langle x(t)\rangle$, for the connectivity realization shown in (a), with each trace initialized from a different initial condition. The histogram on the right shows the distribution of sampled activity values pooled across trajectories after discarding an initial transient of 50 time units. (c) Fraction of 200 sampled networks exhibiting non-unimodal activity, as in (b), as a function of $\taucon$. (d)--(f) Same as (a)--(c), but for networks with purely divergent correlations, i.e. $\taudiv=0.25$ and all other motif correlations set to zero. For the example networks, $N=1000$. For all panels, $J_0=0$, and $g_{\mathrm{eff}}=3$.}
\label{fig:DipTest}
\end{center}

\begin{center}
\includegraphics{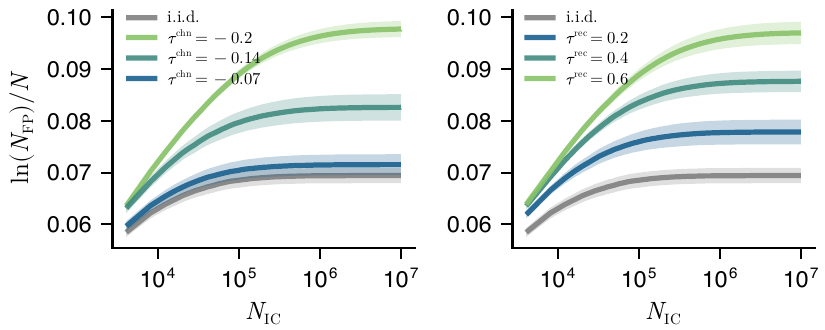}
\captionof{figure}{Empirical fixed-point complexity, $\ln(N_{\mathrm{FP}})/N$, as a function of the number of solver initial conditions, $N_{\mathrm{IC}}$, for networks with negative chain correlations (left) or positive reciprocal correlations (right). The estimate increases with $N_{\mathrm{IC}}$ and approaches a plateau. Parameters are the same as in \protect\refpanels{fig:GlassyPhase}{e,h}. Shaded bands represent $\pm$SEM.}
\label{fig:FPbyIC}
\end{center}

\twocolumngrid
\bibliography{MotifNets}
\end{document}